\documentclass[12pt,twoside,reqno,english]{paper}
\usepackage[T1]{fontenc}
\usepackage[latin9]{inputenc}
\usepackage{geometry}
\usepackage{babel}
\usepackage{calc}
\usepackage{amsmath}
\usepackage{amsthm}
\usepackage{stackrel}
\usepackage{graphicx}
\PassOptionsToPackage{normalem}{ulem}
\usepackage{ulem}
\usepackage[unicode=true,pdfusetitle,
 bookmarks=true,bookmarksnumbered=false,bookmarksopen=false,
 breaklinks=false,pdfborder={0 0 1},backref=false,colorlinks=false]
 {hyperref}

\makeatletter
\numberwithin{equation}{section}
\numberwithin{figure}{section}
\newcommand{\lyxaddress}[1]{
	\par {\raggedright #1
	\vspace{1.4em}
	\noindent\par}
}
\newenvironment{lyxlist}[1]
	{\begin{list}{}
		{\settowidth{\labelwidth}{#1}
		 \setlength{\leftmargin}{\labelwidth}
		 \addtolength{\leftmargin}{\labelsep}
		 }}
	{\end{list}}

\renewcommand\[{\begin{equation}}
\renewcommand\]{\end{equation}}

\makeatother

\begin{document}
\title{Membrane Models as a Means of Propulsion in General Relativity: Super-Luminal
Warp-Drive that Satisfies the Weak Energy Condition}
\author{Greg Huey}
\maketitle

\lyxaddress{Department of Mathematics\\
Sweet Briar College\\
Guion Hall, Sweet Briar, VA 24595 USA}
\begin{abstract}
Presented are toy-models for sub-luminal and super-luminal warp-drives
in 3+1 dimensions. The models are constructed in a chimeric manner
- as different bulk space-times separated by thin membranes. The membranes
contain perfect-fluid-like stress-energy tensors. The Israel junction
conditions relate this stress-energy to a jump in extrinsic curvature
across the brane, which in turn manifests as apparent acceleration
in the bulk space-times. The acceleration on either side of the brane
may be set individually by choice of model parameters. The Weak Energy
Condition~(WEC) is shown to be satisfied everywhere in both models.
Although the branes in these toy models are not compact, it is demonstrated
that super-luminal warp-drive is possible that satisfies the WEC.
Additionally, the nature of these models provides framework for speculation
on a mechanism for transition from sub-luminal to super-luminal warp.
It is shown that the difference in extrinsic curvature across a thin
membrane can yield a positive contribution to the Landau-Raychaudhuri
equation, thus providing a means to evade some super-luminal warp-drive
no-go theorems. Neither quantum effects nor stability of the models
is considered. 
\end{abstract}

\global\long\def\bbR{\mathbb{R}}%

\global\long\def\ccR{\mathcal{R}}%

\global\long\def\dlim{\operatorname{\underrightarrow{{\rm lim}}}}%

\global\long\def\Ker{\operatorname{\rm Ker}}%

\global\long\def\End{\operatorname{\rm End}}%

\global\long\def\myint#1#2#3{\int_{#1}^{#2}\sin#3dx}%

\section{Introduction and Motivation}

In 1994 Miguel Alcubierre presented a model for super-luminal warp-drive
based on classical General Relativity~\cite{bib:Alcub_1994}. While
locally a passenger follows a time-like world-path, from a global
perspective their journey is effectively space-like. This was accomplished
by envisioning a metric that would achieve the desired goal, and then
using the Einstein field equations to determine the necessary stress-energy
tensor. Since then, the topic has captured the interest of many researchers
- there have been a vast array of hurdles identified, and potential
solutions devised. The contributions have been too numerous for brief
complete list. Several researchers have identified challenges faced
by this model: a requirement for negative energy density (large quantities
of negative mass-energy), as well as horizons (with the accompanying
Hawking radiation posing a hazard) and super-luminal flows within
the warp-bubble (hence the critique ``need one to build one'') \cite{bib:Everett_1996,bib:Pfenning-Ford_19967,bib:Olum_1998,bib:Krasnikov_1998,bib:Low_1999,bib:Clark-Hiscock-Larson_1999,bib:VDBroeck_1999,bib:Natario_2002,bib:Lobo-Crawford_2003,bib:Lobo_2007,bib:White-WFM101_2013,bib:Alcub_Lobo_2017,bib:DeBenedictis-Iliji=000107_2018}.
However, several improvements and advancements have also since been
made. In 1999 Chris Van Den Broeck proposed modifications that reduce
the magnitude of the negative mass-energy\cite{bib:VDBroeck_1999}.
In 2002 Jose Natario proposed a volume-element preserving modification
to the Alcubierre warp-drive~\cite{bib:Natario_2002}. In 2018 Andrew
DeBenedictis and Saša Iliji\'{c} proposed a warp-drive model in the
context of Einstein--Cartan gravity~\cite{bib:DeBenedictis-Iliji=000107_2018}.
The spin-density in the source causes a contribution from the torsion
terms that can eliminate WEC violation for some ranges of model parameters.
In 2021 Alexey Bobrick and Gianni Martire further reduced the magnitude
of negative mass-energy required by Alcubierre's super-luminal drive
as well as proposed a model for a sub-luminal warp-drive that utilizes
only positive energy density~\cite{bib:bobrick_martire_subLum_WEC}.
From 2010 to 2023 several researchers have considered the effects
of altering the stress-energy tensor by utilizing various different
fluids: in 2010 Igor Smolyaninov - bi-anisotropic non-reciprocal magnetoelectric
metamaterials~\cite{bib:Smolyaninov_Metamaterials}; in 2020 Willie
Béatrix-Drouhet - exotic equations of state~\cite{bib:Drouhet_ExoticFluids};
in 2023 Gabriel Abellan, Nelson Bolivar and Ivaylo Vasilev - anisotropic
matter~\cite{bib:AbellanBolivarVasilev_AnisotropicMatter}.  Also
in 2021, Erik Lentz proposed a super-luminal warp-drive model that
does not violate the WEC. It utilizes gravitational soliton waves,
potentially sourced by strictly positive energy density classical
matter in the plasma state~\cite{bib:Lentz_soliton_SUPLUM_WEC}.

The propulsion models presented in this paper are constructed in a
chimeric manner by stitching together different bulk space-times
that are separated by a thin membrane (co-dimension~1 hypersurface)
that contains stress-energy. There is translational and rotational
symmetry in the remaining $N=2$~space (parallel) directions. The
Darmois-Israel junction conditions~\cite{bib:IJC,bib:DJC,bib:Poisson_Book_Relativists_Toolkit}
relate stress-energy on the brane to a jump in extrinsic curvature
across the boundary occupied by that brane. The time-components of
the extrinsic curvature of the world-path of the brane in a bulk space-time
manifest as apparent acceleration of the brane in that bulk. For suitable
choices of the parameters of the bulk space-times and the stress-energy
on the brane (such as the equation of state), the acceleration of
the brane on either side may be individually tuned to be positive,
negative or zero. This is the essence of the mechanism of propulsion
utilized in this paper. Fig~\ref{fig:subLum_CrossSect} shows the
structure of the sub-luminal model, and fig~\ref{fig:supLum_CrossSect}
shows the structure of the super-luminal model. There is external
Minkowski space-time on the outside that represents a volume for the
warp-drive to move through, and an interior, static Minkowski region
where a passenger would ride. This interior region is surrounded by
a static layer of perfect-fluid that serves as an interface (hence
the region is called ``Modified-Minkowski''). Between the interior
and exterior Minkowski regions there is a bulk space-time that is
analogous to Schwarzschild-de~Sitter - called $\Lambda-K$. At the
boundary between each of this space-times is a thin membrane (brane)
that contains a stress-energy tensor with delta-function support.
Thus there are four branes in the model - two on each side of the
interior region. Note that the model parameters are tuned such that
the interior pair of branes are static on their Modified-Minkowski
side, yet in motion on their $\Lambda-K$ side. The equation of state
of one of the exterior branes is tuned so that both of these branes
move in unison. The equation of state of the other can be varied to
control the acceleration of the entire model in the exterior bulk
space. The four branes are in motion in the $\Lambda-K$ space - however
the proper distance between adjacent pairs decreases with coordinate
time, never reaching zero.

The reader should be aware that the models presented in section~\ref{sec:Models_presented}
are only toy models - meant as a proof of concept. These would not
be feasible to construct as they are currently. The chief reason is
that the branes are not compact - they extend infinitely in the parallel
directions. A realistic model would need the branes to be closed,
compact boundaries - thin bubbles. Furthermore, incorporating quantum
effects is beyond the scope this paper. The stability of these models
has not be checked. No consideration has been given to how the sub-luminal
warp-drive model of subsection~\ref{subsec:Mdl_subLum} might be
initialized - that is, initiated starting from standard Minkowski
space-time (or any type of space-time that is naturally readily available,
such as Schwarzschild).  

These disclaimers aside, the propulsion models presented in this paper
do illustrate several important and useful concepts. The model of
subsection~\ref{subsec:Mdl_supLum} does demonstrate that super-luminal
warp-drive is possible in classical General Relativity without violating
the Weak Energy Condition. Now that this has been show to be true,
it seems likely that the same should be possible for models with compact
boundaries. However, the super-luminal model does violate the dominant
energy condition at the exterior branes (branes~AD and AB of fig~\ref{fig:supLum_CrossSect}).
This model is locally super-luminal in that an observer in the exterior
Minkowski, at rest in the provided coordinate system, would observe
the boundary (branes~AD or AB) of that space moving toward or away
from them at a speed in excess of the speed of light. This is not
the case for the original Alcubierre model of~\cite{bib:Alcub_1994}
- it is nowhere locally super-luminal. The structural similarity of
the sub-luminal and super-luminal warp-drive models presented in section~\ref{sec:Models_presented}
provide a framework for speculating on a mechanism of transition from
the former to the latter - as is shown in fig~\ref{fig:subLum_to_supLum_transition}.
That speculation is also readily extended to a model with compact
boundaries. Furthermore, the chimeric nature of these models serves
as a demonstration of technique of model construction. The components
and their consequences are simple enough to be easily understandable
by the model-builder - yet the possibilities their chimeric combinations
permit are quite rich. One can easily understand what a given model
is and is not capable of, and how to modify a given model to achieve
desired goals. For example, insight gleaned from the approach of this
paper suggests that the presence of a horizon inside the ``warp-bubble''
- between the inner and outer branes - might be unavoidable with any
model of super-luminal warp-drive that does not utilize negative energy
density. This agrees with previous researchers~\cite{bib:Krasnikov_1998,bib:Low_1999,bib:Clark-Hiscock-Larson_1999}.
Another example, as is discussed in subsections~\ref{subsubsec:modMnk_defn}~and~\ref{subsec:modMnkLK_constr},
is that stitching together $\Lambda-K$ and standard Minkowski bulks
with a brane at the boundary - when a specific one of the two sides
of the $\Lambda-K$ bulk is to be utilized - requires negative energy
density on that brane. However, if one instead modifies the Minkowski
bulk by introducing an interface layer adjacent to the brane of perfect
fluid with appropriately-chosen positive energy density and pressure,
the energy density on that brane may then be made positive. This is
why the models in section~\ref{sec:Models_presented} have standard
Minkowski bulk on the outside, but Modified-Minkowski bulk in the
interior (where a passenger might ride).

In section~\ref{sec:GR_IJC_background} the general background of
thin membranes in General Relativity will be discussed. Subsection~\ref{subsec:GR_BackGrnd_Bulks}
discusses the three different bulk space-times utilized in this paper:
$\Lambda-K$ , standard Minkowski and Modified-Minkowski. Subsection~\ref{subsec:GR_BackGrnd_Branes}
presents the perfect-fluid-like stress-energy tensor of the branes,
and then reviews how the Darmois-Israel junction conditions relate
that stress-energy to the jump in extrinsic curvature across the brane.
Section~\ref{sec:Model_Components} discusses the requirements and
features of placing the different types of bulk space-times on either
side of a brane, where that brane has a stress-energy tensor of of
the type presented in subsection~\ref{subsubsec:GR_BackGrnd_Branes_IJC}.
Section~\ref{sec:Models_presented} presents a sub-luminal and a
super-luminal warp-drive model, based on the components developed
in section~\ref{sec:Model_Components}. In subsection~\ref{subsec:Mdl_supLum_LanRay_mod}
it is shown that the difference in extrinsic curvature across a thin
membrane can yield a positive contribution to the Landau-Raychaudhuri
equation, thus allowing the null geodesic congruence expansion scalar
to increase without a violation of the null energy condition. This
provides a means to evade some super-luminal warp-drive no-go theorems,
such as presented in~\cite{bib:Olum_1998}. This section concludes
with speculation on how these models might be modified to allow transition
from sub-luminal to super-luminal warp. Finally, section~\ref{sec:Conclusions}
reviews and summarizes the results of this paper - the beneficial
novel features of these models as well as the shortcomings - and discusses
prospects for future work. Probably the item of future work of greatest
immediate need is to replace the infinite-extent branes with compact
branes (making them essentially thin-walled closed and compact bubbles
in the bulk space-time).

\subsection{Notation}

Portions of various bulk space-time are utilized in this paper, combined
in a chimeric manner. In space-time~$\left(A\right)$, a codimension-1
surface is identified. Everything on one side of that surface will
be utilized in the model being constructed, with the surface being
a boundary of space-time~$\left(A\right)$. Everything on the other
side of that boundary will be \textbf{\uline{discarded}}, meaning
that it is not utilized in the model. The same is done with space-time~$\left(B\right)$.
The boundaries of $\left(A\right)$ and $\left(B\right)$ are identified,
creating a chimeric space-time that is $\left(A\right)$ on one side
of the boundary, and $\left(B\right)$ on the other side.
\begin{itemize}
\item The metric and coordinate system used to describe the bulk space-time
has the form: $\begin{array}{ccc}
ds^{2} & = & A\left(\left(dx^{0}\right)^{2}-\left(dx^{1}\right)^{2}\right)-Bd\overrightarrow{x}^{2}\end{array}=g_{\mu\nu}dx^{\mu}dx^{\nu}$ with $\mu,\nu\in\left\{ 0,\cdots,N+1\right\} $, $B\ge0$. $\Gamma_{\,\beta\gamma}^{\alpha}$
and $\Gamma_{\alpha\beta\gamma}$ are the Christoffel symbols. 
\item $\tau$ is the proper time of an observe at rest on the membrane (brane).
A membrane is located at the boundary between the bulk space-times.
$\dot{X}\equiv\frac{dX}{d\tau}$ 
\item $N$ is the number of space-like directions parallel to the membrane. 
\item For index $\mu$, $\bar{\mu}\in\left\{ 0,1\right\} $, with $\mu=0$
being coordinate time, $\mu=1$ the direction perpendicular to the
membrane, and $\underline{\mu}\in\left\{ 2,\cdots,N+1\right\} $ being
spatial directions parallel to the brane. 
\item $\rho_{\Lambda}$ is the energy density associated with a bulk cosmological
constant space-time contains a cosmological constant $\kappa\rho_{\Lambda}$
with $\kappa\equiv8\pi G$. 
\item $X_{\left(j\right)}$ is the quantity $X$ on the $j$-th side of
the membrane, or in bulk~$\left(j\right)$ 
\item $S_{\mu\nu}=$stress-energy on the membrane, and $S$ is its trace.
The membrane contains a perfect fluid with delta-function support
in the perpendicular direction, which in the rest frame of an observer
riding on the membrane has energy density $\rho$ and pressure in
the parallel directions $p$. 
\item $\eta_{\left(j\right)}\equiv\text{sign}\left(A_{\left(j\right)}\right)$ 
\item $\chi_{\left(j\right)}^{\mu}\left(\tau\right)$ is the world-path
of the membrane in bulk~$\left(j\right)$, $u_{\left(j\right)}^{\mu}=\dot{\chi}_{\left(j\right)}^{\mu}$ 
\item $u_{\left(j\right)}^{\mu}$ is the relativistic velocity of the membrane
in the bulk. Note that $u_{\left(j\right)}^{\mu}u_{\left(j\right)\mu}=\xi_{\left(j\right)}\eta_{\left(j\right)}$,
where $\xi_{\left(j\right)}=+1$ if $u_{\left(j\right)}^{\mu}$ is
time-like in the frame in which the metric of bulk~$\left(j\right)$
is specified, $\xi_{\left(j\right)}=-1$ if $u_{\left(j\right)}^{\mu}$
is space-like, and $\xi_{\left(j\right)}=-1$ if $u_{\left(j\right)}^{\mu}$
is light-like (not considered in this paper). 
\item $n_{\left(j\right)}^{\mu}$ is the normal to the membrane in bulk~$\left(j\right)$,
pointing from side~$\left(2\right)$ of the membrane to side~$\left(1\right)$.
$n_{\left(j\right)}^{\mu}n_{\left(j\right)\mu}=-\xi_{\left(j\right)}\eta_{\left(j\right)}$
and $u_{\left(j\right)}^{\mu}n_{\left(j\right)\mu}=0$ 
\item $s_{\left(j\right)}\in\left\{ -1,+1\right\} $ in bulk~$\left(j\right)$,
describes which side of the brane world-path is to be utilized in
a model, and which discarded. It is set such that $n_{\left(j\right)}^{\mu}=s_{\left(j\right)}\left(\begin{array}{cc}
0 & 1\\
1 & 0
\end{array}\right)_{\alpha}^{\mu}u_{\left(j\right)}^{\alpha}$ points in the correct direction: $n_{\left(1\right)}^{\mu}$ points
into the side of bulk~$\left(1\right)$ that is discarded, $n_{\left(2\right)}^{\mu}$
points into the side of bulk~$\left(2\right)$ that is kept.
\item $K$ is a constant of integration in the $\Lambda-K$ bulk space-time
considered. It plays a role reminiscent of the central mass in a Schwarzschild
space-time. 
\item $y\equiv\left(-2\lambda\right)\left(\left(x^{0}\right)^{2}-\left(x^{1}\right)^{2}\right)$
is hyperbolic coordinate in the $\Lambda-K$ bulk space-time, with
$\lambda$ a separation constant encountered in the solution of the
bulk Einstein field equations. $\mathcal{Y}_{\left(j\right)}\equiv\left.y_{\left(j\right)}\right|_{x^{\mu}=\chi_{\left(j\right)}^{\mu}}$ 
\item $z$ is a coordinate along which the metric is constant in the $\Lambda-K$
bulk space-time, in an alternate coordinate frame. 
\item $B_{H}$ is the the value of $B$ at the horizon in the $\Lambda-K$
bulk space-time, if $\text{sign}\left(K\right)=-\text{sign}\left(\kappa\rho_{\Lambda}\right)$,
$\mathcal{B}\equiv\sqrt{\frac{B}{B_{H}}}$ 
\item $h_{\nu}^{\mu}\equiv g_{\nu}^{\mu}-\frac{1}{\left(n^{\alpha}n_{\alpha}\right)}n^{\mu}n_{\nu}$
is the induced metric on the membrane. $\xi_{\left(1\right)}\eta_{\left(1\right)}=\xi_{\left(2\right)}\eta_{\left(2\right)}=\xi\eta$ 
\item $\Pi_{\mu\nu}$ is the extrinsic curvature of the world-path of a
membrane. $\Pi_{\mu\nu}\equiv h_{\mu}^{\alpha}n_{\nu;\alpha}$ and
$\Pi\equiv g^{\mu\nu}\Pi_{\mu\nu}=h^{\mu\alpha}n_{\mu;\alpha}$
\item $w_{max}$ is the $w$ coordinate of the location in the modified-Minkowski
coordinate system where that space-time is interfaced to the $\Lambda-K$
space-time
\item When interfacing modified-Minkowski (side~$\left(1\right)$) to $\Lambda-K$
(side~$\left(2\right)$) space-times there is a degree of freedom
resulting from the freedom to rescale the former relative to the later.
This degree of freedom can be expressed as $\beta\equiv-\frac{\partial w}{\partial x^{1}}$,
where $w$ is the spatial direction perpendicular to the brane on
side~$\left(1\right)$ and $x^{1}$ the spatial direction perpendicular
to the brane on side~$\left(2\right)$. This degree of freedom may
also be represented as $\phi\equiv\frac{\left.\ln\left(B_{\left(1\right)}\right)_{,\alpha}n_{\left(1\right)}^{\alpha}\right|_{\chi_{\left(1\right)}^{\mu}\left(\tau\right)}}{\left.\ln\left(B_{\left(2\right)}\right)_{,\alpha}n_{\left(2\right)}^{\alpha}\right|_{\chi_{\left(2\right)}^{\mu}\left(\tau\right)}}$. 
\item $S_{2}$ is a number defined in the design of the Modified-Minkowski
Space-Time of section~\ref{subsubsec:modMnk_defn}. It is defined
as the negative of the ratio of log-derivatives of $A$ and $B$ functions
in the metric, evaluated at the outer boundary of the region. $S_{2}\equiv\left.\frac{-a^{\prime}}{b^{\prime}}\right|_{N=2,w=w_{max}}\simeq1+5.89457\times10^{-8}$.
Generalizing to $N\ne2$, $S_{N}\equiv\left(N-1\right)S_{2}$. 
\end{itemize}

\section{General background for membranes/boundaries in GR: the Israel Junction
Conditions\label{sec:GR_IJC_background}}

\noindent\fbox{\begin{minipage}[t]{1\columnwidth - 2\fboxsep - 2\fboxrule}%
This section:
\begin{lyxlist}{00.00.0000}
\item [{1)}] Introduces the three bulk space-times that will needed
\item [{2)}] Introduces codimension-1 membranes (branes) 
\item [{3)}] Explains that those membranes are boundaries between two bulk
space-times
\item [{4)}] Explains that a codimension-1 stress-energy exists on the
brane, with delta-function support
\item [{5)}] Explains that that brane stress-energy is related to the difference
in extrinsic curvature of the boundary from one side to the other
\end{lyxlist}
\end{minipage}}

The physical system under consideration is two bulk space-times $\mathcal{M}_{\left(1\right)}$
and $\mathcal{M}_{\left(2\right)}$ of $1$~time-like direction~$x^{0}$
and $N+1$~space-like directions~$x^{i}$~($1\le i\le N+1$). $\mathcal{M}_{\left(1\right)}$
and $\mathcal{M}_{\left(2\right)}$ both contain the world-path of
membrane $\Sigma$ (brane for short), which is a $N+1$-dimensional
(co-dimension~one) hypersurface. That co-dimension is one of the
space-like directions, which we will call the perpendicular direction~$x^{1}$.
The entire system is translationally and rotationally symmetric in
the remaining $N$ space-like directions, which are collectively called
parallel directions~$\overrightarrow{x}$. Sometimes it will be useful
to speak of $\Sigma_{\left(1\right)}$ which is in $\mathcal{M}_{\left(1\right)}$
and $\Sigma_{\left(2\right)}$ which is in $\mathcal{M}_{\left(2\right)}$.
There exists a homeomorphism between $\Sigma_{\left(1\right)}$ and
$\Sigma_{\left(2\right)}$, and thus they can identified and jointly
indicated as $\Sigma$. Through this identification of $\Sigma_{\left(1\right)}$
and $\Sigma_{\left(2\right)}$, $\mathcal{M}_{\left(1\right)}$ and
$\mathcal{M}_{\left(2\right)}$ are joined as if by surgery. The direction
within $\mathcal{M}_{\left(j\right)}$ at $\Sigma_{\left(j\right)}$
that is utilized is given by a normal vector field $n_{\left(j\right)}^{\mu}$
defined on $\Sigma_{\left(j\right)}$. Thus at proper time $\tau$,
a point in $\Sigma_{\left(j\right)}$ has velocity $u_{\left(j\right)}^{\mu}\left(\tau\right)$
and normal $n_{\left(j\right)}^{\mu}\left(\tau\right)$. Indices $\mu$
run over the full $1$~time-like direction and $N+1$~space-like
directions, $\bar{\mu}$ run over the time-like and perpendicular
directions only, and $\underline{\mu}$ run over the $N$ parallel
directions only. The $\left(j\right)$~subscripts may be suppressed
in statements that are true for each side of the junction individually,
or when a quantity has the same value on both sides. For travel in
our $3+1$ dimensional Universe, one would generally take $N=2$.
However, we proceed with $N$ as a parameter so that more complex
scenarios may be considered.

The coordinate system is $x^{0}$ for the time direction, $x^{1}$
for the perpendicular space direction, and $x^{k}\ \ 2\le k\le N+1$
(collectively $\overrightarrow{x}$) for the $N$ space directions
parallel to the brane. 
\begin{equation}
\begin{array}{ccc}
ds^{2} & = & A\left(\left(dx^{0}\right)^{2}-\left(dx^{1}\right)^{2}\right)-Bd\overrightarrow{x}^{2}\\
\\
 &  & A=A\left(x^{0},x^{1}\right)\\
 &  & B=B\left(x^{0},x^{1}\right)
\end{array}\label{eq:Bulk_metric_form}
\end{equation}
Note that we do not assume $A$ is positive - instead we define the
sign of $A$: $\eta_{\left(j\right)}\equiv sign\left(A_{\left(j\right)}\right)$
in space-time $\mathcal{M}_{\left(j\right)}$. There is an assumption
of rotational and translational symmetry in the directions parallel
to~$\Sigma$, ie:~$x^{k}\ \ 2\le k\le N+1$ (in the bulk and on
the brane). Thus~$A=A\left(x^{0},x^{1}\right)$ and $B=B\left(x^{0},x^{1}\right)$.

It is assumed that there is stress-energy in the bulk space-time,
as well as stress-energy with delta-function support on the brane.
The metric in a bulk space-time satisfies the $N+2$-dimensional
Einstein field equation
\begin{equation}
E_{\mu\nu}=R_{\mu\nu}-\frac{1}{2}g_{\mu\nu}R=\kappa T_{\mu\nu}\ \ \ \kappa\equiv8\pi G\label{eq:Bulk_Einst_fld_eq}
\end{equation}

\subsection{The Bulk Space-Times: $\Lambda-K$, Minkowski and Modified-Minkowski\label{subsec:GR_BackGrnd_Bulks} }

\noindent\fbox{\begin{minipage}[t]{1\columnwidth - 2\fboxsep - 2\fboxrule}%
This subsection: 
\begin{lyxlist}{00.00.0000}
\item [{1)}] Presents the metric of a space-time containing containing
vacuum energy~$\rho_{\Lambda}$ and possessing an integration constant~$K$
analogous to the mass in the Schwarzschild space-time - henceforth
called $\Lambda-K$. 
\item [{2)}] The metric depends on a single parameter. That parameter may
be taken to be time, or a spatial direction, depending on the sign
of~$\rho_{\Lambda}$.
\item [{3)}] When $\rho_{\Lambda}$ and $K$ have opposing signs the space-time
has a horizon, which is a coordinate singularity. 
\item [{4)}] A coordinate system analogous to Kruskal--Szekeres coordinates
is derived in which the the metric depends on the hyperbolic parameter
$y\equiv\left(-2\lambda\right)\left(\left(x^{0}\right)^{2}-\left(x^{1}\right)^{2}\right)$
\item [{5)}] For a particular value of the constant $\lambda$, the zero
and pole in the $g_{00}$ and $g_{11}$ parts of the metric that occur
at the horizon precisely cancel, removing the horizon coordinate singularity. 
\item [{6)}] This Kruskal--Szekeres-analog coordinate system of the maximal
$\Lambda-K$ space-time is what will be used in the later model-building 
\end{lyxlist}
\end{minipage}}

Consider three possible solutions to eq~\ref{eq:Bulk_Einst_fld_eq}
with the symmetry of eq~\ref{eq:Bulk_metric_form}: $\Lambda-K$
, standard Minkowski and Modified-Minkowski. $\Lambda-K$ is similar
to de~Sitter space - containing vacuum energy $\kappa T_{\mu\nu}=\kappa\rho_{\Lambda}g_{\mu\nu}$
(which will eventually be taken to be positive in the models constructed
later) and an integration constant~$K\ne0$ that is analogous the
central mass in a Schwarzschild space-time. Modified-Minkowski is
a region of standard Minkowski space-time, surrounded by a static
boundary layer in which $\kappa\rho>\kappa p>0$. The boundary layer
smoothly transitions to standard Minkowski in the interior - the derivatives
of the metric vanish to all orders at the interface. The purpose of
this layer is to interface with $\Lambda-K$ space-time at the exterior
such that brane at the junctions contains positive energy density. 

\subsubsection{$\Lambda-K$ Bulk Space-Time\label{subsubsec:LK_defn}}

A suitable bulk space-time to work with contains vacuum energy $\rho_{\Lambda}$
and a constant of integration that will be denoted as $K$. A simple
exact solution is already known for the spherically-symmetric case
- called ``Modified Schwarzschild space'' in Chapter~14.4 of~\cite{bib:Rindler_Relativity}.
The symmetry imposed in this paper is planar rather than spherical,
and this does lead to some important differences. The planar-symmetry
space-time case will be referred to as~$\Lambda-K$ in this paper.
It turns out that the integration constant $K$ plays a role very
similar to that of a mass of a black hole in the Schwarzschild solution~\cite{bib:schw_soln}.
What makes $\Lambda-K$ space-time suitable for the role in which
it will be employed (in sections~\ref{subsec:MnkLK_constr} and~\ref{sec:Models_presented},
regions~$\left(B\right)$~and~$\left(D\right)$ in figs~\ref{fig:subLum_CrossSect}
and~\ref{fig:supLum_CrossSect}) is that it has the right structure
to work, yet is simple enough that there are exact solutions that
are easy to utilize. An example of making the bulk space-times too
simple would be using only flat space. If one makes both sides of
the brane Minkowski space-time, with an assumption of the model of
the brane stress-energy presented in subsection~2.2.1, the junction
conditions of subsection~2.2.2 force the energy density on the brane
to be vanish, and hence also the pressure, and the model becomes trivial.
As one might expect, making one side of the brane Rindler space-time,
and the other Minkowski space-time has the same result. A structural
feature of the $\Lambda-K$ space-time required by the super-luminal
model is the existence of space-like and time-like brane world-paths
of constant induced metric that do not intersect. While the $\Lambda-K$
space-time is not the only space-time that has these features, it
is a space-time with simple exact solutions that has those features. 

For $T_{\mu\nu}=\rho_{\Lambda}g_{\mu\nu}$, a general form of the
solution of eq~\ref{eq:Bulk_Einst_fld_eq} is 
\[
B_{,z}=\frac{2}{N}B^{\frac{2-N}{2}}K+\frac{1}{N+1}B^{\frac{3}{2}}\kappa\rho_{\Lambda}
\]
 where $z$ is a single parameter that $B$ depends on: $B=B\left(z\right)$.
If one further chooses a coordinate frame such that $A=A\left(z\right)$
the result is 
\[
A=\left(const\right)\frac{N}{2}B^{-\frac{1}{2}}B_{,z}e^{2\lambda z}=\left(const\right)N\left(\sqrt{B}\right)_{,z}e^{2\lambda z}
\]
 If $A>0$, one can perform a coordinate transformation such that
$z$ becomes the time coordinate. However, if $\kappa\rho_{\Lambda},K$
are both non-zero and $\text{sign}\left(K\right)=-\text{sign}\left(\kappa\rho_{\Lambda}\right)$
there is a coordinate singularity at 
\begin{equation}
B=B_{H}\equiv\left(\frac{-K}{\kappa\rho_{\Lambda}}\frac{2\left(N+1\right)}{N}\right)^{\frac{2}{N+1}}\label{eq:BH_Horiz_defn}
\end{equation}
 where $B_{,z}$ and hence $A$ vanishes, changing sign. This is analogous
to the horizon of Schwarzschild space-time. The change of sign of
$A$ corresponds to an exchange of the time and the space directions.
 If $A<0$, one can perform a coordinate transformation such that
$z$ becomes the space coordinate analogous to~$x^{1}$, and the
sign of $A$ becomes positive. While $B=B_{H}$ is merely a coordinate
singularity, there is a curvature (physical) singularity at $B=0$.

\subsubsection{Maximal $\Lambda-K$ Space-Time}

One can extend the space-time in a manner reminiscent of Kruskal--Szekeres
coordinates~\cite{bib:kruskal,bib:Szekeres}, eliminating the coordinate
singularity at $B=B_{H}$. This is accomplished by changing the coordinate
dependence of the metric from~$z$~to~$y$: 
\begin{equation}
\begin{array}{ccc}
y & \equiv & \left(-2\lambda\right)\left(\left(x^{0}\right)^{2}-\left(x^{1}\right)^{2}\right)\\
 & = & \frac{1}{-2\lambda\left(const\right)}\exp\left(-2\lambda z\right)\\
z & = & \frac{-1}{2\lambda}\ln\left(-2\lambda y\right)+\left(const\right)\\
dy & = & -2\lambda y\,dz\\
-2\lambda & = & \frac{1}{2}\kappa\rho_{\Lambda}\sqrt{B_{H}}\ >\ 0
\end{array}\label{eq:bulkLK_z_to_y_coord_xform_a}
\end{equation}
\begin{equation}
A=N\left(\sqrt{B}\right)_{,z}\left(\frac{1}{-2\lambda y}\right)=N\left(\sqrt{B}\right)_{,y}=\frac{N}{2}B^{-\frac{1}{2}}B_{,y}\label{eq:bulkLK_z_to_y_coord_xform_b}
\end{equation}
\begin{equation}
\begin{array}{ccccc}
B_{,y} & = & \frac{1}{\left(-2\lambda\right)y}\left(\frac{2}{N}B^{\frac{2-N}{2}}K+\frac{1}{N+1}B^{\frac{3}{2}}\kappa\rho_{\Lambda}\right) & = & \frac{1}{\left(-2\lambda\right)y}B_{,z}\\
A & = & \frac{N}{2}B^{-\frac{1}{2}}B_{,y}=N\left(\sqrt{B}\right)_{,y}
\end{array}\label{eq:bulkLK_z_to_y_coord_xform_c}
\end{equation}

\begin{figure}
\begin{centering}
\includegraphics[scale=0.46]{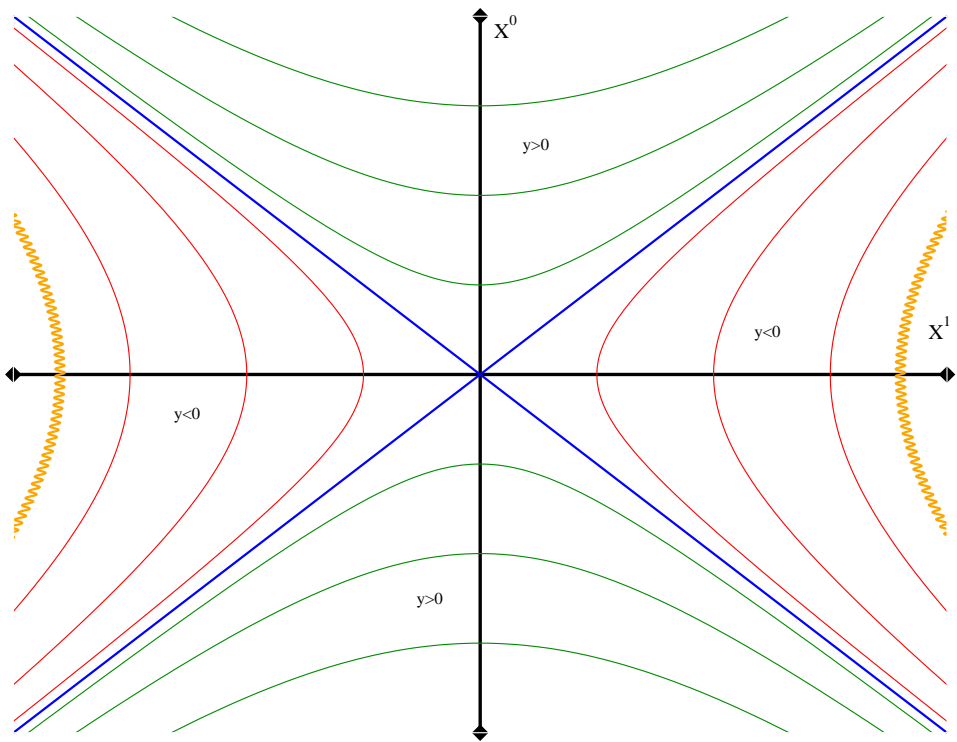}\label{fig:KSanalog_conformal_diagram}\caption{A conformal diagram of the $\Lambda-K$ space-time in an extended
frame that is a close analog to Kruskal-Szekeres coordinates of a
Schwarzschild space-time. The metric is constant along curves of constant
$y\equiv\left(-2\lambda\right)\left(\left(x^{0}\right)^{2}-\left(x^{1}\right)^{2}\right)$,
where $\lambda$ is a constant. Green and red curves are positive
and negative $y$ respectively. The blue $y=0$ lines are a coordinate
singularity in the $z$ coordinate frame, and are similar to the event
horizon of Schwarzschild space-time. Transforming from the $z$ to
$y$ coordinate frame, with $-2\lambda=\frac{1}{2}\kappa\rho_{\Lambda}\sqrt{B_{H}}$,
the metric becomes well-behaved across the blue lines. There is a
physical (curvature) singularity indicated by the jagged orange curves.}
\par\end{centering}
\end{figure}

Note that $B_{,z}$ and $A$ vanish at $B=B_{H}$, hence the coordinate
singularity. However, $B_{,y}$ also contains $y$, which has a zero
at $\left|x^{0}\right|=\left|x^{1}\right|$. If one centers the coordinate
system - or scales $B$ such that $B=B_{H}$ at $y=0$ - then these
zeros in $B_{,y}$ exactly coincide. If one further sets $-2\lambda=\frac{1}{2}\kappa\rho_{\Lambda}\sqrt{B_{H}}$
then the strength of these zeros precisely cancel, and one has a well-behaved
metric outside of the physical singularity at $B=0$. Ignoring for
a moment the parallel directions, this space-time is a $1+1$ analog
of the Schwarzschild solution, and this coordinate frame is the analog
of Kruskal--Szekeres coordinates~\cite{bib:kruskal,bib:Szekeres}.
On a conformal diagram $B=B_{H}$ looks very similar to the event
horizon of a black hole. The swapping of space and time directions
at this $B=B_{H}$ horizon will prove crucial in the model for super-luminal
travel presented in subsection~\ref{subsec:Mdl_supLum}. Thus for
this particular value of $\lambda$ one gets an analog to Kruskal-Szekeres
coordinates: $A>0$ across the horizon and $\left(N+1\right)+1$ dimensional
space-time with metric $ds^{2}=A\left(\left(dx^{0}\right)^{2}-\left(dx^{1}\right)^{2}\right)+Bd\overrightarrow{x}^{2}$,
as summarized in fig~\ref{fig:KSanalog_conformal_diagram}. 

To construct an explicit solution of of eq~\ref{eq:Bulk_Einst_fld_eq}
in this coordinate frame, define the a new quantity $\mathcal{B}$
that is the square root of the ratio of $B$ with its value at the
horizon 
\begin{equation}
\mathcal{B}\equiv\sqrt{\frac{B}{B_{H}}}\label{eq:bulkLK_KScf_scrB_defn}
\end{equation}
 In terms of this variable, eq~\ref{eq:bulkLK_z_to_y_coord_xform_c}
becomes
\begin{equation}
\begin{array}{ccl}
\mathcal{B}_{,y} & = & \frac{1}{-2\lambda y}\left(\frac{\kappa\rho_{\Lambda}}{2\left(N+1\right)}B_{H}^{\frac{1}{2}}\left(\mathcal{B}^{2}-\mathcal{B}^{1-N}\right)\right)\\
 & = & \frac{1}{-2\lambda y}\left(\frac{\kappa\rho_{\Lambda}}{2\left(N+1\right)}B_{H}^{\frac{1}{2}}\left(\mathcal{B}^{N+1}-1\right)\mathcal{B}^{1-N}\right)\\
 & = & \frac{1}{-2\lambda y}\left(\frac{\kappa\rho_{\Lambda}}{2\left(N+1\right)}B_{H}^{\frac{1}{2}}\left(\mathcal{B}-1\right)\left(\frac{\stackrel[k=0]{N}{\sum}\mathcal{B}^{k}}{\mathcal{B}^{N-1}}\right)\right)
\end{array}\label{eq:bulkLK_KScf_scrB_deq_wrt_y}
\end{equation}
 which may be represented as the integral 
\begin{equation}
\begin{array}{ccl}
\int\frac{d\mathcal{B}}{\mathcal{B}^{2}-\mathcal{B}^{1-N}} & = & \frac{1}{-2\lambda}\left(\frac{\kappa\rho_{\Lambda}}{2\left(N+1\right)}B_{H}^{\frac{1}{2}}\right)\int d\ln y\\
\int\frac{\mathcal{B}^{N-1}}{\mathcal{B}^{N+1}-1}d\mathcal{B} & = & \frac{1}{-2\lambda}\left(\frac{\kappa\rho_{\Lambda}}{2\left(N+1\right)}B_{H}^{\frac{1}{2}}\right)\ln y+Const
\end{array}\label{eq:bulkLK_KScf_scrB_integ}
\end{equation}
 The integral on the left-hand side of eq~\ref{eq:bulkLK_KScf_scrB_integ}
diverges at $\mathcal{B}=1$ due to that root in the cyclotomic polynomial,
while the right-hand side diverges at $y=0$. To define a well-behaved
coordinate frame it is necessary to isolate these divergences so that
they may be canceled against one another. This is accomplished by
expanding the reciprocal of the cyclotomic polynomial as 
\begin{equation}
\frac{\mathcal{B}^{N-1}}{\mathcal{B}^{N+1}-1}=\frac{1}{N+1}\frac{1}{\mathcal{B}-1}+\frac{1}{N+1}\frac{-\mathcal{B}^{N-1}+\stackrel[k=0]{N-2}{\sum}\left(k+1\right)\mathcal{B}^{k}}{\stackrel[k=0]{N}{\sum}\mathcal{B}^{k}}\label{eq:bulkLK_KScf_partfracexpan}
\end{equation}
 Substituting this into eq~\ref{eq:bulkLK_KScf_scrB_integ} and integrating
yields 
\begin{equation}
\begin{array}{ccl}
\ln y & = & \frac{\left(-2\lambda\right)2}{\kappa\rho_{\Lambda}\sqrt{B_{H}}}\left(\ln\left(\mathcal{B}-1\right)+\mathcal{H}_{N}\left(\mathcal{B}\right)\right)+Const\\
\mathcal{H}_{N}\left(\mathcal{B}\right) & \equiv & \int\frac{-\mathcal{B}^{N-1}+\stackrel[k=0]{N-2}{\sum}\left(k+1\right)\mathcal{B}^{k}}{\stackrel[k=0]{N}{\sum}\mathcal{B}^{k}}d\mathcal{B}
\end{array}\label{eq:bulkLK_KScf_lny_soln}
\end{equation}
 Note that the denominator of the integrand in $\mathcal{H}_{N}\left(\mathcal{B}\right)$
has no real non-negative roots, hence $\mathcal{H}_{N}\left(\mathcal{B}\right)$
does not diverge. Without loss of generality, set the constant of
integration to zero, and then choose $-2\lambda=\frac{1}{2}\kappa\rho_{\Lambda}\sqrt{B_{H}}$
yielding 
\begin{equation}
\begin{array}{ccl}
y & = & \left(\mathcal{B}-1\right)\exp\left(\mathcal{H}_{N}\left(\mathcal{B}\right)\right)\\
-2\lambda y & = & \left(\frac{1}{2}\kappa\rho_{\Lambda}\sqrt{B_{H}}\right)\left(\mathcal{B}-1\right)\exp\left(\mathcal{H}_{N}\left(\mathcal{B}\right)\right)
\end{array}\label{eq:bulkLK_KScf_n2ly_soln}
\end{equation}
 which yields a solution to eq~\ref{eq:Bulk_Einst_fld_eq} of the
form of eq~\ref{eq:Bulk_metric_form} that is well-behaved at $B=B_{H}$,
although it does possess a physical singularity at $B=0$ for $N>1$
\begin{equation}
\begin{array}{ccl}
\mathcal{B}_{,y} & = & \frac{1}{\left(N+1\right)}\exp\left(-\mathcal{H}_{N}\left(\mathcal{B}\right)\right)\left(\frac{\stackrel[k=0]{N}{\sum}\mathcal{B}^{k}}{\mathcal{B}^{N-1}}\right)\\
A & = & N\sqrt{B_{H}}\mathcal{B}_{,y}=\frac{N\sqrt{B_{H}}}{\left(N+1\right)}\exp\left(-\mathcal{H}_{N}\left(\mathcal{B}\right)\right)\left(\frac{\stackrel[k=0]{N}{\sum}\mathcal{B}^{k}}{\mathcal{B}^{N-1}}\right)
\end{array}\label{eq:bulkLK_KScf_metric_soln}
\end{equation}
 In general, when working with $\Lambda-K$ space, it is this coordinate
frame that will be used. As an example, consider the explicit result
for $N=2$ : 
\begin{equation}
\begin{array}{ccl}
\mathcal{H}_{2}\left(\mathcal{B}\right) & = & \int\frac{-\mathcal{B}+1}{\mathcal{B}^{2}+\mathcal{B}^{1}+1}d\mathcal{B}=\sqrt{3}\arctan\left(\frac{1+2\mathcal{B}}{\sqrt{3}}\right)-\frac{1}{2}\ln\left(\mathcal{B}^{2}+\mathcal{B}^{1}+1\right)\\
\mathcal{B}_{,y} & = & \frac{1}{3}\exp\left(-\mathcal{H}_{2}\left(\mathcal{B}\right)\right)\left(\frac{\mathcal{B}^{2}+\mathcal{B}^{1}+1}{\mathcal{B}}\right)=\frac{1}{3\mathcal{B}}\exp\left(-\sqrt{3}\arctan\left(\frac{1+2\mathcal{B}}{\sqrt{3}}\right)\right)\left(\mathcal{B}^{2}+\mathcal{B}^{1}+1\right)^{\frac{3}{2}}\\
A & = & 2\sqrt{B_{H}}\mathcal{B}_{,y}
\end{array}\label{eq:bulkLK_KScf_metric_soln_Neq2}
\end{equation}

\subsubsection{Modified-Minkowski Space-Time\label{subsubsec:modMnk_defn} }

\noindent\fbox{\begin{minipage}[t]{1\columnwidth - 2\fboxsep - 2\fboxrule}%
This subsection:
\begin{lyxlist}{00.00.0000}
\item [{1)}] Introduces the Modified-Minkowski space-time - a static space
of finite extent in the $w$ (spatial) direction from $w=w_{max}$
to $w=0$. The metric depends only on the variable~$w$. 
\item [{2)}] At $w=w_{max}$ the function $B$ in the metric has a prescribed
value of the derivative $\frac{d\ln\left(B\right)}{dw}$.
\item [{3)}] At $w=0$ the metric smoothly transitions to Minkowski space. 
\item [{4)}] Over the interval $w=w_{max}$ to $w=0$ the metric is the
solution to the Einstein equations with a perfect-fluid stress-energy
tensor - of energy density~$\rho\left(w\right)$, pressure~$p\left(w\right)$. 
\item [{5)}] The metric and stress-energy tensor are symmetric under $w\rightarrow-w$,
thus a symmetric composite space may be formed by ``inserting''
a region of Minkowski space of arbitrary width at $w=0$. The result
is region~C shown in fig~\ref{fig:subLum_CrossSect} and fig~\ref{fig:supLum_CrossSect}. 
\item [{6)}] This boundary layer will allow this Modified-Minkowski space-time
(which is region~C in fig~\ref{fig:subLum_CrossSect} and fig~\ref{fig:supLum_CrossSect})
to be adjoined to $\Lambda-K$ space (regions~B and~D) in the way
shown in those figures, with positive energy density on the branes
(branes~BC and~CD) at the junction. Without this boundary layer
- if region~C were regular Minkowski space-time the junction conditions
would cause those branes to contain negative energy density.
\end{lyxlist}
\end{minipage}}

One wishes to interface the $\Lambda-K$ space-time to an interior
region of Minkowski space. We will use a modified version of Minkowski.
To interface directly from a $\Lambda-K$ bulk to a bulk space-time
that is flat (ie: derivatives of all orders of the metric are zero)
- with the parallel component of the metric ($B$) decreasing in the
direction heading into the boundary - would require a brane with negative
energy density at that boundary. On can intuitively compare this to
concave-up curvature of $B$. The solution is to modify the the Minkowski
bulk so that the boundary region has a negative derivative of $B$
moving away from the boundary with the $\Lambda-K$ bulk. This may
be accomplished by filling the boundary region of the Minkowski space
with a perfect fluid and requiring a certain behavior of~$B$ : 

\begin{equation}
T^{\mu\nu}=\left(\rho+p\right)u^{\mu}u^{\nu}-pg^{\mu\nu}\label{eq:BumpSpc_BulkSE}
\end{equation}
 The bulk metric will be assumed to have the form: 
\begin{equation}
\begin{array}{ccc}
ds^{2} & = & A\left(dt^{2}-dw^{2}\right)-Bd\overrightarrow{x}^{2}\\
 &  & A=A\left(w\right)\\
 &  & B=B\left(w\right)
\end{array}\label{eq:BumpSpc_MetricForm}
\end{equation}

This space-time will vary only in the space direction (thus is static)
- here labeled as $w$. The goal is to pick a profile for $B$ such
that $B\left(w\right)$ has zero derivatives to all orders at the
boundary with the Minkowski region (at $w=0$), and $B_{,\beta}n^{\beta}<0$
at the boundary with $\Lambda-K$, with normal $n^{\mu}$ pointing
away from that boundary. Thus $B\left(w\right)$ will have a positive
curvature in this region. One finds $A$ by solving one of three constraints
coming from the Einstein field equations. The other two constraints
are satisfied by assuming the bulk energy density~$\rho$ and pressure~$p$
have the necessary form as a function of $w$. Because all orders
of derivatives of $A$ and $B$ vanish at $w=0$, it will also be
true that the resulting $\rho$ and $p$ vanish at $w=0$ (and, in
fact, all orders of their derivatives with respect to $w$ vanish
at $w=0$). This will be accomplished by setting $B$ to be a suitable
bump-function $B\left(w\right)\equiv\exp\left(\exp\left(-w^{-2}\right)\right)$.
It will turn out that $A$, $\rho$ and $p$ also will smoothly go
to zero at $w=0$. Thus, at $w=0$ one smoothly transitions from a
bulk containing the above perfect fluid to empty Minkowski space.
Because derivatives of all orders of the metric and stress-energy
vanish at $w=0$, the bulk Riemann tensor is $\mathcal{C}^{\infty}$
across this boundary, the Israel junction conditions are trivially
satisfied and it is consistent to set this boundary to be static,
which we do. Thus, one will have a smooth, static transition from
this perfect-fluid bulk to Minkowski bulk at $w=0$ - without any
brane $\delta\left(w\right)$-supported stress-energy. There is no
brane at the boundary at $w=0$. 

The bulk Einstein field equations become 

\begin{equation}
\begin{array}{ccccc}
E_{00} & = & \frac{1}{8}N\left(N+1\right)\left(\frac{db}{dw}\right)^{2}+\frac{1}{2}N\frac{d^{2}b}{dw^{2}}-\frac{1}{4}N\frac{da}{dw}\frac{db}{dw} & = & A\kappa\rho\\
E_{11} & = & -\frac{1}{8}N\left(N-1\right)\left(\frac{db}{dw}\right)^{2}-\frac{1}{4}N\frac{da}{dw}\frac{db}{dw} & = & A\kappa p\\
E_{01} & = & 0\\
E_{\underline{\mu\nu}} & = & -\frac{B}{A}\left(\frac{1}{2}\frac{d^{2}a}{dw^{2}}+\frac{1}{2}\left(N-1\right)\frac{d^{2}b}{dw^{2}}+\frac{1}{8}N\left(N-1\right)\left(\frac{da}{dw}\right)^{2}\right)\delta_{\underline{\mu\nu}} & = & B\kappa p\\
E_{\overline{\mu}\underline{\nu}} & = & 0\\
 &  & a\equiv\ln A,\ b\equiv\ln B\\
 &  & \eta\equiv\text{sign}\left(A\right)=+1\ \text{assumed}
\end{array}\label{eq:BumpSpc_EinFldEq}
\end{equation}
\begin{equation}
\begin{array}{ccccc}
2\left(E_{11}-\frac{A}{B}E_{\underline{\nu\nu}}\right) & = & \frac{d^{2}a}{dw^{2}}+\left(N-1\right)\frac{d^{2}b}{dw^{2}}-\frac{1}{2}N\frac{da}{dw}\frac{db}{dw} & = & 0\end{array}\label{eq:BumpSpc_abEq}
\end{equation}
 Assume $b\left(w\right)\equiv\exp\left(-w^{-2}\right)$, and then
solving for $\frac{da}{dw}$ and $a$ yields (the subscript indicates
the value of $N$) : 
\begin{equation}
\begin{array}{ccc}
b_{2}\left(w\right) & = & \exp\left(-w^{-2}\right)\\
\frac{da_{2}\left(w\right)}{dw} & = & -\exp\left(b_{2}\left(w\right)\right)\,\stackrel[0]{w}{\int}\frac{d^{2}b_{2}\left(s\right)}{ds^{2}}\exp\left(-b_{2}\left(s\right)\right)\,ds\\
a_{2}\left(w\right) & = & -\stackrel[0]{w}{\int}\exp\left(b_{2}\left(t\right)\right)\,\stackrel[0]{t}{\int}\frac{d^{2}b_{2}\left(s\right)}{ds^{2}}\exp\left(-b_{2}\left(s\right)\right)\,ds\,dt
\end{array}\label{eq:BumpSpc_ab2Soln}
\end{equation}
 One then uses $E_{00}$ to generate a $\rho\left(w\right)$ profile,
and $E_{11}$ to generate a $p\left(w\right)$ profile. Thus all Einstein
field equations are satisfied. One of those constraints is equivalent
to covariant conservation of $T_{\mu\nu}$. It is thus an automatic
consequence of this solution that: 
\begin{equation}
\frac{dp}{dw}=-\frac{1}{2}\frac{da}{dw}\left(\rho+p\right)\label{eq:BumpSpa_dpdw}
\end{equation}
There is an interval centered around $w=0$ for which it is guaranteed
that $\frac{d^{2}b\left(w\right)}{dw^{2}}>0$ and $\frac{da\left(w\right)}{dw}<0$
which implies $E_{00}>0$ which in turn implies $\rho>0$. The choice
of the bump function is not ideal for large~$w$, as it results in
a solution that exhibits undesirable behavior for $w\gg1$. However,
this space will be cut off well below that, so a $b\left(w\right)=\exp\left(-w^{-2}\right)$
is chosen for simplicity. A better choice of bump function would enable
extension to arbitrarily large $w$, while maintaining parameters
within their desired ranges. Profiles for $A\left(w\right)$, $B\left(w\right)$,
$\rho\left(w\right)$, $p\left(w\right)$ as well as the equation
of state~$\frac{p\left(w\right)}{\rho\left(w\right)}$ are shown
for $N=2$ in fig~\ref{fig:BumpSpcProf}. Due to the form of eq~\ref{eq:BumpSpc_abEq},
for $N>2$ 
\begin{equation}
\begin{array}{ccc}
b_{N}\left(w\right) & = & \frac{2}{N}b_{2}\left(w\right)=\frac{2}{N}\exp\left(-w^{-2}\right)\\
a_{N}\left(w\right) & = & \frac{2\left(N-1\right)}{N}a_{2}\left(w\right)\\
 & = & -\frac{2\left(N-1\right)}{N}\stackrel[0]{w}{\int}\exp\left(b_{2}\left(t\right)\right)\,\stackrel[0]{t}{\int}\frac{d^{2}b_{2}\left(s\right)}{ds^{2}}\exp\left(-b_{2}\left(s\right)\right)\,ds\,dt\\
\frac{da_{N}\left(w\right)}{dw} & = & -\frac{2\left(N-1\right)}{N}\exp\left(b_{2}\left(w\right)\right)\,\stackrel[0]{w}{\int}\frac{d^{2}b_{2}\left(s\right)}{ds^{2}}\exp\left(-b_{2}\left(s\right)\right)\,ds\\
\frac{\left(\frac{da_{N}\left(w\right)}{dw}\right)}{\left(\frac{db_{N}\left(w\right)}{dw}\right)} & = & \left(N-1\right)\frac{\left(\frac{da_{2}\left(w\right)}{dw}\right)}{\left(\frac{db_{2}\left(w\right)}{dw}\right)}
\end{array}\label{eq:BumpSpc_abNSoln}
\end{equation}

\begin{figure}
\begin{centering}
\includegraphics[viewport=46bp 28bp 720bp 523bp,scale=0.7]{FIG_Bump-Minkowski_space_profiles}
\par\end{centering}
~

\label{fig:BumpSpcProf}

\caption{Profiles for $\ln\left(A\left(w\right)\right)$, $\ln\left(B\left(w\right)\right)$,
$\rho\left(w\right)$, $p\left(w\right)$ as well as the equation
of state~$\frac{p\left(w\right)}{\rho\left(w\right)}$ are shown
for $N=2$. For convenience of display $\beta=10$, where $w=const-\beta x^{1}$.
This space-time would be split at $w=0$ and an arbitrary breadth
of Minkowski space-time inserted there. Because of the bump function
utilized for $b=\ln\left(B\right)$, derivatives of all orders of
the metric vanish at $w=0$, thus the Riemann tensor will be smooth
at the junction with Minkowski space-time. This space-time will also
be interfaced with $\Lambda-K$ space-time at $w=\pm\frac{1}{4}$.
The discontinuity is such that a brane containing positive energy
density will be located here. In practice $\phi-1$ and hence $\beta$
and the equation of state of the brane would be selected such that
the location of the boundary is stationary in this Modified-Minkowski
space-time (it will not generally be fixed in the $\Lambda-K$ space-time).
In practice, the equation of state may be restricted to the range~$\left[-1,1\right]$
and then $\phi-1$ selected. Note that $\kappa\rho\ge\kappa p\ge0$
and $A,B>0$ everywhere in $-\frac{1}{4}\le w\le\frac{1}{4}$. While
$-\ln\left(A\right)\protect\ne\ln\left(B\right)$ except at $w=0$,
the difference is too small to be visible in this graph. To preempt
any possible confusion with the lower log-plot - note that what is
shown is $10^{-40}\le-\ln\left(A\right),\,\ln\left(B\right)\le10^{-8}$
- the $A,B$ change very little from~$1$ over the region of interest. }
\end{figure}

Thus we define the ``Modified-Minkowski'' bulk (to be used in the
interior of the models) to be this bulk ($0\le w\le w_{max}$) spliced
with regular Minkowski space-time at $w=0$ (the latter is at $w<0$).
There is some freedom of choice of $w_{max}$ - for the proposes of
this paper $w_{max}=\frac{1}{4}$ is chosen. One could then reflect
this entire space-time around, say, $w=-1$ to create a bubble of
bulk space-time that is Minkowski in the interior, and is surrounded
by a layer of this perfect-fluid space to allow for boundary condition
matching while avoiding negative energy density on the brane. This
Modified-Minkowski bulk will then be cut off in the outward direction
$w>w_{max}>0$ before any of the parameters of the space become undesirable.
The intent is for a passenger to be provided with a comfortable ride
in the interior Minkowski region, while no negative energy density
is used in the construction of the models presented later in this
paper. 

\subsection{The branes\label{subsec:GR_BackGrnd_Branes} }

\subsubsection{Brane Stress-Energy\label{subsubsec:GR_BackGrnd_Branes_BSET}}

\noindent\fbox{\begin{minipage}[t]{1\columnwidth - 2\fboxsep - 2\fboxrule}%
In this subsection:
\begin{lyxlist}{00.00.0000}
\item [{1)}] The general stress-energy tensor for a bulk space-time with
a codimension-1 membrane is written as a contribution from the bulk
space $T_{(bulk)\mu\nu}$, and a contribution from the brane~$S_{\mu\nu}$.
\item [{2)}] The $T_{(bulk)\mu\nu}$ may be zero, or vacuum energy, or
the perfect fluid of the boundary layer in subsection~\ref{subsubsec:modMnk_defn}.
\item [{3)}] The brane stress-energy~$S_{\mu\nu}$has delta-funcition
support - it is not a source for the bulk Einstein equations. 
\item [{4)}] The standard stress-energy tensor of a perfect fluid $\left(\rho+p\right)u_{\mu}u_{\nu}-pg_{\mu\nu}$
is modified for use on (possibly super-luminal) branes, yielding $\left(\rho+p\right)u_{\mu}u_{\nu}-\left(u^{\alpha}u_{\alpha}\right)ph_{\mu\nu}$,
where $h_{\mu\nu}$ is the metric induced on the brane, and $u^{\mu}$
is the normalized relativistic velocity of the brane in the bulk.
$u^{\alpha}u_{\alpha}=+1$ for a sub-luminal brane, $u^{\alpha}u_{\alpha}=-1$
for a super-luminal brane.
\end{lyxlist}
\end{minipage}}

Define the codimension~1 boundary surface between two bulks as $\Sigma$.
This is the location of the brane. Let $\Sigma_{\left(1\right)}$
represent one side of the location of the brane, and $\Sigma_{\left(2\right)}$
the other side. In the models to be considered later in this paper,
there is a perfect fluid component of the stress-energy on this brane
(with delta function support): 
\begin{equation}
T_{\mu\nu}=T_{(bulk)\mu\nu}+\left(\text{delta function support on }\Sigma\right)S_{\mu\nu}\label{eq:totSE_eq_bulk_plus_brane_SE}
\end{equation}
where the term $\left(\text{delta function support on }\Sigma\right)$
is such that a $N+2$ dimensional volume (bulk) integral over $T_{\mu\nu}$
becomes the sum of a $N+2$ dimensional integral over $T_{(bulk)\mu\nu}$
plus an $N+1$ dimensional brane hypersurface~$\Sigma$ integral
over $S_{\mu\nu}$, as defined in subsection~3.7.1 of~\cite{bib:Poisson_Book_Relativists_Toolkit}.
The form of the perfect fluid stress-energy tensor on brane is 
\begin{equation}
\begin{array}{ccl}
S_{\nu}^{\mu} & = & \zeta\left\{ \left(\rho+p\right)u^{\mu}u_{\nu}-\left(u^{\alpha}u_{\alpha}\right)ph_{\nu}^{\mu}\right\} \\
h_{\nu}^{\mu} & \equiv & g_{\nu}^{\mu}-\frac{1}{\left(n^{\alpha}n_{\alpha}\right)}n^{\mu}n_{\nu}
\end{array}\label{eq:PerfFlu_SE_form_supLum_brane}
\end{equation}
 where $\chi^{\mu}\left(\tau\right)$ is the world-line of the brane,
$\tau$ the proper time of an observer at rest on the brane, and $u^{\mu}\equiv\frac{d\chi^{\mu}}{d\tau}=\dot{\chi}^{\mu}$
its velocity, and where it is assumed that $\rho$ and $p$ are spatially
homogeneous and isotropic in the rest frame (hence only vary with
$\tau$). $n^{\mu}$ are normal vectors to $\Sigma$. $u^{\mu}u_{\mu}=\xi\eta$,
$\eta\equiv sign\left(A\right)$, $\xi=+1$ for time-like $u^{\mu}$
relative to the frame of the metric eq~\ref{eq:Bulk_metric_form},
$\xi=-1$ for space-like $u^{\mu}$, $\xi=0$ for light-like (not
considered here). Note that $n^{\mu}u_{\mu}=0$ and $n^{\mu}n_{\mu}=-\xi\eta$.
The metric induced on $\Sigma$ is $h_{\nu}^{\mu}\equiv g_{\nu}^{\mu}+\frac{1}{\xi\eta}n^{\mu}n_{\nu}$
which also serves as a projection operator to $\Sigma$. There is
a $\xi\eta$ in the term proportional to $h_{\nu}^{\mu}$ in eq~\ref{eq:PerfFlu_SE_form_supLum_brane}
which would normally be $+1$ in the standard form of the stress-energy
tensor for a (subluminal) perfect fluid. This perfect fluid is generalized
to allow for possibly super-luminal $u^{\mu}$, and thus the standard
conservation equation (originating from $T_{0;\mu}^{\mu}=0$) - with
regard to a change in scale factor~$B$ on the brane - requires that
the $+1$ be replaced with $\xi\eta$. Additionally, the $\rho+p$
in the term proportional to $u^{\mu}u_{\nu}$ will keep the same relative
signs so that a vacuum energy-like fluid ($p=-\rho$) is actually
constant. As is later shown, $\xi_{\left(j\right)}$ and $\eta_{\left(j\right)}$
may assume different values across the brane. However, the product
$\xi\eta$ must be the same on either side. This will turn out to
be a crucial constraint on the possible models that may be constructed.
One does need $\zeta=1$ for a sub-luminal $u^{\mu}$ of the fluid
(or, if one is using the alternate sign convention, then $\zeta=-1$),
so it seems natural to choose $\zeta=1$. However, $\zeta=\xi\eta$
would also be consistent with all that has been done so far. The former
choice is what will be used for the stress-energy tensors on the branes,
thus $\zeta=1$ will be assumed. Thus:
\[
\begin{array}{ccl}
S_{\nu}^{\mu} & = & \left(\rho+p\right)u^{\mu}u_{\nu}-\left(u^{\alpha}u_{\alpha}\right)ph_{\nu}^{\mu}\\
 & = & \left(\rho+p\right)u^{\mu}u_{\nu}-\left(u^{\alpha}u_{\alpha}\right)p\left(g_{\nu}^{\mu}+\frac{1}{\xi\eta}n^{\mu}n_{\nu}\right)\\
\\
S & = & g^{\mu\nu}S_{\mu\nu}=\left(u^{\alpha}u_{\alpha}\right)\left(\rho-Np\right)
\end{array}
\]
 
\begin{equation}
\begin{array}{ccl}
S_{\nu}^{\mu} & = & \left(\rho+p\right)u^{\mu}u_{\nu}-\left(u^{\alpha}u_{\alpha}\right)ph_{\nu}^{\mu}\\
S & = & g^{\mu\nu}S_{\mu\nu}=\left(u^{\alpha}u_{\alpha}\right)\left(\rho-Np\right)
\end{array}\label{eq:brane_SEten_PerfFluid}
\end{equation}
We will ignore the $T_{1;\mu}^{\mu}=0$ constraint - appealing to
the unknown nature of the method of confinement of the stress-energy
to the brane. However, one expects a discontinuity in the perpendicular
derivatives of $A$ and $B$ across the brane. Thus, it is not clear
how to evaluate or enforce $\frac{\partial}{\partial x^{1}}\left(\left|A\right|B^{-N\frac{p}{\rho}}\right)=0$
at the brane. Note that $\rho$ and $p$ implicitly contain a factor
to adjust units such that $\kappa\rho$ and $\kappa p$ are unit-less. 

\subsubsection{The Israel Junction Conditions\label{subsubsec:GR_BackGrnd_Branes_IJC} }

\noindent\fbox{\begin{minipage}[t]{1\columnwidth - 2\fboxsep - 2\fboxrule}%
In this subsection:

1) The Israel junction conditions are presented. These specify how
two bulk space-times may be adjoined, with the boundary membrane between
containing a codimension-1 stress-energy tensor.

2) The Israel junction conditions require the metric induced on the
membrane to be consistent, and the difference in the extrinsic curvature
of the boundary, across the boundary, to be related to the codimension-1
stress-energy tensor on the boundary.

3) A stress-energy tensor of the form of subsection~\ref{subsubsec:GR_BackGrnd_Branes_BSET}
is assumed to be on the branes, and the form of the junction conditions
specialized to this assumption are derived.%
\end{minipage}}

Two different space-times may be joined at the boundary~$\Sigma$,
subject to junction conditions that must be satisfied at that boundary.
The Israel junction conditions were derived in~\cite{bib:IJC} by
integrating the projected Einstein equations (formed from the Gauss-Codazzi
equations) over a infinitesimal volume (neighborhood bisected by~$\Sigma$),
and in~\cite{bib:Poisson_Book_Relativists_Toolkit} by canceling
divergences caused by discontinuities in the metric against the component
of the stress-energy tensor with delta-function support. This amounts
to the set of requirements that are called the Israel Junction Conditions.
There must be a consistent induced metric on the boundary, and the
difference in extrinsic curvature~$\Pi_{\mu\nu}$ across the boundary
must satisfy a relation to the stress-energy on the brane: 
\begin{equation}
\begin{array}{l}
\left(-\left[\Pi_{\mu\nu}\right]+h_{\mu\nu}\left[\Pi\right]\right)\left(n^{\mu}n_{\mu}\right)=\kappa S_{\mu\nu}\\
\Pi_{\mu\nu}\equiv h_{\mu}^{\alpha}n_{\nu;\alpha}\\
\left[X\right]\equiv X_{\left(1\right)}-X_{\left(2\right)}
\end{array}\label{eq:IJC_gen}
\end{equation}
where convention is that the normal~$n_{\left(1\right)}^{\mu}$ in
bulk~$\left(1\right)$ points from the side of the world-path being
discarded to the side of the world-path being utilized. In bulk~$\left(2\right)$
the opposite is true - the normal $n_{\left(2\right)}^{\mu}$ points
from the side of the world-path being utilized to the side of the
world-path being discarded. One can take the trace and rewrite the
junction conditions as 
\begin{equation}
\begin{array}{ccl}
\left[\Pi_{\mu\nu}\right]\left(n^{\mu}n_{\mu}\right) & = & -\kappa\left(S_{\mu\nu}-h_{\mu\nu}\frac{1}{N}S\right)\\
 & = & -\left(\left(\kappa\rho+\kappa p\right)u_{\mu}u_{\nu}-h_{\mu\nu}\xi\eta\frac{1}{N}\kappa\rho\right)
\end{array}\label{eq:IJC_gen_dev_A}
\end{equation}
 It is helpful to consider the parallel space-like directions~($\underline{\mu}\in\left\{ 2,\cdots,2+N-1\right\} $)
separate from the directions that are spanned by the relativistic
velocity of the brane~$u^{\mu}$ and the normal to the brane world-path~$n^{\mu}$~($\bar{\mu}\in\left\{ 0,1\right\} $)
: 
\begin{equation}
\begin{array}{ccl}
\Pi_{\bar{\mu}\bar{\nu}} & = & h_{\bar{\mu}}^{\alpha}n_{\bar{\nu};\alpha}\\
\Pi_{\underline{\mu}\underline{\nu}} & = & h_{\underline{\mu}}^{\alpha}n_{\underline{\nu};\alpha}\\
 & = & \frac{1}{2}g_{\underline{\nu}\underline{\mu},\beta}n^{\beta}\\
 & = & \frac{1}{2}g_{\underline{\nu}\underline{\mu}}\frac{B_{,\beta}}{B}n^{\beta}
\end{array}\label{eq:IJC_gen_dev_B}
\end{equation}
 $g_{\bar{\mu}\underline{\nu}}=0$, $n_{\underline{\nu}}=0$, $g_{\bar{\beta}\bar{\nu},\underline{\mu}}=0$
and $n_{\bar{\nu},\underline{\mu}}=0$ implies: 
\begin{equation}
\begin{array}{ccl}
\Pi_{\underline{\mu}\bar{\nu}} & = & h_{\underline{\mu}}^{\alpha}n_{\bar{\nu};\alpha}\\
 & = & \left(g_{\underline{\mu}}^{\alpha}+\frac{1}{\xi\eta}n_{\underline{\mu}}n^{\alpha}\right)\left(n_{\bar{\nu},\alpha}-\Gamma_{\beta\bar{\nu}\alpha}n^{\beta}\right)\\
 & = & n_{\bar{\nu},\underline{\mu}}-\frac{1}{2}g_{\bar{\beta}\bar{\nu},\underline{\mu}}n^{\bar{\beta}}\\
 & = & 0\\
\Pi_{\mu\nu} & = & \left(\Pi_{\overline{\mu}\overline{\nu}}\right)\bigoplus\left(\Pi_{\underline{\mu\nu}}\right)=\left(\begin{array}{cc}
\left(\Pi_{\overline{\mu}\overline{\nu}}\right) & \begin{array}{ccc}
0 & \cdots & 0\\
0 & \cdots & 0
\end{array}\\
\begin{array}{cc}
0 & 0\\
\vdots & \vdots\\
0 & 0
\end{array} & \left(\Pi_{\underline{\mu\nu}}\right)
\end{array}\right)
\end{array}\label{eq:IJC_gen_X_zero}
\end{equation}

The brane stress-energy viewed from either side of the brane must
match. The easiest approach is to impose the junction conditions in
a frame in which the time is proper time along the brane~$\tau$,
and the perpendicular coordinate is proper distance along the normal
to the brane~$\omega$ (with the brane located at $\omega=0$). The
coordinate transformation matrix evaluated at the brane has columns
that are $u^{\mu}$ and $n^{\mu}$. Thus the time-time component of
the extrinsic curvature to be matched is given by $\Pi_{\mu\nu}u^{\mu}u^{\nu}$.
One must additionally match the extrinsic curvature in spacial directions
parallel to~$\Sigma$, and the induced metric on the brane (up to
an overall constant rescaling). The metric induced on the brane in
this frame has the form $h_{\mu\nu}=diag\left(\xi\eta,0,-B,\cdots,-B\right)_{\mu\nu}$
or $ds^{2}=\xi\eta d\tau^{2}-Bd\overrightarrow{x}^{2}$
\begin{equation}
\begin{array}{ccc}
\left[\Pi_{\mu\nu}u^{\mu}u^{\nu}\right] & = & \xi\eta\left(\kappa\rho\left(\frac{N-1}{N}\right)+\kappa p\right)\\
\left[\Pi_{\underline{\mu}\underline{\nu}}\right] & = & -\frac{1}{N}h_{\underline{\mu}\underline{\nu}}\kappa\rho=-\frac{1}{N}g_{\underline{\mu}\underline{\nu}}\kappa\rho
\end{array}\label{eq:IJC_gen_split}
\end{equation}

For a bulk metric of the form of eq~\ref{eq:Bulk_metric_form}, this
yields the following expression for the junction conditions :

\begin{equation}
\begin{array}{cccccc}
\left(1A\right) & N_{\left(1\right)}=N_{\left(2\right)}\\
\left(1B\right) & \xi_{\left(1\right)}\eta_{\left(1\right)}=\xi_{\left(2\right)}\eta_{\left(2\right)}\\
\left(1C\right) & \left.B_{\left(1\right)}\right|_{\Sigma}=\left.B_{\left(2\right)}\right|_{\Sigma} & \Rightarrow & \left[\frac{1}{B}\dot{B}\right]=\left[\frac{1}{B}B_{,\alpha}u^{\alpha}\right] & = & 0\\
\left(2\right) & \left[\Pi_{\underline{\mu\nu}}\right] & = & \left[\frac{1}{2}g_{\underline{\mu\nu}}\frac{1}{B}B_{,\alpha}n^{\alpha}\right] & = & -\frac{1}{N}g_{\underline{\mu\nu}}\kappa\rho\\
 &  &  & \left[\frac{1}{B}B_{,\alpha}n^{\alpha}\right] & = & -\frac{2}{N}\kappa\rho\\
\left(3\right) & \left[\Pi_{\mu\nu}u^{\mu}u^{\nu}\right] & = & \left[h_{\mu}^{\alpha}n_{\nu;\alpha}u^{\mu}u^{\nu}\right] & = & \xi\eta\left(\left(\frac{N-1}{N}\right)\kappa\rho+\kappa p\right)\\
 &  & = & \left[n_{\nu;\mu}u^{\mu}u^{\nu}\right]\\
 &  & = & \left[-n_{\nu}u^{\mu}u_{\,;\mu}^{\nu}\right]\\
 &  & = & \left[-n_{\bar{\nu}}a^{\bar{\nu}}\right]\\
 &  & = & \left[-n_{\nu}u^{\mu}\left(u_{\,,\mu}^{\nu}+u^{\bar{\rho}}\Gamma_{\,\bar{\rho}\bar{\mu}}^{\bar{\nu}}\right)\right]\\
 &  & = & \left[-\dot{u}^{\bar{\nu}}n_{\bar{\nu}}+\frac{1}{2}u^{\bar{\rho}}u^{\bar{\mu}}g_{\bar{\rho}\bar{\alpha},\bar{\nu}}n^{\bar{\nu}}\right]\\
 &  & = & \left[-\dot{u}^{\bar{\nu}}n_{\bar{\nu}}+\frac{1}{2}\xi\eta\left(\ln A\right)_{,\bar{\nu}}n^{\bar{\nu}}\right]
\end{array}\label{eq:IJC_gen_metric}
\end{equation}
Where $a^{\bar{\nu}}\equiv u^{\mu}u_{\,;\mu}^{\nu}$ is the deviation
of the world-sheet of the brane from a bulk geodesic (on one side).
In general, the path of the brane will not follow a bulk geodesic.
The difference in this deviation from a bulk geodesic across the brane
is proportional to $\xi\eta\left(\left(\frac{N-1}{N}\right)\kappa\rho+\kappa p\right)$.
It is this effect that is the mechanism of propulsion proposed in
this paper. Because the contracted Bianchi identities imply $G_{\ ;\mu}^{\mu\nu}=0$,
stress-energy conservation ($T_{\ ;\mu}^{\mu\nu}=0$) is built into
solutions of the Einstein equations, and hence into solutions of the
junction conditions above (when paired with solutions to the Einstein
equations in the bulk). 

\section{Components of a Propulsion Model\label{sec:Model_Components}}

\noindent\fbox{\begin{minipage}[t]{1\columnwidth - 2\fboxsep - 2\fboxrule}%
In this section:
\begin{lyxlist}{00.00.0000}
\item [{1)}] The components introduced in Section~\ref{sec:GR_IJC_background}
are combined: a bulk space-time is specified as side~$\left(1\right)$
of the membrane and another bulk space-time is specified as side~$\left(2\right)$. 
\item [{2)}] The membrane is assumed to contain a codimension-1 stress-energy
tensor of the nature expressed in subsection~\ref{subsubsec:GR_BackGrnd_Branes_BSET}.
It is explored what constraints are necessary and sufficient for the
junction conditions to be satisfied.
\item [{3)}] Side~$\left(1\right)$ being Minkowski, and Side~$\left(2\right)$
being $\Lambda-K$ is considered. 
\item [{4)}] Side~$\left(1\right)$ and Side~$\left(2\right)$ both being
$\Lambda-K$ is considered, though the values of $\Lambda$ and $K$
may differ. 
\item [{5)}] Side~$\left(1\right)$ being Modified-Minkowski, and Side~$\left(2\right)$
being $\Lambda-K$ is considered. 
\item [{6)}] The nature of geodesics in the $\Lambda-K$ space-time is
considered.
\end{lyxlist}
\end{minipage}}

The first step in using the junction conditions to construct a model
of propulsion is to consider the possibilities for the nature of the
bulk on either side of the brane. There are many possibilities, yet
the models presented in this paper will utilize only Minkowski joined
to $\Lambda-K$; and Modified-Minkowski joined to $\Lambda-K$. The
interface of $\Lambda-K$ joined to $\Lambda-K$ (with possibly different
parameters) will also be discussed in this section. As will be shown,
even with this limited set, profoundly useful models are possible. 

Consider the junction condition~2 in eq~\ref{eq:IJC_gen_metric}.
For Minkowski space-time the expression $\ln\left(B\right)_{,\alpha}n^{\alpha}$
will vanish. Thus if the bulk on both sides of the brane is Minkowski,
it must be true that $\rho=0$ - there can be no stress-energy on
the brane, hence no pressure either. Then $\left[a^{\bar{\nu}}n_{\bar{\nu}}\right]=0$.
There may be a boundary, but there is no physical effect that is useful
for the propulsion models presented in section~\ref{sec:Models_presented}.
Thus, it will be assumed that the bulk on atleast one side of the
brane is not Minkowski. In the analysis that follows, $\chi_{\left(j\right)}^{\mu}\left(\tau\right)$
is the world-path of the brane in coordinates $x^{\mu}$ on side~$\left(j\right)$.
Let $\mathcal{Y}\left(\tau\right)\equiv\left.y\right|_{x^{\mu}=\chi^{\mu}\left(\tau\right)}$.
However, in cases where it is obvious, evaluation of a quantity on
the brane world-path $x^{\mu}=\chi^{\mu}\left(\tau\right)$ will not
be given special notation (ie: $B$ will be used for both $B\left(y\right)$
and $B\left(\mathcal{Y}\left(\tau\right)\right)$). 

The non-zero components of relativistic velocity $u_{\left(j\right)}^{\mu}$
and normal $n_{\left(j\right)}^{\mu}$ of the brane in the bulk on
side~$\left(j\right)$ is given by: 
\begin{equation}
\begin{array}{ccccccc}
u_{\left(j\right)}^{\mu}=u_{\left(j\right)}^{\bar{\mu}} & = & \left(\begin{array}{c}
\dot{\chi}_{\left(2\right)}^{0}\\
\dot{\chi}_{\left(2\right)}^{1}
\end{array}\right)^{\bar{\mu}} & = & \left(\begin{array}{c}
\epsilon_{u0\left(j\right)}\xi_{\left(j\right)}\sqrt{Y_{\left(j\right)}^{2}+\frac{1}{A_{\left(j\right)}}\xi\eta}\\
Y_{\left(j\right)}
\end{array}\right)^{\bar{\mu}} &  & \epsilon_{u0\left(j\right)}=\pm1\\
n_{\left(j\right)}^{\mu}=n_{\left(j\right)}^{\bar{\mu}} & = & s_{\left(j\right)}\left(\begin{array}{c}
u_{\left(j\right)}^{1}\\
u_{\left(j\right)}^{0}
\end{array}\right)^{\bar{\mu}} & = & s_{\left(j\right)}\left(\begin{array}{c}
\dot{\chi}_{\left(2\right)}^{1}\\
\dot{\chi}_{\left(2\right)}^{0}
\end{array}\right)^{\bar{\mu}} &  & s_{\left(j\right)}=\pm1
\end{array}\label{eq:gen_relvel_form}
\end{equation}
 where $s_{\left(j\right)}\in\left\{ -1,+1\right\} $ describes which
side of the brane world-path~$\chi_{\left(j\right)}^{\mu}\left(\tau\right)$
is to be kept, and which is discarded. It is set such that $n_{\left(j\right)}^{\mu}=s_{\left(j\right)}\left(\begin{array}{cc}
0 & 1\\
1 & 0
\end{array}\right)_{\bar{\alpha}}^{\bar{\mu}}u_{\left(j\right)}^{\bar{\alpha}}$ points in the correct direction: $n_{\left(1\right)}^{\mu}$ points
into the side of bulk~$\left(1\right)$ that is discarded, $n_{\left(2\right)}^{\mu}$
points into the side of bulk~$\left(2\right)$ that is kept. Thus
picking a sign for $s_{\left(j\right)}$ is equivalent to choosing,
in bulk~$\left(j\right)$, which side of the brane world-path~$\chi_{\left(j\right)}^{\mu}\left(\tau\right)$
is to be identified as side~$\left(j\right)$ of the brane. When
the space is not symmetric across the $\chi_{\left(j\right)}^{\mu}\left(\tau\right)$
(such as in the $\Lambda-K$ space-time) this is a crucial choice.
The parameterization of $u_{\left(j\right)}^{\mu}$ and $n_{\left(j\right)}^{\mu}$
in eq~\ref{eq:gen_relvel_form} follows from from the constrains:
$u_{\left(j\right)}^{\mu}u_{\left(j\right)\mu}=\xi\eta$, $n_{\left(j\right)}^{\mu}n_{\left(j\right)\mu}=-\xi\eta$
and $u_{\left(j\right)}^{\mu}n_{\left(j\right)\mu}=0$. 

\subsection{The bulk on one side of the brane is Minkowski space-time, the other
is $\Lambda-K$\label{subsec:MnkLK_constr} }

\noindent\fbox{\begin{minipage}[t]{1\columnwidth - 2\fboxsep - 2\fboxrule}%
In this subsection: 
\begin{lyxlist}{00.00.0000}
\item [{1)}] The case is considered in which side~$\left(1\right)$ of
the brane is Minkowski space-time, and side~$\left(2\right)$ is
$\Lambda-K$ 
\item [{2)}] Because the metric induced on the brane from side~$\left(1\right)$
is constant, the brane world-path in side~$\left(2\right)$ must
be $y=const$. This is condition~1 of the junction conditions eq~\ref{eq:IJC_gen_metric}. 
\item [{3)}] Enforcing condition~2 of the junction conditions eq~\ref{eq:IJC_gen_metric}
yields the particular value of $y$ (equivalently $B$) of the $y=const$
world-path. 
\item [{4)}] Enforcing condition~3 of the junction conditions eq~\ref{eq:IJC_gen_metric}
yields the world-path of the brane in side~$\left(1\right)$. Curvature
of this path is the apparent acceleration of the brane as seen by
an observer on side~$\left(1\right)$. 
\item [{5)}] Note that the world-path of the brane in side~$\left(1\right)$
will not transition between sub-luminal and super-luminal. For this
to happen, some parameters treated as static in this model must vary. 
\end{lyxlist}
\end{minipage}}

Let the bulk on side~$\left(1\right)$ of the brane be Minkowski,
and side~$\left(2\right)$ be $\Lambda-K$. Agreement of the induced
brane metric implies that path followed by the brane on the $\Lambda-K$
space-time side must maintain $B=constant$, which is a $\mathcal{Y}_{\left(2\right)}=constant$
path. This significantly simplifies analysis. Minkowski space-time
on side~$\left(1\right)$ implies that $\ln\left(B_{\left(1\right)}\right)_{,\beta}n_{\left(1\right)}^{\beta}=0$,
and thus junction condition~2 of eq~\ref{eq:IJC_gen_metric} then
implies 
\begin{equation}
\begin{array}{cccc}
\left[\Pi_{\underline{\mu\nu}}\right] & = & \left[\frac{1}{2}g_{\underline{\mu\nu}}\ln\left(B\right)_{,\beta}n^{\beta}\right] & =-h_{\underline{\mu\nu}}\frac{1}{N}\kappa\rho\\
 & = & -\frac{1}{2}g_{\underline{\mu\nu}}\ln\left(B_{\left(2\right)}\right)_{,\beta}n_{\left(2\right)}^{\beta}\\
\\
\ln\left(B_{\left(2\right)}\right)_{,\beta}n_{\left(2\right)}^{\beta} & = & \frac{2}{N}\kappa\rho
\end{array}\label{eq:MnkLK_IJC_pp}
\end{equation}

Differentiating $\mathcal{Y}=constant$ with respect to proper time~$\tau$
of an observer at rest with respect to the brane yields $\chi_{\left(2\right)}^{0}u_{\left(2\right)}^{0}-\chi_{\left(2\right)}^{1}u_{\left(2\right)}^{1}=0$.
Thus 
\begin{equation}
\begin{array}{ccccc}
u_{\left(2\right)}^{\mu} & = & u_{\left(2\right)}^{\bar{\mu}} & = & \Psi_{\left(2\right)}\left(\begin{array}{c}
\chi_{\left(2\right)}^{1}\\
\chi_{\left(2\right)}^{0}
\end{array}\right)^{\bar{\mu}}\\
 &  & \Psi_{\left(2\right)} & \equiv & \frac{\dot{\chi}_{\left(2\right)}^{0}}{\chi_{\left(2\right)}^{1}}\\
n_{\left(2\right)}^{\mu} & = & n_{\left(2\right)}^{\bar{\mu}} & = & s_{\left(2\right)}\Psi_{\left(2\right)}\left(\begin{array}{c}
\chi_{\left(2\right)}^{0}\\
\chi_{\left(2\right)}^{1}
\end{array}\right)^{\bar{\mu}}\\
 &  & s_{\left(2\right)} & \in & \left\{ +1,-1\right\} 
\end{array}\label{eq:MnkLK_gen_soln_A}
\end{equation}
   This implies 
\begin{equation}
\begin{array}{cccc}
\frac{2}{N}\kappa\rho & = & \ln\left(B_{\left(2\right)}\right)_{,\alpha}n_{\left(2\right)}^{\alpha}\\
 & = & \ln\left(B_{\left(2\right)}\right)_{,y}y_{,\alpha}n_{\left(2\right)}^{\alpha}\\
 & = & 2\frac{B_{\left(2\right),y}}{B}s_{\left(2\right)}\Psi_{\left(2\right)}\mathcal{Y}_{\left(2\right)}\\
\\
\Psi_{\left(2\right)} & = & s_{\left(2\right)}\frac{B\kappa\rho}{NB_{\left(2\right),y}\mathcal{Y}_{\left(2\right)}}
\end{array}\label{eq:MnkLK_Psi_expr}
\end{equation}
Thus the world-path of the brane in bulk~$\left(2\right)$ obeys:
\begin{equation}
\begin{array}{ccc}
u_{\left(2\right)}^{\bar{\mu}} & = & \Psi_{\left(2\right)}\left(\begin{array}{c}
\chi_{\left(2\right)}^{1}\\
\chi_{\left(2\right)}^{0}
\end{array}\right)^{\bar{\mu}}\\
\epsilon_{u0\left(2\right)}\xi_{\left(2\right)} & = & \text{sign}\left(\Psi_{\left(2\right)}\chi_{\left(2\right)}^{1}\right)\\
 & = & s_{\left(2\right)}\eta_{\left(2\right)}\text{sign}\left(\kappa\rho\right)\text{sign}\left(\mathcal{Y}_{\left(2\right)}\chi_{\left(2\right)}^{1}\right)\\
n_{\left(2\right)}^{\bar{\mu}} & = & s_{\left(2\right)}\Psi_{\left(2\right)}\left(\begin{array}{c}
\chi_{\left(2\right)}^{0}\\
\chi_{\left(2\right)}^{1}
\end{array}\right)^{\bar{\mu}}\\
\Psi_{\left(2\right)} & \equiv & s_{\left(2\right)}\frac{B\kappa\rho}{N\mathcal{Y}_{\left(2\right)}B_{\left(2\right),y}}=s_{\left(2\right)}\frac{\sqrt{B}\kappa\rho}{2\mathcal{Y}_{\left(2\right)}A_{\left(2\right)}}
\end{array}\label{eq:MnkLK_gen_soln_B}
\end{equation}
 $\mathcal{Y}_{\left(2\right)}$, $B_{\left(2\right)}$, $A_{\left(2\right)}$
being constant along the world-path of the brane in bulk~$\left(2\right)$
implies $\rho$ and $\Psi_{\left(2\right)}$ are also constant. Thus
differentiation of $u_{\left(2\right)}^{\mu}$ and $n_{\left(2\right)}^{\mu}$
with respect to $\tau$ yields: 
\begin{equation}
\begin{array}{ccl}
\dot{u}_{\left(2\right)}^{\mu} & = & \Psi_{\left(2\right)}s_{\left(2\right)}n_{\left(2\right)}^{\mu}\\
\dot{n}_{\left(2\right)}^{\mu} & = & \Psi_{\left(2\right)}s_{\left(2\right)}u_{\left(2\right)}^{\mu}\\
\dot{u}_{\left(2\right)}^{\bar{\nu}}n_{\left(2\right)\bar{\nu}} & = & -\xi\eta s_{\left(2\right)}\Psi_{\left(2\right)}
\end{array}\label{eq:MnkLK_u_n_derivs}
\end{equation}
In the models under consideration the stress-energy on the brane
is a perfect fluid, and thus - in the absence of other interactions
- stress-energy conservation requires that the only mechanism of change
of $\rho$ is due to change of the induced scale factor $B$ on the
brane. Thus, in these models, a constant $B$ along the brane world-path
implies a constant $\rho$ along that world-path. The above implies
that $\ddot{u}_{\left(2\right)}^{\mu}=\Psi_{\left(2\right)}^{2}u_{\left(2\right)}^{\mu}$
- a hyperbolic trajectory in $\Lambda-K$ space. This is consistent
with a $\mathcal{Y}_{\left(2\right)}=constant$ path.

From eq~\ref{eq:MnkLK_gen_soln_B} and the requirement that $\dot{\chi}_{\left(2\right)}^{0}\ge0$
one can make a statement about the relation between the sign of $\xi_{\left(2\right)}$
and the region of the $x^{0},x^{1}$ coordinate plane occupied, and
also between the sign of $\xi\eta$ and the sign of $B_{\left(2\right),z}$
(and hence the signs and/or relative dominance of $K$ vs $\kappa\rho_{\Lambda}$
bulk~$\left(2\right)$). For all models considered in this paper,
it will be assumed that $\kappa\rho_{\Lambda}>0$, $K<0$ (for continuity
of $A$ as $y$ changes sign), $-2\lambda>0$ (to avoid coordinate
singularities), $B_{,y}>0,\ \text{sign}\left(\chi_{\left(2\right)}^{1}\right)=\text{sign}\left(\dot{\chi}_{\left(2\right)}^{0}\right)$.
Note the constraint relation between $s_{\left(2\right)}$, $\xi_{\left(2\right)}$
and $\text{sign}\left(\kappa\rho\right)$ : 
\begin{equation}
\begin{array}{ccccc}
\xi_{\left(2\right)}=+1 & \Rightarrow & -2\lambda_{\left(2\right)}\mathcal{Y}_{\left(2\right)}<0 & \Rightarrow & \left|\chi_{\left(2\right)}^{0}\right|<\left|\chi_{\left(2\right)}^{1}\right|\\
\xi_{\left(2\right)}=-1 & \Rightarrow & -2\lambda_{\left(2\right)}\mathcal{Y}_{\left(2\right)}>0 & \Rightarrow & \left|\chi_{\left(2\right)}^{0}\right|>\left|\chi_{\left(2\right)}^{1}\right|\\
\xi\eta & = & -\text{sign}\left(B_{\left(2\right),z}\right)<0\\
s_{\left(2\right)} & = & -\xi_{\left(2\right)}\text{sign}\left(\kappa\rho\right)
\end{array}\label{eq:MnkLK_gen_soln_D}
\end{equation}
Since the models are designed to avoid negative energy density, it
is generally true that $s_{\left(2\right)}=-\xi_{\left(2\right)}$.
If the mandated sign of $s_{\left(2\right)}$ would cause the region
of $\Lambda-K$ space-time that is utilized in a model to be that
of the wrong side of the brane world-path, then

one is forced to modify Minkowski space, as is done in subsection~\ref{subsubsec:modMnk_defn}.
What is meant by ``wrong side of $\Lambda-K$ space

Enforcing the $u_{\left(2\right)}^{\mu}u_{\left(2\right)\mu}=\xi\eta$
condition selects a particular $\mathcal{Y}_{\left(2\right)}=constant$
path for the brane in $\Lambda-K$ space - yielding an expression
for $B$ in terms of the brane energy density $\kappa\rho$ and the
bulk~$\left(2\right)$ quantities $K$, $\kappa\rho_{\Lambda}$:
\begin{equation}
\begin{array}{ccc}
\xi\eta & = & u_{\left(2\right)}^{\mu}u_{\left(2\right)\mu}\\
 & = & A_{\left(2\right)}\Psi_{\left(2\right)}^{2}\left(\left(\chi_{\left(2\right)}^{1}\right)^{2}-\left(\chi_{\left(2\right)}^{0}\right)^{2}\right)\\
 & = & -A_{\left(2\right)}\Psi_{\left(2\right)}^{2}\left(\frac{1}{\left(-2\lambda_{\left(2\right)}\right)}\mathcal{Y}_{\left(2\right)}\right)\\
 & = & \frac{-B\left(\kappa\rho\right)^{2}}{4\left(-2\lambda_{\left(2\right)}\mathcal{Y}_{\left(2\right)}\right)A_{\left(2\right)}}\\
\left(-2\lambda_{\left(2\right)}\mathcal{Y}_{\left(2\right)}\right)A_{\left(2\right)} & = & \frac{-B\left(\kappa\rho\right)^{2}}{4\xi\eta}\\
B^{\frac{3}{2}}\left(\kappa\rho\right)^{2} & = & -\xi\eta2N\left(\frac{2}{N}B^{\frac{2-N}{2}}K_{\left(2\right)}+\frac{1}{N+1}B^{\frac{3}{2}}\kappa\rho_{\Lambda\left(2\right)}\right)\\
-4\xi\eta B^{-\frac{N+1}{2}}K_{\left(2\right)} & = & \xi\eta\frac{2N}{N+1}\kappa\rho_{\Lambda\left(2\right)}+\left(\kappa\rho\right)^{2}\\
B_{B} & \equiv & \left(\frac{-4\xi\eta K_{\left(2\right)}}{\xi\eta\frac{2N}{N+1}\kappa\rho_{\Lambda\left(2\right)}+\left(\kappa\rho\right)^{2}}\right)^{\frac{2}{N+1}}
\end{array}\label{eq:MnkLK_derivof_Bsoln}
\end{equation}
Thus one obtains a solution for $B$, which will be called~$B_{B}$.
This expression for $B_{B}$ can be plugged into the expression for
$y_{\left(2\right)}$ that one obtains from integrating $B_{\left(2\right),y}$
to yield the value of $y$ of the world-path of the brane in bulk~$\left(2\right)$
\begin{equation}
B_{B\left(2\right)}=\left(\frac{-\xi\eta4K_{\left(2\right)}}{\xi\eta\frac{2N}{N+1}\kappa\rho_{\Lambda\left(2\right)}+\left(\kappa\rho\right)^{2}}\right)^{\frac{2}{N+1}}\label{eq:MnkLK_Bsoln}
\end{equation}
Note that eq~\ref{eq:MnkLK_Bsoln} is well-defined: the fraction
on the right-hand side of eq~\ref{eq:MnkLK_Bsoln} is guaranteed
non-negative as a consequence of the constraint $\text{sign}\left(B_{\left(2\right),z}\right)=-\xi\eta$
declared in eq~\ref{eq:MnkLK_gen_soln_D}. That is, if $\xi\eta=-1$
then atleast one of $\left\{ K_{\left(2\right)},\,\kappa\rho_{\Lambda\left(2\right)}\right\} $
must be negative, and the value of $B$ must be such that the dominant
term in $B_{\left(2\right),z}=\frac{2}{N}B^{\frac{2-N}{2}}K_{\left(2\right)}+\frac{1}{N+1}B^{\frac{3}{2}}\kappa\rho_{\Lambda\left(2\right)}$
is negative. If $\xi\eta=+1$, then then atleast one of $\left\{ K_{\left(2\right)},\,\kappa\rho_{\Lambda\left(2\right)}\right\} $
must be positive, and the value of $B$ must be such that the dominant
term in $B_{\left(2\right),z}$ is positive.    Note that $\mathcal{Y}_{\left(2\right)}=constant$
still allows for one degree of freedom in the coordinate position
of the system. The position along the $\mathcal{Y}_{\left(2\right)}=constant$
curve is  set if one knows the coordinate time of the system in bulk~$\left(2\right)$
(possibly up to a reflection symmetry). 

Plugging this value of $B_{B}$ into the expressions for $B_{\left(2\right),y}$,
$A_{\left(2\right)}$ and $\Psi_{\left(2\right)}$ one obtains: 
\begin{equation}
\begin{array}{ccc}
\left.B_{\left(2\right),y}\right|_{B=B_{B}} & = & \frac{-B_{B}^{\frac{3}{2}}\left(\kappa\rho\right)^{2}}{\left(-2\lambda_{\left(2\right)}\mathcal{Y}_{\left(2\right)}\right)2N\xi\eta}\\
\left.A_{\left(2\right)}\right|_{B=B_{B}} & = & \frac{-B_{B}\left(\kappa\rho\right)^{2}}{\left(-2\lambda_{\left(2\right)}\mathcal{Y}_{\left(2\right)}\right)4\xi\eta}\\
\frac{B_{B}}{B_{H}} & = & \left(1+\frac{1}{\xi\eta}\frac{\left(N+1\right)}{2N}\frac{\left(\kappa\rho\right)^{2}}{\kappa\rho_{\Lambda}}\right)^{-\frac{2}{N+1}}\\
\Psi_{\left(2\right)} & = & s_{\left(2\right)}\frac{-\xi\eta\sqrt{B_{H}}}{\sqrt{B_{B}}}\\
 & = & -\xi\eta s_{\left(2\right)}\left(1+\frac{1}{\xi\eta}\frac{\left(N+1\right)}{2N}\frac{\left(\kappa\rho\right)^{2}}{\kappa\rho_{\Lambda}}\right)^{\frac{1}{N+1}}
\end{array}\label{eq:MnkLK_BB_into_A_B_Psi}
\end{equation}

The world-path of the brane in bulk~$\left(2\right)$ is described
in eqs~\ref{eq:MnkLK_gen_soln_B},\ref{eq:MnkLK_Bsoln}. When this
solution is plugged into junction conditions~1~and~2, those conditions
are satisfied. However, the world-path of the brane has in bulk~$\left(1\right)$
- Minkowski space-time - has not yet been described. For this purpose,
one now imposes junction condition~3 and solves for the world-path
of the brane in bulk~$\left(1\right)$. The general form of the
extrinsic curvature that will be used for junction condition~3 is
\begin{equation}
\Pi_{\mu\nu}u^{\mu}u^{\nu}=-\dot{u}^{\bar{\nu}}n_{\bar{\nu}}+\frac{1}{2}\xi\eta\left(\ln A\right)_{,\bar{\nu}}n^{\bar{\nu}}\label{eq:MnkLK_IJC_3_repeat}
\end{equation}
Computing $\Pi_{\mu\nu}u^{\mu}u^{\nu}$ on the Minkowski side of the
boundary yields:
\begin{equation}
\left(\Pi_{\mu\nu}u^{\mu}u^{\nu}\right)_{\left(1\right)}=-\dot{u}_{\left(1\right)}^{\alpha}n_{\left(1\right)\alpha}\label{eq:MnkLK_ExtCurvUU_s1}
\end{equation}
Computing $\Pi_{\mu\nu}u^{\mu}u^{\nu}$ on the $\Lambda-K$ side of
the boundary (using eqs~\ref{eq:MnkLK_gen_soln_B},\ref{eq:MnkLK_gen_soln_D}~and~\ref{eq:MnkLK_Bsoln})
yields: 
\begin{equation}
\begin{array}{ccc}
\left(\Pi_{\mu\nu}u^{\mu}u^{\nu}\right)_{\left(2\right)} & = & -\dot{u}_{\left(2\right)}^{\bar{\nu}}n_{\left(2\right)\bar{\nu}}+\frac{1}{2}\xi\eta\left(\ln A_{\left(2\right)}\right)_{,y}y_{\left(2\right),\bar{\nu}}n_{\left(2\right)}^{\bar{\nu}}\\
 & = & \xi\eta s_{\left(2\right)}\Psi_{\left(2\right)}+\xi\eta\left(\ln A_{\left(2\right)}\right)_{,y}s_{\left(2\right)}\Psi_{\left(2\right)}\mathcal{Y}_{\left(2\right)}\\
 & = & \xi\eta s_{\left(2\right)}\Psi_{\left(2\right)}\mathcal{Y}_{\left(2\right)}\left(\ln\left(A_{\left(2\right)}\left(-2\lambda_{\left(2\right)}y_{\left(2\right)}\right)\right)\right)_{,y}\\
 & = & \xi\eta\frac{1}{N}B\kappa\rho\left(\ln\left(A_{\left(2\right)}\left(-2\lambda_{\left(2\right)}y_{\left(2\right)}\right)\right)\right)_{,B}\\
 & = & \xi\eta\frac{B\kappa\rho}{N}\left(\ln\left(\frac{N}{2\sqrt{B}}B_{\left(2\right),z}\right)\right)_{,B}\\
 & = & -\left(\xi\eta\right)^{2}\frac{1}{2\left(\kappa\rho\right)}\left(\frac{N-1}{N}B^{-\frac{1+N}{2}}\left(-4K\right)+\frac{4}{N+1}\kappa\rho_{\Lambda}\right)\\
 & = & -\xi\eta\kappa\rho\left(\xi\eta\frac{\kappa\rho_{\Lambda\left(2\right)}}{\left(\kappa\rho\right)^{2}}+\frac{N-1}{2N}\right)
\end{array}\label{eq:MnkLK_ExtCurvUU_s2}
\end{equation}
From junction condition~3 of eq~\ref{eq:IJC_gen_metric}
\begin{equation}
\begin{array}{ccc}
\left[\Pi_{\mu\nu}u^{\mu}u^{\nu}\right] & = & \xi\eta\kappa\rho\left(\frac{p}{\rho}+\frac{N-1}{N}\right)\end{array}\label{eq:MnkLK_ExtCurvUU_braneSE}
\end{equation}
Combining eqs~\ref{eq:MnkLK_ExtCurvUU_s1}~and~\ref{eq:MnkLK_ExtCurvUU_s2}
yields:
\begin{equation}
\begin{array}{ccc}
\left[\Pi_{\mu\nu}u^{\mu}u^{\nu}\right] & = & \left(\Pi_{\mu\nu}u^{\mu}u^{\nu}\right)_{\left(1\right)}-\left(\Pi_{\mu\nu}u^{\mu}u^{\nu}\right)_{\left(2\right)}\\
 & = & -\dot{u}_{\left(1\right)}^{\alpha}n_{\left(1\right)\alpha}+\xi\eta\kappa\rho\left(\xi\eta\frac{\kappa\rho_{\Lambda\left(2\right)}}{\left(\kappa\rho\right)^{2}}+\frac{N-1}{2N}\right)
\end{array}\label{eq:MnkLK_ExtCurvUU_discont}
\end{equation}
Thus junction condition~3 implies the world-path of the brane in
bulk~$\left(1\right)$ described by: 
\begin{equation}
\dot{u}_{\left(1\right)}^{\mu}n_{\left(1\right)\mu}=-\xi\eta\kappa\rho\left(\frac{p}{\rho}+\frac{N-1}{2N}-\xi\eta\frac{\kappa\rho_{\Lambda\left(2\right)}}{\left(\kappa\rho\right)^{2}}\right)\label{eq:MnkLK_EoM1_a}
\end{equation}
For operation in our usual $3+1$ Universe, $N=2$. Suppose $\kappa\rho_{\Lambda},\kappa\rho>0$,
yet $\xi\eta=-1$. The constraint regarding $B$ expressed in eq~\ref{eq:MnkLK_Bsoln}
can always be satisfied by making $\kappa\rho$ sufficiently large.
This also insures the $\kappa\rho$ term in eq~\ref{eq:MnkLK_EoM1_a}
dominates, and thus fortunately, the right-hand side of eq~\ref{eq:MnkLK_EoM1_a}
can be made either positive or negative by setting the equation of
state of the stress-energy on the brane $\frac{p}{\rho}$ to something
within the usual range of $\left[-1,1\right]$. 

Using eq~\ref{eq:gen_relvel_form} one finds: 
\begin{equation}
\dot{u}_{\left(1\right)}^{\mu}n_{\left(1\right)\mu}=s_{\left(1\right)}\epsilon_{u0\left(1\right)}\xi_{\left(1\right)}\frac{-\xi\eta}{\sqrt{Y_{\left(1\right)}^{2}+\xi_{\left(1\right)}}}\dot{Y}_{\left(1\right)}\label{eq:MnkLK_udot_n_relvel}
\end{equation}
Combining eq~\ref{eq:MnkLK_EoM1_a} and eq~\ref{eq:MnkLK_udot_n_relvel}
yields: 
\begin{equation}
\frac{1}{\sqrt{Y_{\left(1\right)}^{2}+\xi_{\left(1\right)}}}\dot{Y}_{\left(1\right)}=s_{\left(1\right)}\epsilon_{u0\left(1\right)}\xi_{\left(1\right)}\kappa\rho\left(\frac{p}{\rho}+\frac{N-1}{2N}-\xi\eta\frac{\kappa\rho_{\Lambda\left(2\right)}}{\left(\kappa\rho\right)^{2}}\right)\label{eq:MnkLK_Ydot_expr}
\end{equation}
Note that since $\eta_{\left(1\right)}=+1$ in bulk~$\left(1\right)$
due to it being Minkowski, one may replace $\xi_{\left(1\right)}$
with $\xi\eta$. The left-hand side of eq~\ref{eq:MnkLK_Ydot_expr}
integrates to an $\text{arcsinh}\left(\right)$ or $\text{arccosh}\left(\right)$,
depending on the sign of~$\xi\eta$. Thus one finds an expression
for $u_{\left(1\right)}^{1}$: 
\begin{equation}
Y_{\left(1\right)}=\left\{ \begin{array}{ccccc}
\epsilon_{u0\left(1\right)}\xi_{\left(1\right)}\sinh\left(\int\alpha d\tau+v_{0}\right) &  & \text{for} &  & \xi\eta=+1\\
\epsilon_{u0\left(1\right)}\xi_{\left(1\right)}\cosh\left(\int\alpha d\tau+v_{0}\right) &  & \text{for} &  & \xi\eta=-1\\
\\
\alpha\equiv s_{\left(1\right)}\kappa\rho\left(\frac{p}{\rho}+\frac{N-1}{2N}-\xi\eta\frac{\kappa\rho_{\Lambda\left(2\right)}}{\left(\kappa\rho\right)^{2}}\right)
\end{array}\right\} \label{eq:MnkLK_mnkYaccel}
\end{equation}
 Where $v_{0}$ is an arbitrary constant of integration. Note that
$\alpha$ need not be constant because the equation of state on the
brane~$\frac{p}{\rho}$ may vary. Thus the relativistic velocity
of the brane in bulk~$\left(1\right)$ is 
\begin{equation}
\begin{array}{ccccc}
u_{\left(1\right)}^{\bar{\mu}}=\left(\begin{array}{c}
\epsilon_{u0\left(1\right)}\xi_{\left(1\right)}\cosh\left(\int\alpha d\tau+v_{0}\right)\\
\epsilon_{u0\left(1\right)}\xi_{\left(1\right)}\sinh\left(\int\alpha d\tau+v_{0}\right)
\end{array}\right)^{\bar{\mu}} &  & \text{for} &  & \xi\eta=+1\\
u_{\left(1\right)}^{\bar{\mu}}=\left(\begin{array}{c}
\epsilon_{u0\left(1\right)}\xi_{\left(1\right)}\sinh\left(\int\alpha d\tau+v_{0}\right)\\
\epsilon_{u0\left(1\right)}\xi_{\left(1\right)}\cosh\left(\int\alpha d\tau+v_{0}\right)
\end{array}\right)^{\bar{\mu}} &  & \text{for} &  & \xi\eta=-1
\end{array}\label{eq:MnkLK_mnkrelvel}
\end{equation}
One may compare $\alpha$ to an acceleration, $v_{0}$ to an initial
velocity, and $\int\alpha d\tau+v_{0}$ to an instantaneous velocity
in bulk~$\left(1\right)$. By varying the equation of state on the
brane, the latter may be made arbitrarily positive or negative - or
set to zero. Thus for suitable tuning of the parameters $\frac{p}{\rho}$,
$\kappa\rho$, $\kappa\rho_{\Lambda}$, the brane in bulk~$\left(1\right)$
- the Minkowski side - can be made to accelerate, decelerate or remain
stationary. 

Having $\kappa\rho$ sufficiently larger than $\kappa\rho_{\Lambda}$
insures that the $\left(\kappa\rho\right)^{2}$ term in eq~\ref{eq:MnkLK_EoM1_a}
dominates, which in tern ensures that value of $B$ in eq~\ref{eq:MnkLK_Bsoln}
exists. Additionally, with the $\kappa\rho$ term in eq~\ref{eq:MnkLK_EoM1_a}
dominant, the right-hand side of eq~\ref{eq:MnkLK_EoM1_a} can be
made either positive or negative by setting the equation of state
of the stress-energy on the brane $\frac{p}{\rho}$ to a value within
the range of $\left[-1,1\right]$. Equivalently, the value of the
equation of state on the brane may be tuned to a value within $\left[-1,1\right]$
that will yield $\alpha$ in eqs~\ref{eq:MnkLK_mnkYaccel},\ref{eq:MnkLK_mnkrelvel}
positive, zero or negative. Since $v_{0}$ corresponds to an initial
velocity, this implies that the position of the brane in bulk~$\left(1\right)$
can be made to move in either perpendicular direction or remain stationary
by tuning the equation of state of the stress-energy on the brane.
Since one need not tune $K$ or $\kappa\rho_{\Lambda}$ in the bulk,
in model-building the motion of multiple branes in a single space-time
may be specified independently. This will be important for the propulsion
models considered in section~\ref{sec:Models_presented}. With the
above solutions all of the junction conditions of eq~\ref{eq:IJC_gen_metric}~(1-3)
are satisfied. 

This is a summary of constraints for the case where bulk~$\left(1\right)$
is Minkowski space-time, and bulk~$\left(2\right)$ is $\Lambda-K$:
\begin{equation}
\begin{array}{ccccc}
\xi_{\left(2\right)}=+1 & \Rightarrow & -2\lambda_{\left(2\right)}\mathcal{Y}_{\left(2\right)}<0 & \Rightarrow & \left|\chi_{\left(2\right)}^{0}\right|<\left|\chi_{\left(2\right)}^{1}\right|\\
\xi_{\left(2\right)}=-1 & \Rightarrow & -2\lambda_{\left(2\right)}\mathcal{Y}_{\left(2\right)}>0 & \Rightarrow & \left|\chi_{\left(2\right)}^{0}\right|>\left|\chi_{\left(2\right)}^{1}\right|\\
\xi\eta & = & -\text{sign}\left(B_{\left(2\right),z}\right)=-\text{sign}\left(\frac{2}{N}B^{\frac{2-N}{2}}K_{\left(2\right)}+\frac{1}{N+1}B^{\frac{3}{2}}\kappa\rho_{\Lambda\left(2\right)}\right)\\
 &  & B_{B}=\left(\frac{-\xi\eta4K}{\xi\eta\frac{2N}{N+1}\kappa\rho_{\Lambda}+\left(\kappa\rho\right)^{2}}\right)^{\frac{2}{N+1}}\\
\xi\eta=-1 & \Rightarrow & \left(\kappa\rho\right)^{2}<\frac{2N}{N+1}\kappa\rho_{\Lambda}\\
\epsilon_{u0\left(2\right)}\xi\eta & = & s_{\left(2\right)}\text{sign}\left(\kappa\rho\right)\text{sign}\left(\mathcal{Y}_{\left(2\right)}\chi_{\left(2\right)}^{1}\right)
\end{array}\label{eq:MnkLK_req_smry}
\end{equation}

\subsection{The bulk on both sides of the brane is $\Lambda-K$ space-time}

\noindent\fbox{\begin{minipage}[t]{1\columnwidth - 2\fboxsep - 2\fboxrule}%
In this subsection:
\begin{lyxlist}{00.00.0000}
\item [{1)}] The case is considered in which side~$\left(1\right)$ and
side~$\left(2\right)$ are both $\Lambda-K$ space-time is considered,
though the values of $\Lambda$ and $K$ may differ. 
\item [{2)}] The metric induced on the brane from side~$\left(1\right)$
is must match that induced from side~$\left(2\right)$, though it
need not be constant along the brane world-path. This is condition~1
of the junction conditions eq~\ref{eq:IJC_gen_metric}.
\item [{3)}] Enforcing condition~2 of eq~\ref{eq:IJC_gen_metric} yields
a 2-dimensional vector equation. The solution is the instantaneous
relativistic velocity of the world-path of the brane on either side. 
\item [{4)}] The solution may be understood graphically as a finding points
in the $\left(X,Y\right)$~plane along two hyperbole n such that
the $Y$-coordinates are the same and the $X$-coordinates differ
by a prescribed amount. 
\item [{5)}] Enforcing condition~3 of the junction conditions eq~\ref{eq:IJC_gen_metric}
yields the standard conservation equations of the fluid on the brane.
This is to be expected since the junction conditions are derived from
the higher-dimensional Einstein equations, and the covariant divergence
of the Einstein equations vanishes. Never the less, this result is
a reassuring sanity-check.
\end{lyxlist}
\end{minipage}}

Consider the case where the bulk on both sides of the brane is $\Lambda-K$,
each with possibly different parameters. The junction conditions of
eq~\ref{eq:IJC_gen_metric} become: 

\begin{equation}
\begin{array}{cccccc}
\left(1A\right) & N_{\left(1\right)}=N_{\left(2\right)}\\
\left(1B\right) & \xi_{\left(1\right)}\eta_{\left(1\right)}=\xi_{\left(2\right)}\eta_{\left(2\right)}\\
\left(1C\right) & \left.B_{\left(1\right)}\right|_{\Sigma}=\left.B_{\left(2\right)}\right|_{\Sigma} & \Rightarrow & \left[\frac{1}{B}\dot{B}\right]=\left[\frac{1}{B}B_{,\alpha}u^{\alpha}\right]=\left[\frac{B_{,y}}{B}y_{,\alpha}u^{\alpha}\right] & = & 0\\
\left(2\right) & \left[\Pi_{\underline{\mu\nu}}\right] & = & \left[\frac{1}{2}g_{\underline{\mu\nu}}\frac{1}{B}B_{,\alpha}n^{\alpha}\right] & = & -\frac{1}{N}g_{\underline{\mu\nu}}\kappa\rho\\
 &  &  & \left[\frac{1}{B}B_{,\alpha}n^{\alpha}\right]=\left[\frac{1}{B}B_{,y}y_{,\alpha}n^{\alpha}\right] & = & -\frac{2}{N}\kappa\rho\\
\left(3\right) & \left[\Pi_{\mu\nu}u^{\mu}u^{\nu}\right] & = & \left[-\dot{u}^{\bar{\nu}}n_{\bar{\nu}}+\frac{1}{2}\xi\eta\left(\ln A\right)_{,\bar{\nu}}n^{\bar{\nu}}\right] & = & \xi\eta\left(\left(\frac{N-1}{N}\right)\kappa\rho+\kappa p\right)\\
 &  & = & \left[-\dot{u}^{\bar{\nu}}n_{\bar{\nu}}+\frac{1}{2}\xi\eta\left(\ln A\right)_{,y}y_{,\bar{\nu}}n^{\bar{\nu}}\right]
\end{array}\label{eq:IJC_gen_LKLK_y}
\end{equation}
 Due to the fact that $B$ depends solely on $y$, derivatives of
$B$ can be re-written as derivatives with respect to $y$ : 
\begin{equation}
\begin{array}{ccc}
\frac{1}{B}\left(\begin{array}{c}
B_{,\alpha}u^{\alpha}\\
B_{,\alpha}n^{\alpha}
\end{array}\right) & = & \frac{1}{B}\left(\begin{array}{c}
B_{,y}y_{,\alpha}u^{\alpha}\\
B_{,y}y_{,\alpha}n^{\alpha}
\end{array}\right)=\left(-4\lambda\right)\frac{B_{,y}}{B}\left(\begin{array}{c}
\chi^{0}u^{0}-\chi^{1}u^{1}\\
-s\chi^{1}u^{0}+s\chi^{0}u^{1}
\end{array}\right)\\
 & = & \left(-4\lambda\right)\frac{B_{,y}}{B}\left(\begin{array}{cc}
1 & 0\\
0 & s
\end{array}\right)\left(\begin{array}{cc}
\chi^{0} & -\chi^{1}\\
-\chi^{1} & \chi^{0}
\end{array}\right)\left(\begin{array}{c}
u^{0}\\
u^{1}
\end{array}\right)
\end{array}\label{eq:LKLK_B_y_gradient}
\end{equation}
Thus junction conditions~1C~and~2 of eq~\ref{eq:IJC_gen_LKLK_y}
can be combined into a vector equation: 
\begin{equation}
\left[\left(-4\lambda\right)\frac{1}{B}B_{,y}\left(\begin{array}{cc}
1 & 0\\
0 & s
\end{array}\right)\left(\begin{array}{cc}
\chi^{0} & -\chi^{1}\\
-\chi^{1} & \chi^{0}
\end{array}\right)\left(\begin{array}{c}
u^{0}\\
u^{1}
\end{array}\right)\right]=\frac{-2}{N}\kappa\rho\left(\begin{array}{c}
0\\
1
\end{array}\right)\label{eq:LKLK_vector_IJC_a}
\end{equation}
 The matrix $\left(\begin{array}{cc}
\chi^{0} & -\chi^{1}\\
-\chi^{1} & \chi^{0}
\end{array}\right)$ looks like a coordinate transformation - and indeed it is. It is
a transformation to a coordinate frame in which the gradient of $y$
and contours of constant $y$ become coordinate directions (this is
the $z$ coordinate frame of section~\ref{subsubsec:LK_defn}). Define
a frame-transformed version of the relativistic velocity: 
\begin{equation}
\begin{array}{ccl}
\widetilde{u}^{\bar{\mu}} & = & \frac{\left(-2\lambda\right)}{\sqrt{\left|-2\lambda\mathcal{Y}\right|}}\left(\begin{array}{cc}
\chi^{0} & -\chi^{1}\\
-\chi^{1} & \chi^{0}
\end{array}\right)_{\bar{\alpha}}^{\bar{\mu}}\left(\begin{array}{c}
u^{0}\\
u^{1}
\end{array}\right)^{\bar{\alpha}}=\left(\begin{array}{c}
\widetilde{\epsilon}_{u0}\xi\sqrt{\widetilde{Y}^{2}+\frac{1}{A}\xi\eta\text{sign}\left(-2\lambda\mathcal{Y}\right)}\\
\widetilde{Y}
\end{array}\right)^{\bar{\mu}}\\
\widetilde{u}^{\mu}\widetilde{u}_{\mu} & = & \xi\eta\text{sign}\left(-2\lambda\mathcal{Y}\right)\\
\widetilde{Y} & = & \frac{\left(-2\lambda\right)}{\sqrt{\left|-2\lambda\mathcal{Y}\right|}}\left(-\chi^{1}u^{0}+\chi^{0}u^{1}\right)=\frac{\left(-2\lambda\right)}{\sqrt{\left|-2\lambda\mathcal{Y}\right|}}\left(-\chi^{1}\epsilon_{u0}\sqrt{Y^{2}+\frac{1}{A}\xi\eta}+\chi^{0}Y\right)
\end{array}\label{eq:LKLK_xformed_u}
\end{equation}
 One then has junction conditions~1C~\&~2 expressed in terms
of this~$\widetilde{u}^{\bar{\mu}}$: 
\begin{equation}
\left[\sqrt{\left|-2\lambda\mathcal{Y}\right|}A\left(\begin{array}{cc}
1 & 0\\
0 & s
\end{array}\right)\widetilde{u}\right]=\left[\sqrt{\left|-2\lambda\mathcal{Y}\right|}A\left(\begin{array}{c}
\widetilde{\epsilon}_{u0}\xi\sqrt{\widetilde{Y}^{2}+\frac{1}{A}\xi\eta\text{sign}\left(-2\lambda\mathcal{Y}\right)}\\
s\widetilde{Y}
\end{array}\right)\right]=\frac{-1}{2}\sqrt{B}\kappa\rho\left(\begin{array}{c}
0\\
1
\end{array}\right)\label{eq:LKLK_vector_IJC_b}
\end{equation}
\begin{equation}
\left[\left(\begin{array}{c}
\widetilde{\epsilon}_{u0}\xi\eta\sqrt{\left(\widetilde{Y}A\sqrt{\left|-2\lambda\mathcal{Y}\right|}\right)^{2}+A\left(-2\lambda\mathcal{Y}\right)\xi\eta}\\
s\left(\widetilde{Y}A\sqrt{\left|-2\lambda\mathcal{Y}\right|}\right)
\end{array}\right)\right]=\frac{-1}{2}\sqrt{B}\kappa\rho\left(\begin{array}{c}
0\\
1
\end{array}\right)\label{eq:LKLK_HypDiff_a}
\end{equation}
 It is immediately obvious that $\widetilde{\epsilon}_{u0\left(1\right)}=\widetilde{\epsilon}_{u0\left(2\right)}$.
Eq~\ref{eq:LKLK_HypDiff_a} is essentially a statement about the
difference of the graphs of two hyperbolas 
\begin{equation}
\begin{array}{c}
\left(\begin{array}{c}
U_{\left(2\right)}^{0}\\
U_{\left(2\right)}^{1}
\end{array}\right)-\left(\begin{array}{c}
U_{\left(1\right)}^{0}\\
U_{\left(1\right)}^{1}
\end{array}\right)=\frac{1}{2}\sqrt{B}\kappa\rho\left(\begin{array}{c}
0\\
1
\end{array}\right)\\
U_{\left(j\right)}^{\bar{\mu}}\equiv\sqrt{\left|-2\lambda_{\left(j\right)}\mathcal{Y}_{\left(j\right)}\right|}A_{\left(j\right)}\left(\begin{array}{cc}
1 & 0\\
0 & s_{\left(j\right)}
\end{array}\right)_{\bar{\alpha}}^{\bar{\mu}}\widetilde{u}_{\left(j\right)}^{\bar{\alpha}}\\
U_{\left(j\right)}^{1}=s_{\left(j\right)}\left(\widetilde{Y}_{\left(j\right)}A_{\left(j\right)}\sqrt{\left|-2\lambda_{\left(j\right)}\mathcal{Y}_{\left(j\right)}\right|}\right)\\
U_{\left(j\right)}^{0}=\sqrt{\left(U_{\left(j\right)}^{1}\right)^{2}+\xi\eta A_{\left(j\right)}\left(-2\lambda_{\left(j\right)}\mathcal{Y}_{\left(j\right)}\right)}\\
\left(U_{\left(j\right)}^{0}\right)^{2}-\left(U_{\left(j\right)}^{1}\right)^{2}=\xi\eta A_{\left(j\right)}\left(-2\lambda_{\left(j\right)}\mathcal{Y}_{\left(j\right)}\right)
\end{array}\label{eq:LKLK_HypDiff_b}
\end{equation}

\begin{figure}
\begin{centering}
\includegraphics[scale=0.6]{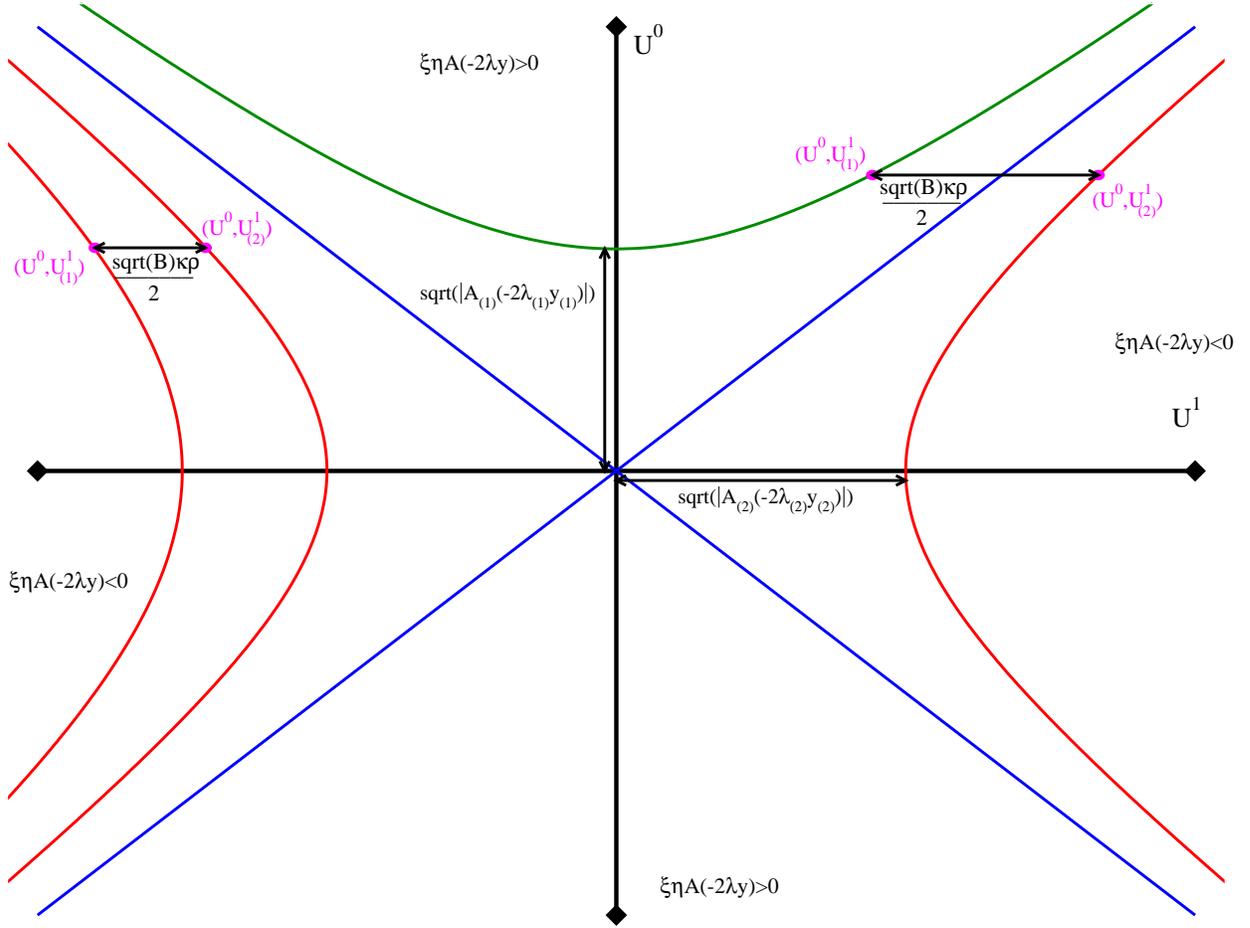}\label{fig:LK-LK_space_hyp_u_comp}
\par\end{centering}
\caption{This is a graphical representation of how the world-paths of either
side of the boundary between two $\Lambda-K$ spaces are determined.
The junction conditions~1C~and~2 of eq~\ref{eq:IJC_gen_LKLK_y}
can be combined to imply that the rescaled relativistic velocities
$U_{\left(j\right)}^{\bar{\mu}}$ have the same $0$ components, and
their~$1$ components differ by~$\frac{1}{2}\sqrt{B}\kappa\rho$.
From those two constraints the spatial velocities $\widetilde{Y}_{\left(1\right)}$
and $\widetilde{Y}_{\left(2\right)}$ may be computed, and from this
the $\widetilde{u}_{\left(j\right)}^{\mu}$, and then the $u_{\left(j\right)}^{\mu}$
in the original frame.Combining instantaneous $u_{\left(j\right)}^{\mu}$
with standard perfect-fluid evolution of $\rho$ due to $B$ and $p$,
and integrating along the trajectory yields $u^{\mu}\left(\tau\right)$
(or $u^{\mu}\left(B\left(\mathcal{Y}\left(\tau\right)\right)\right)$).
If one then plugs this into the left-hand side of junction condition~3
of eq~\ref{eq:IJC_gen_LKLK_y}, one gets an expression for the derivation
of $\rho$ with respect to $B$. The equality of this expression to
the right-hand side is the standard perfect-fluid evolution equation
of $\rho$ due to $p$ and change of volume element.}
\end{figure}

Given the parameters $A_{\left(j\right)}$, $B$, $\kappa\rho$, $-2\lambda_{\left(j\right)}\mathcal{Y}_{\left(j\right)}$,
$\xi\eta$, $s_{\left(j\right)}$ one uses the two constraints of
eq~\ref{eq:LKLK_HypDiff_b} to solve for $U_{\left(j\right)}^{1}$,
and hence $\widetilde{Y}_{\left(j\right)}$. The dependence of this
solution on the input parameter values can be understood graphically,
as shown in fig~\ref{fig:LK-LK_space_hyp_u_comp}. The solution yields
the relativistic velocity of the world-path of the brane in both bulks:
\begin{equation}
\begin{array}{c}
\widetilde{Y}_{\left(1\right)}=s_{\left(1\right)}\frac{\xi\eta\left[A\left(-2\lambda\mathcal{Y}\right)\right]-\frac{1}{4}B\left(\kappa\rho\right)^{2}}{A_{\left(1\right)}\sqrt{B}\left(\kappa\rho\right)\sqrt{\left|-2\lambda_{\left(1\right)}\mathcal{Y}_{\left(1\right)}\right|}}=s_{\left(1\right)}\frac{\xi\eta\left[B^{\frac{2-N}{2}}K+\frac{N}{2\left(N+1\right)}B^{\frac{3}{2}}\kappa\rho_{\Lambda}\right]-\frac{1}{4}B^{\frac{3}{2}}\left(\kappa\rho\right)^{2}}{A_{\left(1\right)}B\left(\kappa\rho\right)\sqrt{\left|-2\lambda_{\left(1\right)}\mathcal{Y}_{\left(1\right)}\right|}}\\
\widetilde{Y}_{\left(2\right)}=s_{\left(2\right)}\frac{\xi\eta\left[A\left(-2\lambda\mathcal{Y}\right)\right]+\frac{1}{4}B\left(\kappa\rho\right)^{2}}{A_{\left(2\right)}\sqrt{B}\left(\kappa\rho\right)\sqrt{\left|-2\lambda_{\left(2\right)}\mathcal{Y}_{\left(2\right)}\right|}}=s_{\left(2\right)}\frac{\xi\eta\left[B^{\frac{2-N}{2}}K+\frac{N}{2\left(N+1\right)}B^{\frac{3}{2}}\kappa\rho_{\Lambda}\right]+\frac{1}{4}B^{\frac{3}{2}}\left(\kappa\rho\right)^{2}}{A_{\left(2\right)}B\left(\kappa\rho\right)\sqrt{\left|-2\lambda_{\left(2\right)}\mathcal{Y}_{\left(2\right)}\right|}}
\end{array}\label{eq:LKLK_Ysolns_b}
\end{equation}
 The one inserts those expressions for $\widetilde{Y}_{\left(j\right)}$
into the expression for $\widetilde{u}^{\mu}$ 
\begin{equation}
\widetilde{u}_{\left(j\right)}^{\bar{\mu}}=\left(\begin{array}{c}
\widetilde{\epsilon}_{u0}\xi_{\left(j\right)}\sqrt{\widetilde{Y}_{\left(j\right)}^{2}+\frac{1}{A_{\left(j\right)}}\xi\eta\text{sign}\left(-2\lambda_{\left(j\right)}\mathcal{Y}_{\left(j\right)}\right)}\\
\widetilde{Y}_{\left(j\right)}
\end{array}\right)^{\bar{\mu}}\label{eq:LKLK_utilde_soln}
\end{equation}
Then the relativistic velocity $u_{\left(j\right)}^{\bar{\mu}}$ of
the brane in bulk~$\left(j\right)$ is given by transforming back
to the original frame: 
\begin{equation}
u_{\left(j\right)}^{\bar{\mu}}=\frac{\sqrt{\left|-2\lambda_{\left(j\right)}\mathcal{Y}_{\left(j\right)}\right|}}{\mathcal{Y}_{\left(j\right)}}\left(\begin{array}{cc}
\chi_{\left(j\right)}^{0} & \chi_{\left(j\right)}^{1}\\
\chi_{\left(j\right)}^{1} & \chi_{\left(j\right)}^{0}
\end{array}\right)_{\bar{\alpha}}^{\bar{\mu}}\widetilde{u}_{\left(j\right)}^{\bar{\alpha}}\label{eq:LKLK_u_soln}
\end{equation}
 The above solution satisfies the junction conditions~(1-2) of eq~\ref{eq:IJC_gen_LKLK_y}.
The additional information of junction conditions~3 of eq~\ref{eq:IJC_gen_LKLK_y}
is equivalent to stress-energy conservation on the brane. Since the
contracted Bianchi identities imply $G_{\ ;\mu}^{\mu\nu}=0$, stress-energy
conservation ($T_{\ ;\mu}^{\mu\nu}=0$) is built into solutions of
the Einstein equations. Thus, in general, one of the constraints imposed
by the various components of the Einstein equations can be ignored
if one imposes $T_{\ ;\mu}^{\mu\nu}=0$. This will be the case with
regard to the above junction conditions - we will work with junction
conditions~1~\&~2. Condition~3 would then be satisfied if one
assumes that the stress-energy on the brane is conserved:~$h_{\mu}^{\alpha}\nabla_{\alpha}S^{\mu0}=0$.
Since the stress-energy on the brane is assumed to be a perfect fluid,
this is effectively the familiar (from Friedmann-Robertson-Walker-Lemaitre
cosmology, for example) statement that the energy density on the brane
evolves due to dilution/concentration and work done, in turn due to
scale factor expansion/contraction in the induced metric~$h_{\mu\nu}$.

\subsection{The bulk on one side of the brane is Modified-Minkowski space-time,
the other is $\Lambda-K$\label{subsec:modMnkLK_constr} }

\noindent\fbox{\begin{minipage}[t]{1\columnwidth - 2\fboxsep - 2\fboxrule}%
In this subsection:
\begin{lyxlist}{00.00.0000}
\item [{1)}] The case is considered in which side~$\left(1\right)$ of
the brane is Modified-Minkowski space-time, and side~$\left(2\right)$
is $\Lambda-K$
\item [{2)}] The metric of side~$\left(1\right)$ is static, and for the
purpose of the models of sec~\ref{sec:Models_presented}, the location
of the boundary in side~$\left(1\right)$ should remain fixed. Thus
the brane world-path in side~$\left(2\right)$ must be $y=const$,
as in subsection~\ref{subsec:MnkLK_constr}. This is condition~1
of the junction conditions eq~\ref{eq:IJC_gen_metric}.
\item [{3)}] Enforcing condition~2 of the junction conditions eq~\ref{eq:IJC_gen_metric}
yields the particular value of $y$ (equivalently $B$) of the $y=const$
world-path.
\item [{4)}] Enforcing condition~3 of the junction conditions eq~\ref{eq:IJC_gen_metric}
would yield the world-path of the brane in side~$\left(1\right)$.
However, location of the boundary in side~$\left(1\right)$ should
remain fixed, thus one requires that acceleration to vanish. 
\item [{5)}] Vanishing acceleration of side~$\left(1\right)$ leads to
a constraint relating parameters describing the boundary layer of
the Modified-Minkowski space-time to the energy and pressure on the
brane and $\kappa\rho_{\Lambda}$ on side~$\left(2\right)$. This
constrain is used to fix the value of $\phi$. 
\end{lyxlist}
\end{minipage}}

Let the bulk on side~$\left(1\right)$ of the brane be Modified-Minkowski,
and side~$\left(2\right)$ be $\Lambda-K$. It is no longer true
that agreement of the induced brane metric implies that path followed
by the brane on the $\Lambda-K$ space-time side must maintain $B=constant$
(ie: a $\mathcal{Y}_{\left(2\right)}=constant$ path). However, as
the goal for the use of Modified-Minkowski space-time is to create
a static interior region without use of negative energy density on
the brane, never the less parameters will be tuned so that the metric
induced on the brane remains static along the world-path of that brane.
Thus, $\mathcal{Y}_{\left(2\right)}=constant$ will be applied.

Differentiating $\mathcal{Y}=constant$ with respect to proper time~$\tau$
of an observer at rest with respect to the brane yields $\chi_{\left(2\right)}^{0}u_{\left(2\right)}^{0}-\chi_{\left(2\right)}^{1}u_{\left(2\right)}^{1}=0$.
Thus 
\begin{equation}
\begin{array}{ccccc}
u_{\left(2\right)}^{\mu} & = & u_{\left(2\right)}^{\bar{\mu}} & = & \Psi_{\left(2\right)}\left(\begin{array}{c}
\chi_{\left(2\right)}^{1}\\
\chi_{\left(2\right)}^{0}
\end{array}\right)^{\bar{\mu}}\\
 &  & \Psi_{\left(2\right)} & \equiv & \frac{\dot{\chi}_{\left(2\right)}^{0}}{\chi_{\left(2\right)}^{1}}\\
n_{\left(2\right)}^{\mu} & = & n_{\left(2\right)}^{\bar{\mu}} & = & s_{\left(2\right)}\Psi_{\left(2\right)}\left(\begin{array}{c}
\chi_{\left(2\right)}^{0}\\
\chi_{\left(2\right)}^{1}
\end{array}\right)^{\bar{\mu}}\\
 &  & s_{\left(2\right)} & \in & \left\{ +1,-1\right\} 
\end{array}\label{eq:modMnkLK_gen_soln_A}
\end{equation}
 where $s_{\left(2\right)}=+1$ indicates that the side of the boundary
corresponding to $n_{\left(2\right)}^{\bar{\mu}}=\left(\begin{array}{c}
u_{\left(2\right)}^{1}\\
u_{\left(2\right)}^{0}
\end{array}\right)^{\bar{\mu}}$ is being utilized for a path where the direction of proper time~$\tau$
coincides with that of coordinate time~$x^{0}$, while the other
side is discarded (right side of the boundary of Fig~\ref{fig:KSanalog_conformal_diagram}).The
opposite is true for $s_{\left(2\right)}=-1$. 

To define the junction between the Modified-Minkowski space-time of
subsection~\ref{subsubsec:modMnk_defn} and the $\Lambda-K$ space
of subsection~\ref{subsubsec:LK_defn}, let $w=const-\beta x^{1},\ \beta>0$
where $\beta$ amounts to a freedom to rescale the coordinate system
of subsection~\ref{subsubsec:modMnk_defn}. With $w_{max}=\frac{1}{4}$
fixed as the world-path of the brane in the Modified-Minkowski bulk,
and numerically computing the metric and its derivatives at this location,
one then has: 
\[
\begin{array}{ccc}
\ln\left(B_{\left(1\right)}\right)_{,\alpha}n_{\left(1\right)}^{\alpha} & = & \ln\left(B_{\left(1\right)}\right)_{,1}n_{\left(1\right)}^{1}\\
 & = & -\beta\frac{2}{N}\left(128e^{-16}\right)u_{\left(1\right)}^{0}\\
 & = & \frac{2}{N}\frac{\beta}{\sqrt{\left|A_{\left(1\right)}\right|}}\left(128e^{-16}\right)\\
 & \simeq & -\frac{2}{N}\frac{\beta}{\sqrt{A_{\left(1\right)0}}}\left(\frac{128e^{-16}}{\exp\left(-5.62676\times10^{-8}\right)}\right)\\
 & \simeq & -\frac{2}{N}\frac{\beta}{\sqrt{A_{\left(1\right)0}}}\left(1.44045\times10^{-5}\right)
\end{array}
\]
where $A_{\left(1\right)0}\equiv A_{\left(1\right)}(w=0$) may be
rescaled as desired. One is free to choose any $\beta>0$. That freedom
will be used to establish the desired relation between the derivatives
of the metric on either side of the brane. Define the parameter~$\phi$
such that 
\[
\begin{array}{ccc}
\phi\left.\ln\left(B_{\left(2\right)}\right)_{,\alpha}n_{\left(2\right)}^{\alpha}\right|_{\chi_{\left(2\right)}^{\mu}\left(\tau\right)} & = & \left.\ln\left(B_{\left(1\right)}\right)_{,\alpha}n_{\left(1\right)}^{\alpha}\right|_{\chi_{\left(1\right)}^{\mu}\left(\tau\right)}\\
 & \simeq & -\frac{2}{N}\frac{\beta}{\sqrt{A_{\left(1\right)0}}}\left(1.44045\times10^{-5}\right)
\end{array}
\]
 Thus the freedom to rescale the coordinate $w$ of subsection~\ref{subsubsec:modMnk_defn}
becomes the freedom to set $\phi$. For the cases of interest $\left.\ln\left(B_{\left(2\right)}\right)_{,\alpha}n_{\left(2\right)}^{\alpha}\right|_{\chi_{\left(2\right)}^{\mu}\left(\tau\right)}<0$,
and thus $\phi>0$ (the further constraint of $\phi>1$ will be imposed).
The extrinsic curvature tensor in the parallel directions becomes:
\[
\Pi_{\left(1\right)\underline{\mu\nu}}=\frac{1}{2}g_{\underline{\mu\nu}}\ln\left(B_{\left(1\right)}\right)_{,\alpha}n_{\left(1\right)}^{\alpha}=\frac{1}{2}g_{\underline{\mu\nu}}\phi\ln\left(B_{\left(2\right)}\right)_{,\alpha}n_{\left(2\right)}^{\alpha}\ \ \ \phi>1
\]
 which is evaluated at the boundary. Junction condition~2 of eq~\ref{eq:IJC_gen_metric}
then implies 
\[
\begin{array}{ccc}
\left[\Pi_{\underline{\mu\nu}}\right] & = & \left[\frac{1}{2}g_{\underline{\mu\nu}}\ln\left(B\right)_{,\beta}n^{\beta}\right]\\
 & = & \frac{1}{2}g_{\underline{\mu\nu}}\left(\phi-1\right)\ln\left(B_{\left(2\right)}\right)_{,\alpha}n_{\left(2\right)}^{\alpha}=-h_{\underline{\mu\nu}}\frac{1}{N}\kappa\rho\\
\\
\frac{2}{N}\kappa\rho & = & -\left(\phi-1\right)\ln\left(B_{\left(2\right)}\right)_{,\alpha}n_{\left(2\right)}^{\alpha}
\end{array}
\]
 Tus if Modified-Minkowski space-time is employed in place of standard
Minkowski space-time when $\left.\ln\left(B_{\left(2\right)}\right)_{,\alpha}n_{\left(2\right)}^{\alpha}\right|_{\chi_{\left(2\right)}^{\mu}\left(\tau\right)}<0$,
a choice of $\phi>1$ corresponds to positive energy density $\kappa\rho$
on the brane. If standard Minkowski space-time was used instead ($\phi=0$)
the result would be negative $\kappa\rho$. The earlier results for
interfacing directly between Minkowski and $\Lambda-K$ space-times
can be obtained by taking $\phi\rightarrow0$.One generally does
not want to take~$\phi\rightarrow1$ because then the junction condition~2
of eq~\ref{eq:IJC_gen_metric} would imply that $\kappa\rho=0$ -
which would generally be problematic for several reasons, including
difficulties in being able to make the acceleration of the brane world
path in the Modified-Minkowski space zero (ie: static situation on
side~$\left(1\right)$). 

Combining junction conditions~1~and~2 of eq~\ref{eq:IJC_gen_metric}
yields 
\begin{equation}
\begin{array}{ccc}
-\frac{2}{N}\kappa\rho & = & \left(\phi-1\right)\ln\left(B_{\left(2\right)}\right)_{,\alpha}n_{\left(2\right)}^{\alpha}\\
 & = & \left(\phi-1\right)\ln\left(B_{\left(2\right)}\right)_{,y}y_{,\alpha}n_{\left(2\right)}^{\alpha}\\
 & = & \left(\phi-1\right)\frac{1}{B}B_{\left(2\right),y}2s_{\left(2\right)}\Psi_{\left(2\right)}\mathcal{Y}_{\left(2\right)}\\
\\
\Psi_{\left(2\right)} & = & s_{\left(2\right)}\frac{B\kappa\rho}{N\left(\phi-1\right)B_{\left(2\right),y}\left(-\mathcal{Y}_{\left(2\right)}\right)}
\end{array}\label{eq:modMnkLK_Psi_expr}
\end{equation}
\begin{equation}
\begin{array}{ccccc}
u_{\left(2\right)}^{\bar{\mu}} & = & \Psi_{\left(2\right)}\left(\begin{array}{c}
\chi_{\left(2\right)}^{1}\\
\chi_{\left(2\right)}^{0}
\end{array}\right)^{\bar{\mu}}\\
\epsilon_{u0\left(2\right)}\xi_{\left(2\right)} & = & \text{sign}\left(\Psi_{\left(2\right)}\chi_{\left(2\right)}^{1}\right)\\
 & = & s_{\left(2\right)}\eta_{\left(2\right)}\text{sign}\left(\kappa\rho\right)\text{sign}\left(\mathcal{Y}_{\left(2\right)}\chi_{\left(2\right)}^{1}\right)\\
\\
n_{\left(2\right)}^{\bar{\mu}} & = & s_{\left(2\right)}\Psi_{\left(2\right)}\left(\begin{array}{c}
\chi_{\left(2\right)}^{0}\\
\chi_{\left(2\right)}^{1}
\end{array}\right)^{\bar{\mu}}\\
\Psi_{\left(2\right)} & \equiv & s_{\left(2\right)}\frac{B\kappa\rho}{N\left(\phi-1\right)\left(-\mathcal{Y}_{\left(2\right)}\right)B_{\left(2\right),y}} & = & s_{\left(2\right)}\frac{\sqrt{B}\kappa\rho}{2\left(\phi-1\right)\left(-\mathcal{Y}_{\left(2\right)}\right)A_{\left(2\right)}}
\end{array}\label{eq:modMnkLK_gen_soln_B}
\end{equation}
Note that $\mathcal{Y}_{\left(2\right)}$, $B_{\left(2\right)}$,
$A_{\left(2\right)}$ being constant along the world-path of the brane
in bulk~$\left(2\right)$ implies $\rho$ and $\Psi_{\left(2\right)}$
are also constant. Thus differentiation of $u_{\left(2\right)}^{\mu}$
and $n_{\left(2\right)}^{\mu}$ with respect to $\tau$ yields:
\begin{equation}
\begin{array}{ccc}
\dot{u}_{\left(2\right)}^{\mu} & = & s_{\left(2\right)}\Psi_{\left(2\right)}n_{\left(2\right)}^{\mu}\\
\dot{n}_{\left(2\right)}^{\mu} & = & s_{\left(2\right)}\Psi_{\left(2\right)}u_{\left(2\right)}^{\mu}\\
\dot{u}_{\left(2\right)}^{\bar{\nu}}n_{\left(2\right)\bar{\nu}} & = & -\xi\eta s_{\left(2\right)}\Psi_{\left(2\right)}\\
\ddot{u}_{\left(2\right)}^{\mu} & = & \Psi_{\left(2\right)}^{2}u_{\left(2\right)}^{\mu}
\end{array}\label{eq:modMnkLK_gen_soln_C}
\end{equation}
From eq~\ref{eq:modMnkLK_gen_soln_B} and the requirement that
$\dot{\chi}_{\left(2\right)}^{0}\ge0$ one can make statements similar
to those of eq~\ref{eq:MnkLK_gen_soln_D} about the relation between
the sign of $\xi_{\left(2\right)}$ and the region of the $x^{0},x^{1}$
coordinate plane occupied, and also between the sign of $\xi\eta$
and the sign of $B_{\left(2\right),z}$. The new feature is that
the constraint relation between $s_{\left(2\right)}$, $\xi_{\left(2\right)}$
and $\text{sign}\left(\kappa\rho\right)$ in eq~\ref{eq:modMnkLK_gen_soln_D}
are modified by the presence of $\text{sign}\left(\phi-1\right)$.
 
\begin{equation}
\begin{array}{ccccc}
\xi_{\left(2\right)}=+1 & \Rightarrow & -2\lambda_{\left(2\right)}\mathcal{Y}_{\left(2\right)}<0 & \Rightarrow & \left|\chi_{\left(2\right)}^{0}\right|<\left|\chi_{\left(2\right)}^{1}\right|\\
\xi_{\left(2\right)}=-1 & \Rightarrow & -2\lambda_{\left(2\right)}\mathcal{Y}_{\left(2\right)}>0 & \Rightarrow & \left|\chi_{\left(2\right)}^{0}\right|>\left|\chi_{\left(2\right)}^{1}\right|\\
\xi\eta & = & -\text{sign}\left(B_{\left(2\right),z}\right)<0\\
s_{\left(2\right)} & = & \xi_{\left(2\right)}\text{sign}\left(\kappa\rho\right)\text{sign}\left(\phi-1\right)
\end{array}\label{eq:modMnkLK_gen_soln_D}
\end{equation}

Similarly to eq~\ref{eq:MnkLK_derivof_Bsoln}, enforcing the $u_{\left(2\right)}^{\mu}u_{\left(2\right)\mu}=\xi\eta$
condition selects a particular $\mathcal{Y}_{\left(2\right)}=constant$
path for the brane in $\Lambda-K$ space - yielding an expression
for $B$ in terms of the brane energy density $\kappa\rho$ and the
bulk~$\left(2\right)$ quantities $K$, $\kappa\rho_{\Lambda}$,
as well as $\phi-1$ : 
\begin{equation}
\begin{array}{ccc}
\xi\eta & = & u_{\left(2\right)}^{\mu}u_{\left(2\right)\mu}\\
 & = & A_{\left(2\right)}\Psi_{\left(2\right)}^{2}\left(\left(\chi_{\left(2\right)}^{1}\right)^{2}-\left(\chi_{\left(2\right)}^{0}\right)^{2}\right)\\
 & = & -A_{\left(2\right)}\Psi_{\left(2\right)}^{2}\left(\frac{1}{\left(-2\lambda_{\left(2\right)}\right)}\mathcal{Y}_{\left(2\right)}\right)\\
 & = & \frac{-B\left(\kappa\rho\right)^{2}}{4\left(\phi-1\right)^{2}\left(-2\lambda_{\left(2\right)}\mathcal{Y}_{\left(2\right)}\right)A_{\left(2\right)}}\\
\left(-2\lambda_{\left(2\right)}\mathcal{Y}_{\left(2\right)}\right)A_{\left(2\right)} & = & \frac{-B\left(\kappa\rho\right)^{2}}{4\left(\phi-1\right)^{2}\xi\eta}\\
\frac{B^{\frac{3}{2}}\left(\kappa\rho\right)^{2}}{\left(\phi-1\right)^{2}} & = & -\xi\eta2N\left(\frac{2}{N}B^{\frac{2-N}{2}}K_{\left(2\right)}+\frac{1}{N+1}B^{\frac{3}{2}}\kappa\rho_{\Lambda\left(2\right)}\right)\\
-4\xi\eta B^{-\frac{N+1}{2}}K_{\left(2\right)} & = & \xi\eta\frac{2N}{N+1}\kappa\rho_{\Lambda\left(2\right)}+\frac{\left(\kappa\rho\right)^{2}}{\left(\phi-1\right)^{2}}\\
B_{B} & \equiv & \left(\frac{-4\xi\eta K_{\left(2\right)}}{\xi\eta\frac{2N}{N+1}\kappa\rho_{\Lambda\left(2\right)}+\frac{\left(\kappa\rho\right)^{2}}{\left(\phi-1\right)^{2}}}\right)^{\frac{2}{N+1}}
\end{array}\label{eq:modMnkLK_derivof_Bsoln}
\end{equation}
Thus one obtains a modified version of~$B_{B}$ of eq~\ref{eq:MnkLK_Bsoln}.
This is the value of $B$ in the metric induced on the brane. 
\begin{equation}
B_{B\left(2\right)}=\left(\frac{-\xi\eta4K_{\left(2\right)}}{\xi\eta\frac{2N}{N+1}\kappa\rho_{\Lambda\left(2\right)}+\frac{\left(\kappa\rho\right)^{2}}{\left(\phi-1\right)^{2}}}\right)^{\frac{2}{N+1}}\label{eq:modMnkLK_Bsoln}
\end{equation}
Note that eq~\ref{eq:modMnkLK_Bsoln} is well-defined by a similar
argument to that made for eq~\ref{eq:MnkLK_Bsoln}. 

Plugging this value of $B_{B}$ into the expressions for $B_{\left(2\right),y}$,
$A_{\left(2\right)}$ and $\Psi_{\left(2\right)}$ one obtains a modified
version of eq~\ref{eq:MnkLK_BB_into_A_B_Psi} : 
\begin{equation}
\begin{array}{ccc}
\left.B_{\left(2\right),y}\right|_{B=B_{B}} & = & \frac{-B_{B\left(2\right)}^{\frac{3}{2}}\left(\kappa\rho\right)^{2}}{2N\xi\eta\left(-2\lambda_{\left(2\right)}\mathcal{Y}_{\left(2\right)}\right)\left(\phi-1\right)^{2}}\\
\left.A_{\left(2\right)}\right|_{B=B_{B}} & = & \frac{-B_{B\left(2\right)}\left(\kappa\rho\right)^{2}}{\left(-2\lambda_{\left(2\right)}\mathcal{Y}_{\left(2\right)}\right)\left(\phi-1\right)^{2}4\xi\eta}\\
\frac{B_{B}}{B_{H}} & = & \left(1+\frac{1}{\xi\eta}\frac{\left(N+1\right)}{2N}\frac{\left(\kappa\rho\right)^{2}}{\kappa\rho_{\Lambda\left(2\right)}\left(\phi-1\right)^{2}}\right)^{-\frac{2}{N+1}}\\
\Psi_{\left(2\right)} & = & s_{\left(2\right)}\left(\phi-1\right)\xi\eta\frac{\kappa\rho_{\Lambda\left(2\right)}}{\kappa\rho}\sqrt{\frac{B_{H\left(2\right)}}{B_{B\left(2\right)}}}\\
 & = & \xi\eta s_{\left(2\right)}\left(\phi-1\right)\frac{\kappa\rho_{\Lambda\left(2\right)}}{\kappa\rho}\left(1+\frac{1}{\xi\eta}\frac{\left(N+1\right)}{2N}\frac{\left(\kappa\rho\right)^{2}}{\kappa\rho_{\Lambda\left(2\right)}\left(\phi-1\right)^{2}}\right)^{\frac{1}{N+1}}
\end{array}\label{eq:modMnkLK_BB_into_A_B_Psi}
\end{equation}

The world-path of the brane in bulk~$\left(2\right)$ is described
in eqs~\ref{eq:modMnkLK_gen_soln_B},\ref{eq:modMnkLK_Bsoln}. When
this solution is plugged into junction conditions~1~and~2, those
conditions are satisfied. However, the world-path of the brane has
in bulk~$\left(1\right)$ - Modified-Minkowski space-time - has not
yet been described. For this purpose, one now imposes junction condition~3
and solves for the world-path of the brane in bulk~$\left(1\right)$.
 The general form of the extrinsic curvature that will be used for
junction condition~3 is
\begin{equation}
\Pi_{\mu\nu}u^{\mu}u^{\nu}=-\dot{u}^{\bar{\nu}}n_{\bar{\nu}}+\frac{1}{2}\xi\eta\left(\ln A\right)_{,\bar{\nu}}n^{\bar{\nu}}\label{eq:modMnkLK_IJC_3_repeat}
\end{equation}
Computing $\Pi_{\mu\nu}u^{\mu}u^{\nu}$ on the Modified-Minkowski
side of the boundary yields:
\begin{equation}
\begin{array}{ccc}
\left(\Pi_{\mu\nu}u^{\mu}u^{\nu}\right)_{\left(1\right)} & = & -\dot{u}_{\left(1\right)}^{\alpha}n_{\left(1\right)\alpha}+\frac{1}{2}\xi\eta\left(\ln A_{\left(1\right)}\right)_{,\alpha}n_{\left(1\right)}^{\alpha}\\
\\
\left(\ln A_{\left(1\right)}\right)_{,\alpha}n_{\left(1\right)}^{\alpha} & = & \left(\ln A_{\left(1\right)}\right)_{,1}n_{\left(1\right)}^{1}\\
 & = & \frac{da}{dx^{1}}n_{\left(1\right)}^{1}\\
 & = & -\left(\frac{-a^{\prime}}{b^{\prime}}\right)\frac{db}{dx^{1}}n_{\left(1\right)}^{1}\\
 & = & -S_{N}\left(\ln B_{\left(1\right)}\right)_{,1}n_{\left(1\right)}^{1}\\
 & = & -\left(N-1\right)S_{2}\left(\ln B_{\left(1\right)}\right)_{,\alpha}n_{\left(1\right)}^{\alpha}\\
 & = & -\left(N-1\right)S_{2}\phi\left.\ln\left(B_{\left(2\right)}\right)_{,\beta}n_{\left(2\right)}^{\beta}\right|_{\chi_{\left(2\right)}^{\mu}\left(\tau\right)}\\
 & = & \left(N-1\right)S_{2}\frac{2\phi}{N\left(\phi-1\right)}\kappa\rho\\
\\
\left(\Pi_{\mu\nu}u^{\mu}u^{\nu}\right)_{\left(1\right)} & = & -\dot{u}_{\left(1\right)}^{\alpha}n_{\left(1\right)\alpha}+\xi\eta\left(N-1\right)S_{2}\frac{\phi}{N\left(\phi-1\right)}\kappa\rho
\end{array}\label{eq:modMnkLK_ExtCurvUU_s1}
\end{equation}
where $S_{2}\equiv\left.\frac{-a^{\prime}}{b^{\prime}}\right|_{N=2,w=w_{max}}\simeq1+5.89457\times10^{-8}$,
and $S_{N}\equiv\left(N-1\right)S_{2}$. Computing $\Pi_{\mu\nu}u^{\mu}u^{\nu}$
on the $\Lambda-K$ side of the boundary yields:
\begin{equation}
\begin{array}{ccc}
\left(\Pi_{\mu\nu}u^{\mu}u^{\nu}\right)_{\left(2\right)} & = & -\dot{u}_{\left(2\right)}^{\nu}n_{\left(2\right)\nu}+\frac{1}{2}\xi\eta\left(\ln A_{\left(2\right)}\right)_{,y}y_{\left(2\right),\bar{\nu}}n_{\left(2\right)}^{\bar{\nu}}\\
 & = & \xi\eta s_{\left(2\right)}\Psi_{\left(2\right)}+\xi\eta\left(\ln A_{\left(2\right)}\right)_{,y}s_{\left(2\right)}\Psi_{\left(2\right)}\mathcal{Y}_{\left(2\right)}\\
 & = & \xi\eta s_{\left(2\right)}\Psi_{\left(2\right)}\mathcal{Y}_{\left(2\right)}\left(\ln\left(A_{\left(2\right)}\left(-2\lambda_{\left(2\right)}y_{\left(2\right)}\right)\right)\right)_{,y}\\
 & = & -\xi\eta\frac{B\kappa\rho}{N\left(\phi-1\right)}\left(\ln\left(\frac{N}{2\sqrt{B}}B_{\left(2\right),z}\right)\right)_{,B}\\
 & = & \left(\xi\eta\right)^{2}\frac{\left(\phi-1\right)}{2\kappa\rho}\left(\frac{N-1}{N}B^{-\frac{1+N}{2}}\left(-4K\right)+\frac{4}{N+1}\kappa\rho_{\Lambda}\right)\\
 & = & \xi\eta\frac{\left(\phi-1\right)}{2\left(\kappa\rho\right)}\left(\xi\eta\left(\frac{2\left(N-1\right)}{N+1}+\frac{4}{N+1}\right)\kappa\rho_{\Lambda\left(2\right)}+\frac{N-1}{N}\frac{\left(\kappa\rho\right)^{2}}{\left(\phi-1\right)^{2}}\right)\\
 & = & \xi\eta\frac{\kappa\rho}{\left(\phi-1\right)}\left(\xi\eta\frac{\kappa\rho_{\Lambda\left(2\right)}\left(\phi-1\right)^{2}}{\left(\kappa\rho\right)^{2}}+\frac{N-1}{2N}\right)
\end{array}\label{eq:modMnkLK_ExtCurvUU_s2}
\end{equation}
  From junction condition~3 of eq~\ref{eq:IJC_gen_metric}
\begin{equation}
\begin{array}{ccc}
\left[\Pi_{\mu\nu}u^{\mu}u^{\nu}\right] & = & \xi\eta\kappa\rho\left(\frac{p}{\rho}+\frac{N-1}{N}\right)\\
 & = & \xi\eta\frac{\kappa\rho}{\left(\phi-1\right)}\left(\frac{p}{\rho}\left(\phi-1\right)+\frac{N-1}{N}\left(\phi-1\right)\right)
\end{array}\label{eq:modMnkLK_ExtCurvUU_braneSE}
\end{equation}
Combining eqs~\ref{eq:modMnkLK_ExtCurvUU_s1}~and~\ref{eq:modMnkLK_ExtCurvUU_s2}
yields:
\begin{equation}
\begin{array}{ccc}
\left[\Pi_{\mu\nu}u^{\mu}u^{\nu}\right] & = & \left(\Pi_{\mu\nu}u^{\mu}u^{\nu}\right)_{\left(1\right)}-\left(\Pi_{\mu\nu}u^{\mu}u^{\nu}\right)_{\left(2\right)}\\
 & = & -\dot{u}_{\left(1\right)}^{\alpha}n_{\left(1\right)\alpha}-\xi\eta\frac{\kappa\rho}{\left(\phi-1\right)}\left(-S_{2}\frac{N-1}{N}\phi+\xi\eta\frac{\kappa\rho_{\Lambda\left(2\right)}\left(\phi-1\right)^{2}}{\left(\kappa\rho\right)^{2}}+\frac{N-1}{2N}\right)
\end{array}\label{eq:modMnkLK_ExtCurvUU_discont}
\end{equation}
Enforcing junction condition~3 by equating eqs~\ref{eq:modMnkLK_ExtCurvUU_braneSE}~and~\ref{eq:modMnkLK_ExtCurvUU_discont}
yields: 
\begin{equation}
\dot{u}_{\left(1\right)}^{\mu}n_{\left(1\right)\mu}=\xi\eta\frac{\kappa\rho}{\left(\phi-1\right)}\left(S_{2}\frac{N-1}{N}\phi-\xi\eta\left(\phi-1\right)^{2}\frac{\kappa\rho_{\Lambda\left(2\right)}}{\left(\kappa\rho\right)^{2}}-\frac{N-1}{2N}-\left(\phi-1\right)\left(\frac{p}{\rho}+\frac{N-1}{N}\right)\right)\label{eq:modMnk_LK_EoM1_a}
\end{equation}
This leads to a modified expression for $\alpha$ of eq~\ref{eq:MnkLK_mnkYaccel}
\begin{equation}
\alpha\equiv s_{\left(1\right)}\frac{\kappa\rho}{\left(\phi-1\right)}\left(\left(\phi-1\right)\left(\frac{p}{\rho}+\frac{N-1}{N}\right)+\frac{N-1}{2N}-S_{2}\frac{N-1}{N}\phi-\xi\eta\left(\phi-1\right)^{2}\frac{\kappa\rho_{\Lambda\left(2\right)}}{\left(\kappa\rho\right)^{2}}\right)\label{eq:modMnkLK_mnkYaccel}
\end{equation}
For the models considered in this paper, the case of interest will
be when the brane is stationary in the Modified-Minkowski region and
$\xi\eta=+1$. This requires $\alpha=0$. Eq~\ref{eq:modMnkLK_mnkYaccel}
is a quadratic equation in $\phi-1$. Because it has been assumed
$\kappa\rho_{\Lambda\left(2\right)}>0$, there are always two real
roots of $\alpha=0$ - one with $\phi>1$ and one with $\phi<1$.
The latter leads to either negative $\kappa\rho$ or proper time on
the brane increasing as $x^{0}$ decreases. Thus the root of $\alpha=0$
with $\phi<1$ is discarded, and the acceptable value of $\phi$ for
which the brane is stationary in the Modified-Minkowski region is
\begin{equation}
\begin{array}{ccc}
\\
\phi-1 & = & \frac{\left(\kappa\rho\right)^{2}}{2\xi\eta\kappa\rho_{\Lambda\left(2\right)}}\left(\left(\left(S_{2}-1\right)\left(\frac{N-1}{N}\right)-\frac{p}{\rho}\right)\right.\\
 &  & \left.+\sqrt{\left(\left(S_{2}-1\right)\left(\frac{N-1}{N}\right)-\frac{p}{\rho}\right)^{2}+4\xi\eta\left(S_{2}-\frac{1}{2}\right)\left(\frac{N-1}{N}\right)\frac{\kappa\rho_{\Lambda\left(2\right)}}{\left(\kappa\rho\right)^{2}}}\right)>0
\end{array}\label{eq:modMnkLK_phiSoln}
\end{equation}

\subsection{What geodesics in $\Lambda-K$ bulk space look like }

\noindent\fbox{\begin{minipage}[t]{1\columnwidth - 2\fboxsep - 2\fboxrule}%
In this section: 
\begin{lyxlist}{00.00.0000}
\item [{1)}] The nature of geodesics in a $\Lambda-K$ space-time are examined.
The behavior of a test particle on the geodesic may be understood
in analogy to a ball rolling on a hill, with kinetic energy $KE$
and potential energy $PE$. 
\item [{2)}] Equivalently, the analogy may be represented as the equality
of a force $F$ to mass $m$ times acceleration $a$. 
\item [{3)}] A $y=const$ geodesic of a test particle of relativistic velocity
$v^{\mu}$ being stable requires $\text{sign}\left(K\right)=\text{sign}\left(\kappa\rho_{\Lambda}\right)=-v^{\mu}v_{\mu}$
\item [{4)}] Recall the existence of a horizon in $\Lambda-K$ space-time
requires $\text{sign}\left(K\right)=-\text{sign}\left(\kappa\rho_{\Lambda}\right)$,
thus a stable geodesic can not exist if there is a horizon. 
\item [{5)}] A $y=const$ stable geodesic would be desired for a patch
of space that could approximate Minkowski - for a passenger traveling
inside $\Lambda-K$ space-time. The unavailability of a horizon in
such a space has important implications for model-building such as
those in section~\ref{sec:Models_presented}. 
\end{lyxlist}
\end{minipage}}

Returning attention to the $\Lambda-K$ bulk space-time, for the purpose
of future model-building it is worth considering what geodesics look
like. Since a geodesic might be an ideal world-path for a passenger
in a propulsion model, one would be particularly interested in geodesics
that would not collide with the brane(s), and for which the metric
would change little or none with respect to proper time of the passenger,
and for which variation of the metric around the geodestic would not
be extreme. To that end, particular attention will paid to the existence
of a geodesic of constant~$y$. To a passenger along a $y=const$
geodesic the metric would be constant. When $A$ and $B$ are constant
with respect to the proper time of a passenger, it is not required,
of course, that they are equal. A coordinate rescaling by a constant
factor could transform the metric, evaluated along such a geodesic,
to that of Minkowski space-time. Of course one would still need to
be concerned with variation of the metric with respect to slight variations
of the world-path around this geodesic. 

Let $\chi^{\mu}\left(\tau\right)$ be a world-path of the geodesic
as a function of proper time~$\tau$, and $v^{\mu}\equiv\dot{\chi}^{\mu}$
be its relativistic velocity. For computational simplicity, it will
be assumed that $v^{\mu}$ has no component in the parallel directions
(ie: $v^{\mu}=0$ for $2\le\mu\le N+1$, or $v^{\underline{\mu}}=0$).
To derive the geodesic world-path, consider the deviation $a_{v}^{\mu}$
of $v^{\mu}$ away from parallel-transport, and set this to zero:
\begin{equation}
\begin{array}{ccccc}
a_{v}^{\mu} & = & \frac{dv^{\mu}}{d\tau}+v^{\alpha}v^{\beta}g^{\mu\nu}\Gamma_{\nu\alpha\beta} & = & 0\\
 & = & \dot{v}^{\mu}+v^{\alpha}v^{\beta}g^{\mu\nu}\left(g_{\nu\alpha}\frac{A_{,\beta}}{A}-\frac{1}{2}g_{\alpha\beta}\frac{A_{,\nu}}{A}\right)\\
 & = & \frac{1}{A}\left(\dot{v}^{\mu}A+v^{\mu}\dot{A}-\frac{\xi\eta}{2}g^{\mu\beta}A_{,\beta}\right)
\end{array}\label{eq:gen_V_accel}
\end{equation}
Note that $\xi\eta=v^{\mu}v_{\mu}$ is that of the geodesic relativistic
velocity itself - it is independent and unrelated to the value of
$\xi\eta$ of any branes that might be in this bulk..
\begin{equation}
\begin{array}{ccccc}
a_{v}^{\mu}A\eta_{\mu\nu}\chi^{\nu} & = & \ddot{\chi}^{\mu}\eta_{\mu\nu}\chi^{\nu}A+\dot{\chi}^{\mu}\eta_{\mu\nu}\chi^{\nu}\dot{A}-\frac{\xi\eta}{2}\frac{1}{A}A_{,\nu}\chi^{\nu} & = & 0\\
 & = & \frac{d}{d\tau}\left(\dot{\chi}^{\mu}\eta_{\mu\nu}\chi^{\nu}A\right)-\dot{\chi}^{\mu}\eta_{\mu\nu}\dot{\chi}^{\nu}A-\frac{\xi\eta}{2}\frac{1}{A}A_{,\nu}\chi^{\nu}\\
 & = & \frac{d}{d\tau}\left(\dot{\chi}^{\mu}\eta_{\mu\nu}\chi^{\nu}A\right)-\xi\eta-\frac{1}{2}\xi\eta\frac{1}{A}A_{,\nu}\chi^{\nu}\\
 & = & \frac{1}{-4\lambda}\frac{d}{d\tau}\left(\dot{y}A\right)-\frac{\xi\eta}{A}\left(Ay\right)_{,y}
\end{array}\label{eq:LK_Geod_deriv_B}
\end{equation}
where we have used $\dot{\chi}^{\mu}=v^{u}$ and $y\equiv\left(-2\lambda\right)\eta_{\mu\nu}\chi^{\mu}\chi^{\nu}$,
where $\eta_{\mu\nu}\equiv diag\left(1,-1,-1,\cdots,-1\right)$ is
the Minkowski metric.
\begin{equation}
\begin{array}{ccccc}
a_{v}^{\mu}\eta_{\mu\nu}\chi^{\nu}A^{2}\dot{y} & = & \frac{1}{-4\lambda}\frac{d}{d\tau}\left(\dot{y}A\right)A\dot{y}-\xi\eta\left(Ay\right)_{,y}\dot{y} & = & 0\\
a_{v}^{\mu}\eta_{\mu\nu}\chi^{\nu}A^{2}\left(-2\lambda\dot{y}\right) & = & \frac{1}{4}\frac{d}{d\tau}\left(\dot{y}^{2}A^{2}\right)-\xi\eta\frac{d}{d\tau}\left(A\left(-2\lambda y\right)\right) & = & 0
\end{array}\label{eq:LK_Geod_deriv_C}
\end{equation}
Integrate with respect to proper time $\tau$ along the world-path
to get:
\begin{equation}
\begin{array}{c}
\frac{1}{4}\left(\dot{y}A\right)^{2}-\xi\eta A\left(-2\lambda y\right)+Const=0\\
\frac{N^{2}}{4}\left(\frac{d\sqrt{B}}{d\tau}\right)^{2}-\xi\eta A\left(-2\lambda y\right)+Const=0\\
KE\sim\frac{1}{4}\left(\dot{y}A\right)^{2}=\frac{N^{2}}{4}\left(\frac{d\sqrt{B}}{d\tau}\right)^{2}\ \ \ PE\sim-\xi\eta A\left(-2\lambda y\right)+Const\\
KE+PE=0\\
\\
\frac{N^{2}}{2}\left(\frac{d^{2}\sqrt{B}}{d\tau^{2}}\right)=\xi\eta\left(A\left(-2\lambda y\right)\right)_{,\sqrt{B}}\\
ma\sim\frac{N^{2}}{2}\left(\frac{d^{2}\sqrt{B}}{d\tau^{2}}\right)\ \ \ F\sim\xi\eta\left(A\left(-2\lambda y\right)\right)_{,\sqrt{B}}\\
ma=F
\end{array}\label{eq:LK_Geod_Kine_Dynam_equiv}
\end{equation}
where $A=N\left(\sqrt{B}\right)_{,y}$ has been used, and $\frac{d\sqrt{B}}{d\tau}$
is the rate of change of $\sqrt{B}$ with respect to $\tau$ along
the geodesic path. Thus the evolution of the system along a geodesic
can be understood in analogy to a ball rolling on a hill. $KE$ represents
the kinetic energy of such a rolling ball, and $PE$ its potential
energy. Equivalently, a dynamical perspective may be adopted, with
a ``mass~$\times$~acceleration'' being equal to a ``force''.
This is true for an arbitrary geodesic in $\Lambda-K$ space-time.
To find a $y=constant$ geodesic at some $B=B_{E}$, the intuition
of eq~\ref{eq:LK_Geod_Kine_Dynam_equiv} tells us that that it is
necessary and sufficient to require an extremum of $PE$ at $B_{E}$
with respect to $y$ (or equivalently~$\sqrt{B}$). Furthermore,
stability of such a geodesic requires that the extremum is also a
minimum.
\begin{equation}
\begin{array}{ccccc}
\\
PE_{,y} & = & \left(-\xi\eta A\left(-2\lambda y\right)\right)_{,y} & = & 0\\
 & = & \left(-\xi\eta\frac{N}{2\sqrt{B}}B_{,z}\right)_{,y}\\
 & = & -\xi\eta\left(B^{\frac{1-N}{2}}K+\frac{N}{2\left(N+1\right)}B\kappa\rho_{\Lambda}\right)_{,y}\\
 & = & -\xi\eta\left(B^{\frac{1-N}{2}}K+\frac{N}{2\left(N+1\right)}B\kappa\rho_{\Lambda}\right)_{,B}B_{,y}\\
 & = & -\xi\eta\left(\frac{1-N}{2}B^{\frac{-1-N}{2}}K+\frac{N}{2\left(N+1\right)}\kappa\rho_{\Lambda}\right)B_{,y}
\end{array}\label{eq:eq:LK_Geod_PE_extremum}
\end{equation}
This is satisfied if $N\ge2$ and $\text{sign}\left(K\right)=\text{sign}\left(\kappa\rho_{\Lambda}\right)$
for 
\begin{equation}
B_{E}\equiv\left(\frac{\left(N+1\right)\left(N-1\right)}{N}\frac{K}{\kappa\rho_{\Lambda}}\right)^{\frac{2}{N+1}}\label{eq:eq:LK_Geod_BE_soln}
\end{equation}
 The same result can be obtained by plugging in the $y=constant$
expression for $u^{\mu}$ and $n^{\mu}$ from eq~\ref{eq:MnkLK_gen_soln_B}
into eq~\ref{eq:gen_V_accel}. Recall that existence of the horizon
at $B_{H}$ requires $\text{sign}\left(K\right)=-\text{sign}\left(\kappa\rho_{\Lambda}\right)$.
Thus a $\Lambda-K$ space-time can not have both a horizon at $B_{H}$
and a $y=constant$ geodesic at $B_{E}$ (for $0<B<\infty$). For
this geodesic to be stable, the extremum must also be a minimum of
$PE$, which requires $\text{sign}\left(K\right)=\text{sign}\left(\kappa\rho_{\Lambda}\right)=-\xi\eta=-v^{\mu}v_{\mu}$.

\section{Results - Specific Models: Sub-Luminal and Super-Luminal\label{sec:Models_presented}}

\noindent\fbox{\begin{minipage}[t]{1\columnwidth - 2\fboxsep - 2\fboxrule}%
In this section: 
\begin{lyxlist}{00.00.0000}
\item [{1)}] The components introduced in section~\ref{sec:Model_Components}
are combined to create propulsion models. 
\item [{2)}] One desires the passenger to ride in a patch of Minkowski
space-time. The space-time external to the model is assumed to be
Minkowski. However, one gains no propulsion effects if Minkowski space
is present on both sides of a boundary. Thus one must layer the branes
as in fig~\ref{fig:supLum_CrossSect} and fig~\ref{fig:supLum_CrossSect},
with curved space-time in region~$\left(B\right)$ and region~$\left(D\right)$
between the inner and outer branes.
\item [{3)}] A sub-luminal model is constructed in section~\ref{subsec:Mdl_subLum}
where the outer branes follow a time-like trajectory in the external
Minkowski space-time. 
\item [{4)}] A super-luminal model is constructed in section~\ref{subsec:Mdl_supLum}
where the outer branes follow a space-like trajectory in the external
Minkowski space-time. This model satisfies the weak and null energy
conditions, but fails the dominant energy condition at the external
branes. Additionally, a mechanism of transition from the sub-luminal
to super-luminal model is speculated on.
\end{lyxlist}
\end{minipage}}

In this section, models will be constructed from the components of
section~\ref{sec:Model_Components} and analyzed. Note that although
these models of propulsion do accelerate in the ambient bulk space-time,
stress-energy is everywhere conserved. This must be true because the
junction conditions come from the Einstein field equations, and those
have vanishing covariant divergence (and thus must always be coupled
to a conserved stress-energy tensor. For that reason, these models
are guaranteed to conserve energy and momentum. If one attempted to
have such a model do work against a measurement apparatus, one would
find the work done exactly canceled loss of energy somewhere in the
model - perhaps by deformation leading to a change of volume, or reshifting
due to changing scale factor, etc.

Two types of propulsion models will be considered - sub-luminal and
super-luminal. The former is fairly straight-forward, requiring only
tuning to maintain the constant size of the interior Minkowski space-time
region intended for a passenger. The super-luminal models, however,
are much more involved to construct. The essence of the difficulty
is that one is trying to match space-like to time-like world-paths.
With the components developed in this paper, a horizon is apparently
required. This would significantly complicate operation of any realistic
spaceship using such a propulsion model. Thus, it seem that a horizon
is an unavoidable consequence of super-luminal travel. However, all
of the models presented in this section are constructed with non-negative
energy density. It is shown that both the sub-luminal and super-luminal
warp models satisfy the weak and null energy conditions. The sub-luminal
model also satisfies the dominant energy condition, while the super-luminal
model is guaranteed to violate it. 

\subsection{Sub-luminal warp\label{subsec:Mdl_subLum} }

\noindent\fbox{\begin{minipage}[t]{1\columnwidth - 2\fboxsep - 2\fboxrule}%
In this subsection:
\begin{lyxlist}{00.00.0000}
\item [{1)}] A sub-luminal model is constructed where the outer branes
follow a time-like trajectory in the external Minkowski space-time,
while the inner branes are static in region~$\left(C\right)$. 
\item [{2)}] The weak, null and dominant energy conditions are satisfied
with all energy density positive ($\text{sign}\left(K\right)=-1=-\text{sign}\left(\kappa\rho_{\Lambda}\right)$)
and the equations of state in the interval~$\left[-1,1\right]$. 
\item [{3)}] Model parameter values or ranges are determined as several
constraints are imposed: that the inner branes be stationary in region~$\left(C\right)$,
that region~$\left(B\right)$ and region~$\left(D\right)$ have
the same $\kappa\rho_{\Lambda}$, that the outer branes accelerate
in the same direction (one towards the Minkowski side, the other away
from it), with equal magnitude. 
\end{lyxlist}
\end{minipage}}

\begin{figure}

\begin{centering}
\includegraphics[scale=0.55]{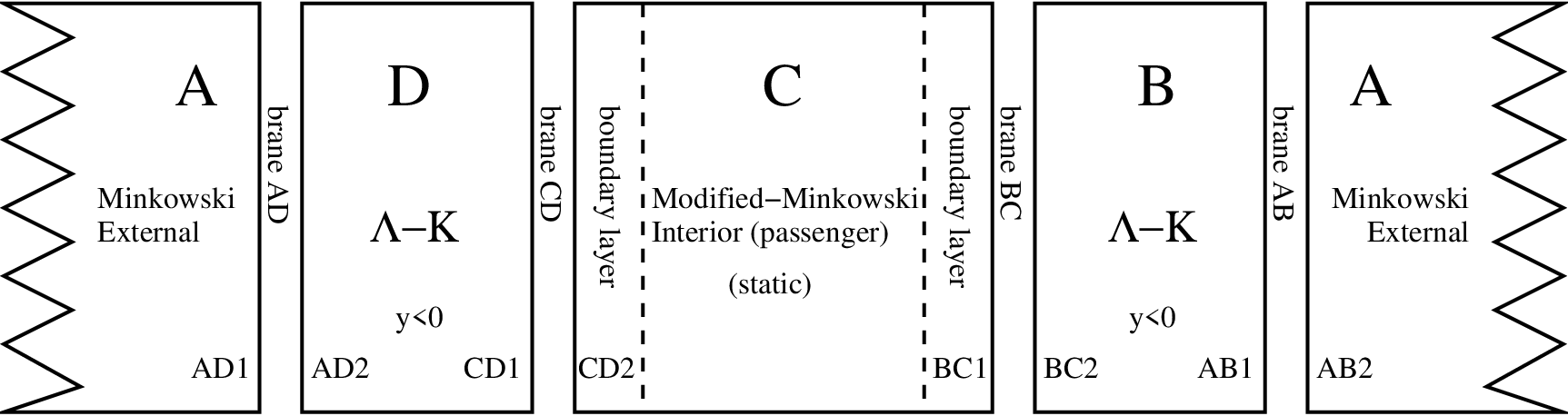}\label{fig:subLum_CrossSect}\caption{ This figure shows the cross-section of a possible sub-luminal propulsion
model, constructed according to the constraints explored in sub-sections~\ref{subsec:MnkLK_constr}
and \ref{subsec:modMnkLK_constr}. Minkowski region~$\left(A\right)$
(or bulk~$\left(A\right)$) is the external Universe. A passenger
rides in Modified-Minkowski region~$\left(C\right)$. Regions~$\left(B\right)\,\&\,\left(D\right)$
are $\Lambda-K$ space-times. Four branes at the interface between
these regions are labeled $AB$, $BC$, $CD$, $AD$. The opposite
sides of those branes are labeled $AB1$~\&~$AB2$; $BC1$~\&~$BC2$;
$CD1$~\&~$CD2$; $AD1$~\&~$AD2$. }
\par\end{centering}
\end{figure}

The first model to be considered will be that of sub-luminal warp.
The chimeric space-time is constructed to have a Modified-Minkowski
interior (bulk~$\left(C\right)$~=~region~$\left(C\right)$, where
a passenger would ride), with an exterior Minkowski space-time (region~$\left(A\right)$)
shared with an observer at infinity. Sandwiched between the Minkowski
regions is, as shown in fig~\ref{fig:subLum_CrossSect}, are $\Lambda-K$
space-time regions~$\left(B\right)$~\&~$\left(D\right)$. It is
not required in this model for region~$\left(B\right)$ and region~$\left(D\right)$
to have the same values of $K<0$ and $\kappa\rho_{\Lambda}>0$. However,
a more realistic warp-drive model would have a single $\Lambda-K$~bulk
(or its equivalent) enclosing region~$\left(C\right)$. Thus, looking
forward to future model development, one should require $K_{\left(B\right)}=K_{\left(D\right)}$
and $\kappa\rho_{\Lambda\left(B\right)}=\kappa\rho_{\Lambda\left(D\right)}$.
Furthermore, an additional requirement will be imposed: no negative
energy densities - in a bulk or on a brane. Although it could be speculated
that quantum effects such as the Casimir effect could generate effective
negative energy densities, all quantum effects are outside of the
scope of this paper, and thus will not be considered. Thus $\kappa\rho_{\Lambda\left(B\right)}=\kappa\rho_{\Lambda\left(D\right)}\ge0$
and $\kappa\rho_{\left(j\right)}\ge0\ \ \forall j$ will be imposed.
It will then be necessary to have $K_{\left(B\right)},\,K_{\left(D\right)}<0$
to satisfy $B_{\left(B\right),z}<0$ and $B_{\left(D\right),z}<0$
in eq~\ref{eq:MnkLK_gen_soln_D}. One might argue that although this
$K$ is merely an integration constant from solving the bulk Einstein
field equations, it has an origin similar to the central mass in the
Schwarzschild black hole solution, and thus selecting a particular
sign for $K$ might be equivalent to imposing the existence of negative
mass. However, that argument is countered by the fact that such an
equivalent to the Schwarzschild black hole mass would be located at
the physical singularity at~$B=0$. Due to the chimeric nature of
this model's space-time surgery, the $B=0$ location does not exist
- it has been discarded by cutting out that part of regions~$\left(B\right)$~\&~$\left(D\right)$,
and replacing it with region~$\left(C\right)$. This is one of the
advantages of utilizing chimeric space-times in model-building - one
may create models that involve a space-time which would normally include
an undesirable region, but then excise that region via surgery. One
must still insure that the boundary conditions that are necessary
are not themselves forbidden in any way. In the models presented in
this paper all energy densities are non-negative, and the equations
of state are in the interval~$\left[-1,1\right]$. 

\begin{figure}

\centering{}\includegraphics[viewport=318bp 4bp 789bp 608bp,clip,scale=0.9]{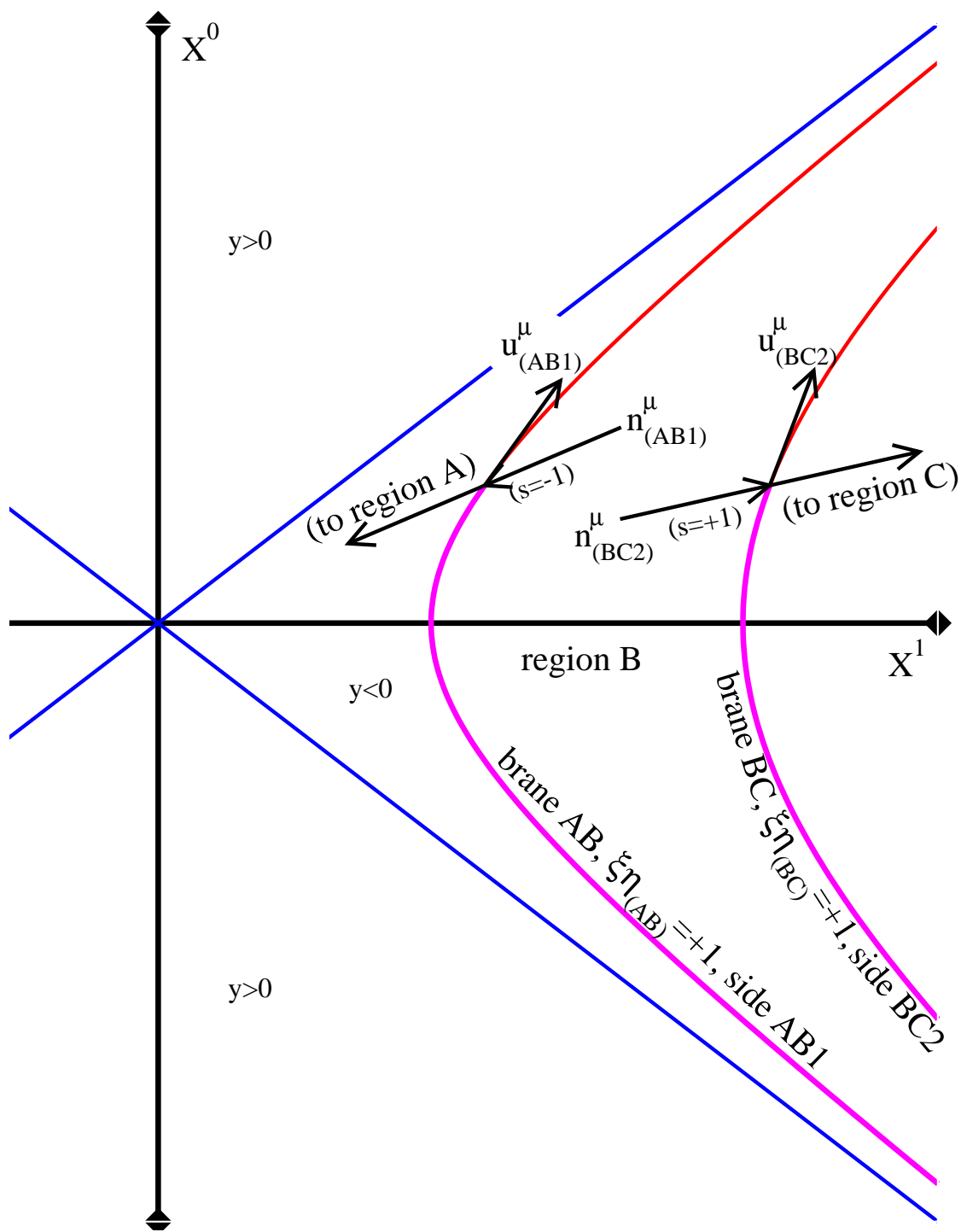}\label{fig:subLum_ConfDiag}\caption{This figure shows the conformal structure of region~$\left(B\right)$
($\Lambda-K$ space-time) of the sub-luminal propulsion model shown
in fig~\ref{fig:subLum_CrossSect}. Region~$\left(D\right)$ has
an identical structure. The red curves are the $y=const$ world-paths
of the branes, and the magenta segments of those curves are the history
of those branes at some coordinate time~$x^{0}$. The bulk between
those red curves is utilized - it exists in this model. The remainder
of the $\Lambda-K$ space-time - including the curvature singularity
at $B=0$ and the horizon at $B=B_{H}$ - is discarded, and does not
exist in this model. }
\end{figure}

Brane~AB requires $s_{\left(AB1\right)}=-1$ for region~$\left(B\right)$
to be the correct side of the world-path of brane~AB in the $\Lambda-K$
bulk (similarly $s_{\left(AD2\right)}=-1$ and thus $s_{\left(AB1\right)}y>0$
and $s_{\left(AD2\right)}y>0$). The time-like trajectory of brane~AB
implies that~$\xi_{\left(AB1\right)}=\xi_{\left(AB2\right)}=+1$,
and because $A>0$ everywhere, one also has $\left(\xi\eta\right)_{AB}=+1$.
Considering eq~\ref{eq:MnkLK_gen_soln_D}, this is consistent with
$\kappa\rho_{AB}>0$. However, brane~BC requires $s_{\left(BC2\right)}=+1$
for region~$\left(B\right)$ to be the correct side of the world-path
of brane~BC in that same $\Lambda-K$ bulk. If region~$\left(C\right)$
were merely standard Minkowski space-time, by eq~\ref{eq:MnkLK_gen_soln_D}
this would require $\kappa\rho_{BC}<0$, which is to be avoided. Thus
region~$\left(C\right)$ must instead be Modified-Minkowski space-time.
One sets $\kappa\rho_{BC}=\kappa\rho_{CD}$, $p_{BC}=p_{CD}$ and
chooses~$\phi_{\left(C\right)}=\phi_{\left(BC1\right)}=\phi_{\left(CD2\right)}>1$,
then by eq~\ref{eq:modMnkLK_gen_soln_D} $\kappa\rho_{BC}>0$ and
$\kappa\rho_{CD}>0$. 

It will be assumed that the model has reflection symmetry in the $x^{1}$
direction around region~$\left(C\right)$, except for the equation
of state (pressure) on brane~AB vs brane~AD. That is, pick a point
in region~$\left(C\right)$ such that the space-time in an open neighborhood
around it is standard Minkowski. Then apply reflection symmetry in
the $x^{1}$ direction around that point - except for the pressure~$p_{\left(AB\right)}$
vs $p_{\left(AD\right)}$. This specifies all of the internal parameters
of the model. Because all derivatives of all orders of the metric
vanish inside this neighborhood, reflecting around a point within
it in this manner is a trivial matter. Setting the pressure on brane~AB
and brane~AD then specifies the acceleration of those branes in the
external (standard) Minkowski space-time - relative to an external
observer.

\subsubsection{Energy Conditions in the Sub-Luminal Model: Dominant, Weak, Null
and Strong}

A natural question is what energy conditions (dominant, weak, null
and strong) do the propulsion models presented in this paper satisfy,
and which do they violate. For a time-like world-path brane ($u^{\mu}$
a time-like vector, $u^{\mu}u_{\mu}=\eta=sign\left(A\right)$, ie:
$\xi\eta=+1$) eq~\ref{eq:brane_SEten_PerfFluid} indicates that
the stress-energy on the brane is a perfect fluid, and thus one might
expect the conditions under which the various energy conditions hold
to be a well-known matter. However, there is also the fact that the
brane is a co-dimension~1 hypersuface with a normal vector that yields
zero when contracted with the brane stress-energy tensor. Thus a detailed
investigation is warranted. 

Consider an arbitrary bulk vector positioned on the brane world-path
- it can be divided into three components - a component collinear
with the brane relativistic velocity~$u^{\mu}$, a component collinear
with the brane normal vector $n^{\mu}$ and a component in the parallel
space-like directions of the brane. 
\begin{equation}
X^{\mu}=\alpha_{u}u^{\overline{\mu}}+\alpha_{n}n^{\overline{\mu}}+\alpha_{p}v^{\underline{\mu}}\label{eq:gen_RelVec_decomp}
\end{equation}
 Note that $X^{\mu}$ need not be tangent to the world-sheet of the
brane. To investigate the weak and null energy conditions, consider
under which conditions $S_{\mu\nu}X^{\mu}X^{\nu}\ge0$. This model
is sub-luminal - thus $\xi\eta=+1$ for all branes 
\begin{equation}
\begin{array}{ccc}
S_{\mu\nu} & = & \left(\rho+p\right)u_{\mu}u_{\nu}-\xi\eta p\left(g_{\mu\nu}+\frac{1}{\xi\eta}n_{\mu}n_{\nu}\right)\\
 & = & \left(\rho+p\right)u_{\mu}u_{\nu}-p\left(g_{\mu\nu}+n_{\mu}n_{\nu}\right)\\
S_{\mu\nu}X^{\mu}X^{\nu} & = & \rho\alpha_{u}^{2}+pB\alpha_{p}^{2}v^{2}\\
v^{2} & \equiv & \delta_{\underline{\mu\nu}}v^{\underline{\mu}}v^{\underline{\nu}}\\
X^{\mu}X_{\mu} & = & g_{\mu\nu}X^{\mu}X^{\nu}\\
 & = & \alpha_{u}^{2}-\alpha_{n}^{2}-B\alpha_{p}^{2}v^{2}
\end{array}\label{eq:gen_RelVec_prods}
\end{equation}
 where $X^{\mu}X_{\mu}>0$ for the weak energy condition and $X^{\mu}X_{\mu}=0$
for the null energy condition. 
\begin{equation}
\begin{array}{cccccc}
Weak & : & S_{\mu\nu}X^{\mu}X^{\nu}\ge0 & \forall X^{\mu}X_{\mu}>0 & \Rightarrow & B\alpha_{p}^{2}v^{2}<\alpha_{u}^{2}\\
\\
Null & : & S_{\mu\nu}X^{\mu}X^{\nu}\ge0 & \forall X^{\mu}X_{\mu}=0 & \Rightarrow & B\alpha_{p}^{2}v^{2}\le\alpha_{u}^{2}\\
\\
S_{\mu\nu}X^{\mu}X^{\nu}\ge0 & \Rightarrow & \rho\alpha_{u}^{2}+pB\alpha_{p}^{2}v^{2}\ge0\\
 & \Rightarrow & \rho+p\ge0
\end{array}\label{eq:gen_RelVec_Wk_Nl_ECs}
\end{equation}
Because $S_{\mu\nu}n^{\mu}=0$, it possible for $\alpha_{p}^{2}v^{2}\rightarrow0$
while $\alpha_{u}^{2}>0$ even in the null case. Thus both the weak
and null energy conditions require $\rho\ge0$, which is already assumed
for all energy densities (bulk and brane). Additionally, both the
weak and null energy conditions require $\rho+p\ge0$. This is satisfied
if one demands that the equation of state always fall in the interval
$\left[-1,1\right]$, which is done for all models. Thus the sub-luminal
warp model satisfies the weak and null energy conditions. 

The dominant energy condition requires the weak energy condition be
satisfied, and additionally that, given a future-pointing time/light-like
$X^{\mu}$, that $Y^{\mu}\equiv S_{\nu}^{\mu}X^{\nu}$ is a future-pointing
time/light-like vector. Using the same vector decomposition, one finds:
\[
Y^{\mu}=\rho\alpha_{u}u^{\mu}-p\alpha_{p}v^{\underline{\mu}}
\]
The condition on $X^{\mu}$ implies $X^{\mu}X_{\mu}=\alpha_{u}^{2}-\alpha_{n}^{2}-B\alpha_{p}^{2}v^{2}\ge0$
which in turn implies $\alpha_{u}^{2}\ge\alpha_{n}^{2}+B\alpha_{p}^{2}v^{2}\ge B\alpha_{p}^{2}v^{2}$
which yields: 
\[
Y^{\alpha}Y_{\alpha}=\rho^{2}\alpha_{u}^{2}-p^{2}B\alpha_{p}^{2}v^{2}\ge\rho^{2}\alpha_{u}^{2}-p^{2}\alpha_{u}^{2}=\alpha_{u}^{2}\left(\rho^{2}-p^{2}\right)
\]
 Thus $Y^{\mu}$ is time/light-like if $\rho\ge\left|p\right|$ in
addition to the weak energy condition being satisfied. In addition,
$X^{\mu}$ being future-pointing implies $\alpha_{u}u^{0}+\alpha_{n}n^{0}=\alpha_{u}u^{0}+\alpha_{n}su^{1}>0$
(recall $s=\pm1$), while $X^{\mu}$ being non-space-like implies
$\left|\alpha_{u}\right|\ge\left|\alpha_{n}\right|$ and $u^{\mu}$
future-pointing and time-like implies $u^{0}>0$ and $u^{0}>\left|n^{0}\right|$,
which implies $\alpha_{u}>0$ and hence $Y^{0}=\rho\alpha_{u}u^{0}>0$.
Thus all brane and bulk stress-energy tensors in the sub-luminal warp
model satisfy the dominant energy condition. Note that the weak, null
and dominant energy conditions hold for all branes in this model,
since all branes have world-paths such that $\xi\eta=+1$. 

The strong energy condition is already violated for the bulk $\Lambda-K$
space-times, as they contain a cosmological constant. One could impose
the strong energy condition for the branes merely by increasing the
lower limit imposed on the equation of state. However, the value of
doing so is unclear. Thus, the models presented in this paper will
be allowed to violate the strong energy condition.

\subsubsection{Determining the Model Parameters}

Because this is a model of sub-luminal propulsion, $\xi_{\left(AB2\right)}=\xi_{\left(AD1\right)}=+1$.
Because the branes around region~$\left(C\right)$ should be stationary
in region~$\left(C\right)$, $\xi_{\left(BC1\right)}=\xi_{\left(CD2\right)}=+1$.
$\eta\equiv\text{sign}\left(A\right)=+1$ in the Minkowski bulk space-time,
and in the Kruskal--Szekeres-analog coordinate frame in the $\Lambda-K$
bulk, $\eta=+1$ in region~$\left(B\right)$ and region~$\left(C\right)$.
Thus due to $\eta=+1$ everywhere, and the fact that $\xi\eta$ must
have the same value on either side of the brane, it must be true that
$\xi\eta=+1$ and $\xi=+1$ everywhere. By the conditions summarized
in eq~\ref{eq:MnkLK_req_smry}, all of the world-paths of the branes
in the $\Lambda-K$ bulk satisfy $\left|\chi^{0}\right|<\left|\chi^{1}\right|$.
Eq~\ref{eq:MnkLK_req_smry} also requires that $B_{,z}<0$ in the
region $\left|\chi^{0}\right|<\left|\chi^{1}\right|$, which in turn
implies that $\text{sign}\left(K\right)=-\text{sign}\left(\kappa\rho_{\Lambda}\right)$.
Due to the insistence that $\kappa\rho_{\Lambda}>0$, $K<0$ and $K$
is dominant in $\left|\chi^{0}\right|<\left|\chi^{1}\right|$. The
value of $B$ is smaller in $\left|\chi^{0}\right|<\left|\chi^{1}\right|$,
passes through $B_{H}$ at $y=0$, and becomes larger in $\left|\chi^{1}\right|<\left|\chi^{0}\right|$.
There is a curvature singularity at $B=0$ along the curve $y=-\exp\left(\mathcal{H}_{N}\left(0\right)\right)$
in $\left|\chi^{0}\right|<\left|\chi^{1}\right|$ - however this region
does not exist in this model. 

The branes~$BC$ and $CD$ should be stationary in region~$\left(C\right)$
(however, in regions~$\left(B\right)$~and~$\left(D\right)$ they
will follow a hyperbolic trajectory - one of constant~$y$). Thus,
recalling eq~\ref{eq:modMnkLK_phiSoln}, $\alpha_{\left(BC1\right)}=\alpha_{\left(CD2\right)}=0$
is satisfied if one sets: 
\[
\begin{array}{ccc}
\phi_{\left(C\right)}-1 & = & \frac{\left(\kappa\rho_{\left(BC\right)}\right)^{2}}{2\xi\eta\kappa\rho_{\Lambda\left(B\right)}}\left(\left(\left(S_{2}-1\right)\left(\frac{N-1}{N}\right)-\frac{p_{\left(BC\right)}}{\rho_{\left(BC\right)}}\right)\right.\\
 &  & \left.+\sqrt{\left(\left(S_{2}-1\right)\frac{N-1}{N}-\frac{p_{\left(BC\right)}}{\rho_{\left(BC\right)}}\right)^{2}+4\xi\eta\left(S_{2}-\frac{1}{2}\right)\left(\frac{N-1}{N}\right)\frac{\kappa\rho_{\Lambda\left(B\right)}}{\left(\kappa\rho_{BC}\right)^{2}}}\right)
\end{array}
\]

Consider now the outer branes. In the general case the outer branes
would have accelerations in the external Minkowski space 
\begin{equation}
\begin{array}{ccc}
\alpha_{\left(AB2\right)} & = & s_{\left(AB2\right)}\kappa\rho_{\left(AB\right)}\left(\frac{p_{\left(AB\right)}}{\rho_{\left(AB\right)}}+\frac{N-1}{2N}-\left(\xi\eta\right)_{\left(AB\right)}\frac{\kappa\rho_{\Lambda\left(B\right)}}{\left(\kappa\rho_{\left(AB\right)}\right)^{2}}\right)\\
\alpha_{\left(AD1\right)} & = & s_{\left(AD1\right)}\kappa\rho_{\left(AD\right)}\left(\frac{p_{\left(AD\right)}}{\rho_{\left(AD\right)}}+\frac{N-1}{2N}-\left(\xi\eta\right)_{\left(AD\right)}\frac{\kappa\rho_{\Lambda\left(D\right)}}{\left(\kappa\rho_{\left(AD\right)}\right)^{2}}\right)
\end{array}\label{eq:Mdl_subLum_OBRawAccel}
\end{equation}
 However, it would be preferable for branes $AB$ and $AD$ to have
acceleration of equal magnitude and opposite direction in the external
Minkowski space-time (region~$\left(A\right)$). This is both for
the aesthetics of the model and to improve the expected applicability
of these results to a more realistic model in which the boundaries
$AB$ and $AD$ are replaced by a single compact boundary - such as
an ellipsoid. In such a case the Minkowski space-time of region~$\left(A\right)$
would actually be enclosing the 'warp-bubble'. Set $\left(\xi\eta\right)_{\left(AB\right)}=\left(\xi\eta\right)_{\left(AD\right)}=+1$,
$\kappa\rho_{\left(AB\right)}=\kappa\rho_{\left(AD\right)}$, $s_{\left(AB2\right)}=-s_{\left(AD1\right)}$
and $\kappa\rho_{\Lambda\left(B\right)}=\kappa\rho_{\Lambda\left(D\right)}$.
The only difference is the equation of states $\frac{p_{\left(AB\right)}}{\rho_{\left(AB\right)}}$
and $\frac{p_{\left(AD\right)}}{\rho_{\left(AD\right)}}$ - or equivalently
the pressures. Setting $\alpha_{\left(AD1\right)}=\alpha_{\left(AB2\right)}$
yields 
\[
\frac{N-1}{2N}+\frac{p_{\left(AB\right)}}{\rho_{\left(AB\right)}}-\frac{\kappa\rho_{\Lambda\left(B\right)}}{\left(\kappa\rho_{\left(AB\right)}\right)^{2}}=-\frac{N-1}{2N}-\frac{p_{\left(AD\right)}}{\rho_{\left(AB\right)}}+\frac{\kappa\rho_{\Lambda\left(B\right)}}{\left(\kappa\rho_{\left(AB\right)}\right)^{2}}
\]
 It is natural to constrain the equations of state to fall within
the range~$\left[-1,1\right]$, which requires 
\[
-2\le2\frac{\kappa\rho_{\Lambda\left(B\right)}}{\left(\kappa\rho_{\left(AB\right)}\right)^{2}}-\frac{N-1}{N}\le2
\]
specializing to $N=2$, $\left(\xi\eta\right)_{\left(AB\right)}=+1$
and $\kappa\rho_{\Lambda\left(B\right)}>0$ yields 
\[
\frac{p_{\left(AB\right)}}{\rho_{\left(AB\right)}}+\frac{p_{\left(AD\right)}}{\rho_{\left(AB\right)}}=2\frac{\kappa\rho_{\Lambda\left(B\right)}}{\left(\kappa\rho_{\left(AB\right)}\right)^{2}}-\frac{1}{2}
\]
\[
0<\frac{\kappa\rho_{\Lambda\left(B\right)}}{\left(\kappa\rho_{\left(AB\right)}\right)^{2}}\le\frac{5}{4}
\]

However, the full range of $\left[-1,1\right]$ is not open to both
equations of state once the accelerations are balanced:~$\alpha_{\left(AD1\right)}=\alpha_{\left(AB2\right)}$.
Instead, the the equations of state are confined to the line
\begin{equation}
\frac{p_{\left(AB\right)}}{\rho_{\left(AB\right)}}+\frac{p_{\left(AD\right)}}{\rho_{\left(AB\right)}}=2\frac{\kappa\rho_{\Lambda\left(B\right)}}{\left(\kappa\rho_{\left(AB\right)}\right)^{2}}-\frac{1}{2}\label{eq:Mdl_subLum_OBMatAccEos_line}
\end{equation}
 Subdividing the allowed range of $\frac{\kappa\rho_{\Lambda\left(B\right)}}{\left(\kappa\rho_{\left(AB\right)}\right)^{2}}$:
For $0<\frac{\kappa\rho_{\Lambda\left(B\right)}}{\left(\kappa\rho_{\left(AB\right)}\right)^{2}}\le\frac{1}{4}$
the equation of state on brane~AB is confined to the range 
\[
-1\le\frac{p_{\left(AB\right)}}{\rho_{\left(AB\right)}}\le2\frac{\kappa\rho_{\Lambda\left(B\right)}}{\left(\kappa\rho_{\left(AB\right)}\right)^{2}}+\frac{1}{2}
\]
 with $\frac{p_{\left(AD\right)}}{\rho_{\left(AB\right)}}$ given
by 
\begin{equation}
\frac{p_{\left(AD\right)}}{\rho_{\left(AB\right)}}=2\frac{\kappa\rho_{\Lambda\left(B\right)}}{\left(\kappa\rho_{\left(AB\right)}\right)^{2}}-\frac{1}{2}-\frac{p_{\left(AB\right)}}{\rho_{\left(AB\right)}}\label{eq:Mdl_subLum_OBMatAccEos_AD}
\end{equation}
 while for $\frac{1}{4}\le\frac{\kappa\rho_{\Lambda\left(B\right)}}{\left(\kappa\rho_{\left(AB\right)}\right)^{2}}<\frac{5}{4}$
the equation of state on brane~AB is confined to the range 
\[
2\frac{\kappa\rho_{\Lambda\left(B\right)}}{\left(\kappa\rho_{\left(AB\right)}\right)^{2}}-\frac{3}{2}\le\frac{p_{\left(AB\right)}}{\rho_{\left(AB\right)}}\le1
\]
 with $\frac{p_{\left(AD\right)}}{\rho_{\left(AB\right)}}$ again
given by eq~\ref{eq:Mdl_subLum_OBMatAccEos_AD}. For $\frac{p_{\left(AB\right)}}{\rho_{\left(AB\right)}}>\frac{p_{\left(AD\right)}}{\rho_{\left(AB\right)}}$
(with the constraint eq~\ref{eq:Mdl_subLum_OBMatAccEos_line} assumed
to hold) the entire model will accelerate in the $s_{\left(AB2\right)}x^{1}$
direction, while for $\frac{p_{\left(AB\right)}}{\rho_{\left(AB\right)}}<\frac{p_{\left(AD\right)}}{\rho_{\left(AB\right)}}$
it will accelerate in the opposite direction and for $\frac{p_{\left(AB\right)}}{\rho_{\left(AB\right)}}=\frac{p_{\left(AD\right)}}{\rho_{\left(AB\right)}}$
the model will not accelerate (this can be seen via reflection symmetry
in the perpendicular spatial coordinate). 

Once the equation of state $\frac{p_{\left(AB\right)}}{\rho_{\left(AB\right)}}$
is chosen, the entire model will have 
\[
\alpha_{\left(AB2\right)}=s_{\left(AB2\right)}\kappa\rho_{\left(AB\right)}\left(\frac{p_{\left(AB\right)}}{\rho_{\left(AB\right)}}+\frac{1}{4}-\frac{\kappa\rho_{\Lambda\left(B\right)}}{\left(\kappa\rho_{\left(AB\right)}\right)^{2}}\right)
\]
 and the equation of state on brane~AD will be determined by the
constraint eq~\ref{eq:Mdl_subLum_OBMatAccEos_line}: 
\[
\frac{p_{\left(AD\right)}}{\rho_{\left(AB\right)}}=2\frac{\kappa\rho_{\Lambda\left(B\right)}}{\left(\kappa\rho_{\left(AB\right)}\right)^{2}}-\frac{1}{2}-\frac{p_{\left(AB\right)}}{\rho_{\left(AB\right)}}
\]

Note that the $B_{B}$ of eq~\ref{eq:MnkLK_Bsoln} should be the
same value for branes $BC$ and $CD$ (because in a model with compact
branes $BC$ and $CD$ would be the same brane). Since $K_{\left(B\right)}=K_{\left(D\right)}$,
$\kappa\rho_{\Lambda\left(B\right)}=\kappa\rho_{\Lambda\left(D\right)}$,
$\kappa\rho_{\left(BC\right)}=\kappa\rho_{\left(CD\right)}$ and $\phi_{\left(C\right)}=\phi_{\left(BC1\right)}=\phi_{\left(CD2\right)}$
this is constraint is satisfied.  A potential avenue by which this
model could fail would be if the required value of $B_{B}$ from eq~\ref{eq:MnkLK_Bsoln}
was inconsistent with the requirement that $\left|\chi^{0}\right|<\left|\chi^{1}\right|$.
That is, if $B_{B}=B_{H}$, then for the entire world-path of the
brane $\left|\chi^{0}\right|=\left|\chi^{1}\right|$. If $B_{B}>B_{H}$,
then $\left|\chi^{1}\right|<\left|\chi^{0}\right|$. It is thus instructive
to compute the ratio
\[
\frac{B_{B}}{B_{H}}=\left(1+\frac{1}{\xi\eta}\frac{\left(N+1\right)}{2N}\frac{\left(\kappa\rho\right)^{2}}{\kappa\rho_{\Lambda}}\right)^{-\frac{2}{N+1}}
\]

Thus with the assumption $\kappa\rho_{\Lambda}>0$ and $\kappa\rho>0$,
one has $\xi\eta=+1\Rightarrow B_{B}<B_{H}$ and $\xi\eta=-1\Rightarrow B_{B}>B_{H}$.
This model passes this self-consistency test. Thus the conditions
of eq~\ref{eq:MnkLK_req_smry} are satisfied and this model is consistent
with all conditions discussed in subsection~\ref{subsec:MnkLK_constr}.

\subsection{Super-Luminal Warp\label{subsec:Mdl_supLum}}

\noindent\fbox{\begin{minipage}[t]{1\columnwidth - 2\fboxsep - 2\fboxrule}%
In this subsection:
\begin{lyxlist}{00.00.0000}
\item [{1)}] A super-luminal model is constructed where the outer branes
follow a space-like trajectory in the external Minkowski space-time,
while the inner branes are static in region~$\left(C\right)$. 
\item [{2)}] The portion of region~$\left(B\right)$ and region~$\left(D\right)$
utilized in this model is by necessity such that there is a horizon
between the inner and outer branes. Because the time-like trajectories
of the inner branes must co-exist and not intersect the space-like
trajectories of the outer branes, the appearance of a horizon in models
such as this may be unavoidable. 
\item [{3)}] The weak and null energy conditions are satisfied with all
energy density positive ($\text{sign}\left(K\right)=-1=-\text{sign}\left(\kappa\rho_{\Lambda}\right)$)
and the equations of state in the interval~$\left[-1,1\right]$.
However, the equation of state of the outer branes must be restricted
to~$\left[-1,0\right]$. The dominant energy condition is not satisfied
at the outer branes.
\item [{4)}] Model parameter values or ranges are determined as several
constraints are imposed: that the inner branes be stationary in region~$\left(C\right)$,
that region~$\left(B\right)$ and region~$\left(D\right)$ have
the same $\kappa\rho_{\Lambda}$, that the outer branes accelerate
in the same direction (one towards the Minkowski side, the other away
from it), with equal magnitude.
\item [{5)}] The Landau-Raychaudhuri equation is re-derived in the presence
of a thin-shell membrane. A modification is found that can provide
a positive contribution to the expansion scalar without NEC violation.
\item [{6)}] The discontinuity in extrinsic curvature at the brane may
de-focus a geodesic congruence. This allows for certain super-luminal
warp-drive no-go theorems to be evaded by models with thin-shell membranes. 
\item [{7)}] A proper distance between the inner and outer branes is defined
in two possible ways. With both definitions, that distance is decreasing
with late-time increasing proper time of an observer riding on the
branes. 
\item [{8)}] A mechanism of transition from the sub-luminal to the super-luminal
model is speculated on. Parameter values that are treated as constants
in these models would need to vary. The metric induced on the outer
branes due to region~$\left(A\right)$ would be held fixed, while
the value of $B$ at which the horizon occurs would decrease, becoming
less than that at the brane. Thus the outer branes would cross the
horizon in the $\Lambda-K$ space-time and their trajectory would
necessarily become space-like. 
\end{lyxlist}
\end{minipage}}

As one can see in fig~\ref{fig:supLum_ConfDiag}, the primary difference
in the structure of the super-luminal warp model versus that of sub-luminal
warp of fig~\ref{subsec:Mdl_subLum} is that the outer branes~AB~and~AD
follow space-like world-paths and are on the other side of the horizon
in $\Lambda-K$ space. For regions A,B,C,D and the inner branes~BC~and~CD
everything is the same as the sub-luminal warp model of subsection~\ref{subsec:Mdl_subLum}.
The super-luminal nature of this model is provided by the outer branes
following a space-like trajectory relative to an observer in the outer
Minkowski space-time, while the proper distance between the inner
and outer branes (along constant coordinate lines) is not increasing
faster than the speed of light, while the passenger in region~$\left(C\right)$
is at rest relative to the inner branes. Said another way, the passenger
in region~$\left(C\right)$ is moving on a trajectory that is locally
time-like, yet - relative to an external observer at infinity (in
the $x^{1}$ direction) - the passenger departs the forward light-cone
(what it would be without the warp-drive) in region~$\left(A\right)$
 originating from their initial location. The utility of the $\Lambda-K$
bulk between the inner and outer branes is that it allows space-like
and time-like trajectories to coexist and not intersect - that is,
the $\xi\eta$ of the outer branes need not be the same as that of
the inner branes, and thus of an observer riding as a passenger. 

This model differs from that of subsection~\ref{subsec:Mdl_subLum}
in that $\left(\xi\eta\right)_{\left(AB\right)}=\left(\xi\eta\right)_{\left(AD\right)}=-1$,
and hence $\mathcal{Y}_{\left(AB1\right)},\mathcal{Y}_{\left(AD2\right)}>0$
and $B_{B\left(AB\right)}=B_{B\left(AD\right)}>B_{H\left(B\right)}$.
The horizon of the $\Lambda-K$ space-time (region~$\left(B\right)$
and region~$\left(D\right)$) is located between the external branes
(AB, AD) and the internal branes (BC, CD). This could prove problematic
for control of the propulsion in reaction to external conditions,
however it might be possible for the trajectory of the external branes
to be made to conform to pre-planned, pre-set parameters. The existence
of such a horizon between the interior (a neighborhood in which the
passenger follows a locally time-like trajectory) and the exterior
(with respect to which that trajectory is 'globally space-like') of
a super-luminal warp model might be impossible to avoid due the necessity
to - in an intuitive sense - swap space-like and time-like directions
between the interior and exterior. While in the coordinate frame used
in this paper, the time coordinate is the same in every region ($A$
never goes to zero), the trajectories of $y=constant$ do necessarily
have their time-like versus space-like nature swapped as one crosses
the $y=0$ lines. 

\begin{figure}
\centering{}\includegraphics[scale=0.55]{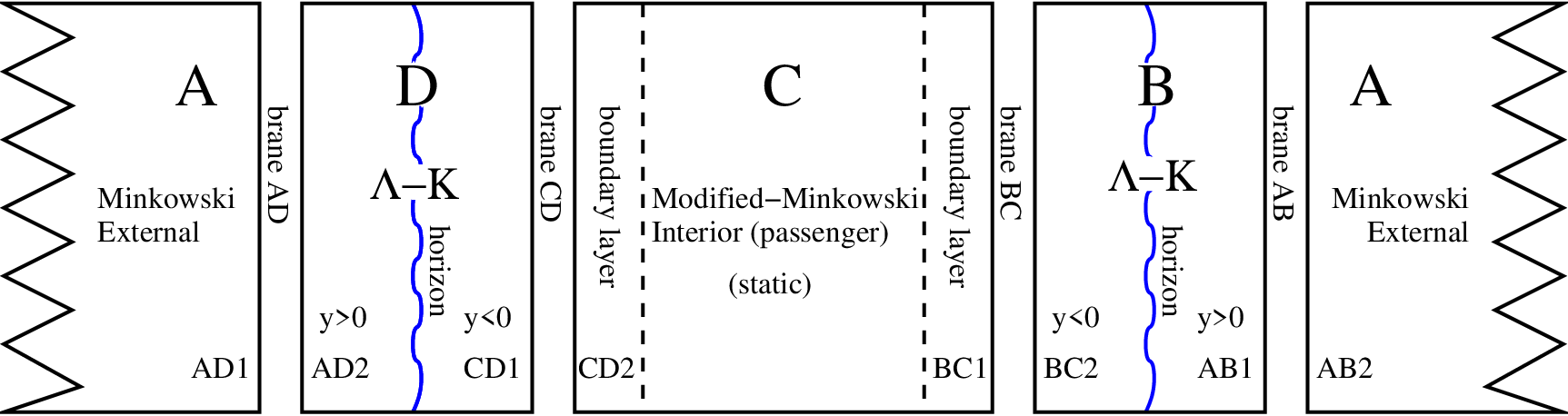}\label{fig:supLum_CrossSect}\caption{ This figure shows the cross-section of a possible super-luminal
propulsion model, constructed according to the constraints explored
in sub-sections~\ref{subsec:MnkLK_constr} and \ref{subsec:modMnkLK_constr}.
Minkowski region~$\left(A\right)$ (or bulk~$\left(A\right)$) is
the outside Universe. A passenger rides in Modified-Minkowski region~$\left(C\right)$.
Region~$\left(B\right)$ and region~$\left(D\right)$ are $\Lambda-K$
space-times. These regions contain horizons separating the space-like
and time-like brains. Four branes at the interface between these regions
are labeled $AB$, $BC$, $CD$, $AD$. The opposite sides of those
branes are labeled $AB1$~\&~$AB2$; $BC1$~\&~$BC2$; $CD1$~\&~$CD2$;
$AD1$~\&~$AD2$. }
\end{figure}

\begin{figure}
\centering{}\includegraphics[viewport=201bp 4bp 792bp 608bp,clip,scale=0.75]{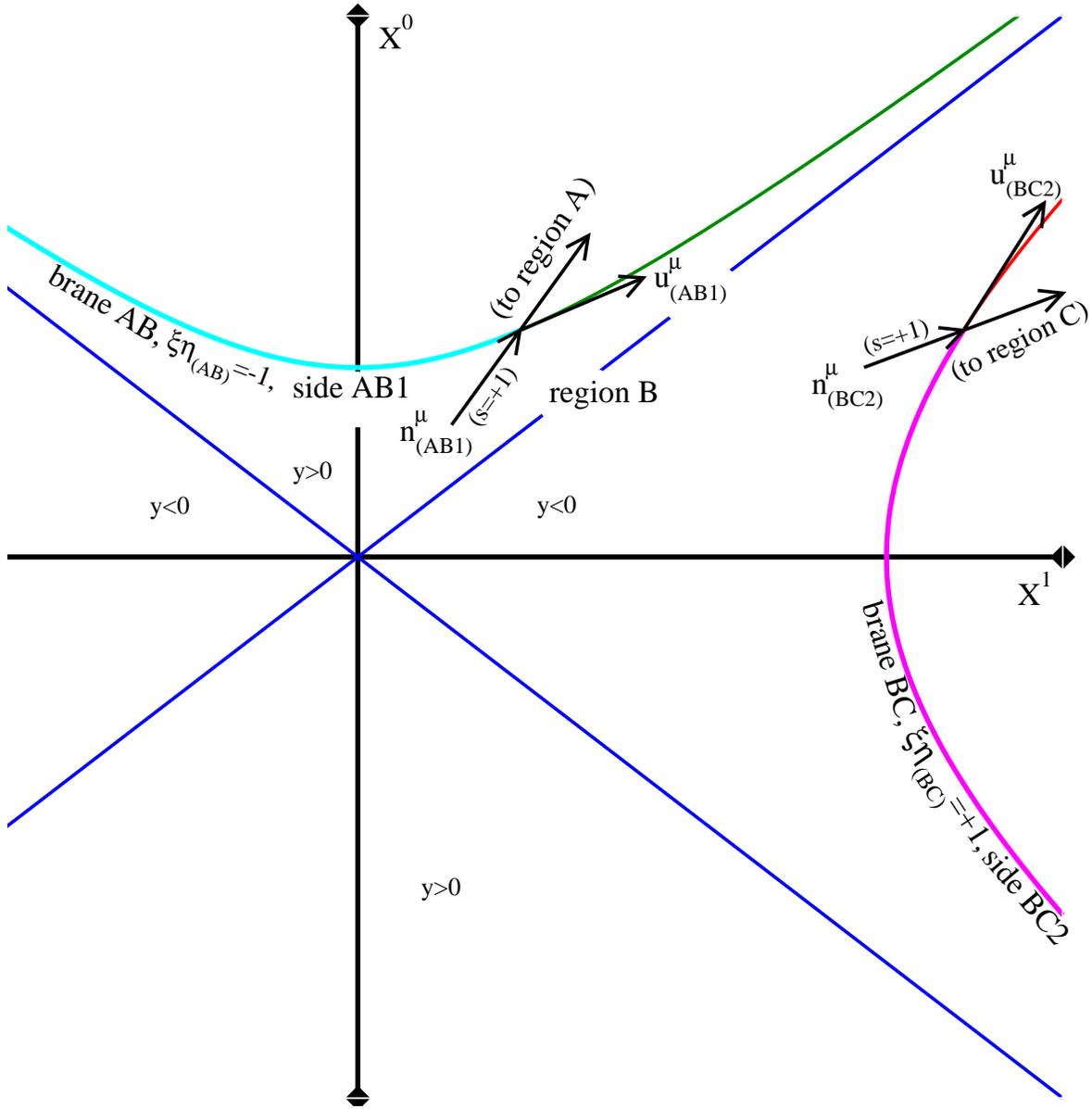}\label{fig:supLum_ConfDiag}\caption{This figure shows the conformal structure in the $\Lambda-K$ space-time
of region~$\left(B\right)$ of the super-luminal propulsion model
shown in fig~\ref{fig:supLum_CrossSect}. Region~$\left(D\right)$
has an identical structure. The green and red curves are the $y=const$
world-paths of the branes, and the cyan and magenta segments respectively
of those curves are the history of those branes at some coordinate
time~$x^{0}$. The bulk between the green and red curves - including
the horizon (blue $45^{\circ}$ lines) at $B=B_{H}$ - is utilized
- it exists in this model. The remainder of the $\Lambda-K$ space-time
- including the curvature singularity at $B=0$ (to the right of the
magenta/red hyperbola) - is discarded, and does not exist in this
model. Note that the world-path of brane~AB is space-like, with $\xi\eta_{\left(AB\right)}=-1$,
while the world-path of brane~BC is time-like, with $\xi\eta_{\left(BC\right)}=+1$. }
\end{figure}

\subsubsection{Energy Conditions in the Super-Luminal Model: Dominant, Weak, Null
and Strong}

As with the sub-luminal warp model, it is worth considering what
energy conditions (dominant, weak, null and strong) the super-luminal
warp model satisfies. While the inner branes (branes~BC~and~CD)
still follow time-like world-paths as in the sub-luminal warp model
(and hence the energy-condition results of the sub-luminal model still
apply to these), now the outer branes (branes~AB~and~AD) follow
a space-like trajectory. For a space-like world-path brane ($u^{\mu}$
a space-like vector, $u^{\mu}u_{\mu}=-\eta=-\text{sign}\left(A\right)$,
ie: $\xi\eta=-1$) eq~\ref{eq:brane_SEten_PerfFluid} indicates that
the stress-energy on the brane is a modification of perfect fluid,
which might have quite different conditions under which the various
energy conditions hold or are violated.  Since the results derived
for the sub-luminal model still hold for the time-like inner branes
(branes~BC~and~CD), these will not be considered further. Thus
only the energy conditions regarding the space-like outer branes (branes~AB~and~AD)
will discussed in what follows. 

Let bulk vector $X^{\mu}$ positioned on the brane world-path have
a form given by eq~\ref{eq:gen_RelVec_decomp}. Consider $S_{\mu\nu}$
when $\xi\eta=-1$ and the conditions under which $S_{\mu\nu}X^{\mu}X^{\nu}\ge0$
\[
\begin{array}{cccc}
S_{\mu\nu} & = & \left(\rho+p\right)u_{\mu}u_{\nu}-\xi\eta p\left(g_{\mu\nu}+\frac{1}{\xi\eta}n_{\mu}n_{\nu}\right)\\
 & = & \left(\rho+p\right)u_{\mu}u_{\nu}+p\left(g_{\mu\nu}+n_{\mu}n_{\nu}\right)\\
S_{\mu\nu}X^{\mu}X^{\nu} & = & \rho\alpha_{u}^{2}-pB\alpha_{p}^{2}v^{2}\\
v^{2} & \equiv & \delta_{\underline{\mu\nu}}v^{\underline{\mu}}v^{\underline{\nu}}\\
X^{\mu}X_{\mu} & = & g_{\mu\nu}X^{\mu}X^{\nu}\\
 & = & \alpha_{n}^{2}-\alpha_{u}^{2}-B\alpha_{p}^{2}v^{2}
\end{array}
\]
Regardless of the value of $\alpha_{u}$ and $\alpha_{p}$, the value
of $\alpha_{n}$ may always be selected such that $X^{\mu}X_{\mu}>0$
(for the weak energy condition), or $X^{\mu}X_{\mu}=0$ (for the null
energy condition). If $\alpha_{p}=0$ then $S_{\mu\nu}X^{\mu}X^{\nu}=\rho\alpha_{u}^{2}\ge0\Leftrightarrow\rho\ge0$.
If $\alpha_{u}=0$ then $S_{\mu\nu}X^{\mu}X^{\nu}=-pB\alpha_{p}^{2}v^{2}\ge0\Leftrightarrow p\le0$.
If $p\le0$ then $S_{\mu\nu}X^{\mu}X^{\nu}\ge0$ is guaranteed. 
Thus the weak and null energy conditions are satisfied for $\xi\eta=-1$
if and only if: $\rho\ge0$ and $p\le0$. Due to this, the condition
$p\le0$ will be imposed for the outer branes (branes~AB~and~AD). 

The dominant energy condition requires the weak energy condition be
satisfied, and additionally that, given a future-pointing time/light-like
$X^{\mu}$ that $Y^{\mu}=S_{\nu}^{\mu}X^{\nu}$ is a future-pointing
time/light-like vector. Using the vector decomposition of eq~\ref{eq:gen_RelVec_decomp},
one finds: 
\[
Y^{\mu}=-\rho\alpha_{u}u^{\mu}+p\alpha_{p}v^{\underline{\mu}}
\]
 One can set $\alpha_{p}=0$ and specify a time-like $X^{\mu}$ by
$\left|\alpha_{n}\right|>\left|\alpha_{u}\right|$, or a light-like
$X^{\mu}$ by $\left|\alpha_{n}\right|=\left|\alpha_{u}\right|$.
In either case one obtains $Y^{\mu}=-\rho\alpha_{u}u^{\mu}$ which
is clearly space-like because $u^{\mu}$ is space-like (ie: $Y^{\mu}Y_{\mu}=-\rho^{2}\alpha_{u}^{2}<0$).
Thus the dominant energy condition is guaranteed to not be satisfied
for branes~AB~and~AD. It will still be satisfied for branes~BC~and~CD,
however.

One issue with this model is that the induced metric on the external
branes~AB and AD is Euclidean. Never the less, the stress-energy
tensor on those branes does satisfy the Weak Energy Condition. Note
also that there has been no attempt to enforce $T_{\,;\mu}^{\mu\nu}=0$
of these branes in the normal direction, as this would involve derivatives
of the delta-function support. In physical terms, attempting to do
so would involve specifying the mechanism that confines the brane
stress-energy to the boundary, and this is beyond the scope of this
paper. Furthermore, $T_{\,;\mu}^{\mu\nu}=0$ in the normal direction
it is an issue that every model involving a brane has - it is not
unique to this model of super-luminal propulsion.

\subsubsection{Determining the Model Parameters }

Brane~AB requires $s_{\left(AB1\right)}=s_{\left(AD2\right)}=+1$
for region~$\left(B\right)$ to be the correct side of the world-path
of brane~AB, and for region~$\left(D\right)$ to be the correct
side of the world-path of brane~AD in the $\Lambda-K$ bulk (similarly
$s_{\left(AD2\right)}=-1$ and thus $s_{\left(AB1\right)}y>0$ and
$s_{\left(AD2\right)}y>0$). ``Correct'' in this sense means that
fig~\ref{fig:supLum_CrossSect} and fig~\ref{fig:supLum_ConfDiag}
are accurate. The space-like trajectory of brane~AB implies that~$\xi_{\left(AB1\right)}=\xi_{\left(AB2\right)}=-1$,
and because $A>0$ everywhere, one also has $\left(\xi\eta\right)_{AB}=\left(\xi\eta\right)_{AD}=-1$.
Considering eq~\ref{eq:MnkLK_gen_soln_D}, this is consistent with
$\kappa\rho_{AB}>0$. 

In general, the outer branes have acceleration in the external Minkowski
space determined by eq~\ref{eq:MnkLK_mnkYaccel} with $\xi\eta=\left(\xi\eta\right)_{\left(AB\right)}=-1$
\begin{equation}
\begin{array}{ccc}
\alpha_{\left(AB2\right)} & = & s_{\left(AB2\right)}\kappa\rho_{\left(AB\right)}\left(\frac{p_{\left(AB\right)}}{\rho_{\left(AB\right)}}+\frac{N-1}{2N}+\frac{\kappa\rho_{\Lambda\left(B\right)}}{\left(\kappa\rho_{\left(AB\right)}\right)^{2}}\right)\\
\alpha_{\left(AD1\right)} & = & s_{\left(AD1\right)}\kappa\rho_{\left(AD\right)}\left(\frac{p_{\left(AD\right)}}{\rho_{\left(AD\right)}}+\frac{N-1}{2N}+\frac{\kappa\rho_{\Lambda\left(D\right)}}{\left(\kappa\rho_{\left(AD\right)}\right)^{2}}\right)
\end{array}\label{eq:Mdl_supLum_OBRawAccel}
\end{equation}
 However, it would be preferable for branes $AB$ and $AD$ to have
acceleration of equal magnitude and opposite direction in the external
Minkowski space-time (region~$\left(A\right)$). This is both for
the aesthetics of the model and to improve the expected applicability
of these results to a more realistic model in which the branes $AB$
and $AD$ are replaced by a single compact boundary - such as something
like an ellipsoid. In such a case the Minkowski space-time of region~$\left(A\right)$
would actually be enclosing the 'warp-bubble'. Set $\left(\xi\eta\right)_{\left(AB\right)}=\left(\xi\eta\right)_{\left(AD\right)}=-1$,
$\kappa\rho_{\left(AB\right)}=\kappa\rho_{\left(AD\right)}$, $s_{\left(AB2\right)}=-s_{\left(AD1\right)}$
and $\kappa\rho_{\Lambda\left(B\right)}=\kappa\rho_{\Lambda\left(D\right)}$.
The only difference is the equation of states $\frac{p_{\left(AB\right)}}{\rho_{\left(AB\right)}}$
and $\frac{p_{\left(AD\right)}}{\rho_{\left(AD\right)}}$ - or equivalently
the pressures. Setting $\alpha_{\left(AD1\right)}=\alpha_{\left(AB2\right)}$
yields 
\[
\frac{N-1}{2N}+\frac{p_{\left(AB\right)}}{\rho_{\left(AB\right)}}+\frac{\kappa\rho_{\Lambda\left(B\right)}}{\left(\kappa\rho_{\left(AB\right)}\right)^{2}}=-\frac{N-1}{2N}-\frac{p_{\left(AD\right)}}{\rho_{\left(AB\right)}}-\frac{\kappa\rho_{\Lambda\left(B\right)}}{\left(\kappa\rho_{\left(AB\right)}\right)^{2}}
\]
 It is natural to constrain the equations of state to fall within
the range~$\left[-1,0\right]$, which requires 
\[
-1+\frac{N-1}{2N}\le-\frac{\kappa\rho_{\Lambda\left(B\right)}}{\left(\kappa\rho_{\left(AB\right)}\right)^{2}}\le\frac{N-1}{2N}
\]
Specializing to $N=2$ and $\kappa\rho_{\Lambda\left(B\right)}>0$
yields 
\[
0<\frac{\kappa\rho_{\Lambda\left(B\right)}}{\left(\kappa\rho_{\left(AB\right)}\right)^{2}}\le\frac{3}{4}
\]
 However, the full range of $\left[-1,0\right]$ is not open to both
equations of state once the accelerations are balanced:~$\alpha_{\left(AD1\right)}=\alpha_{\left(AB2\right)}$.
Instead, the the equations of state are confined to the line
\begin{equation}
\frac{p_{\left(AB\right)}}{\rho_{\left(AB\right)}}+\frac{p_{\left(AD\right)}}{\rho_{\left(AB\right)}}=-2\frac{\kappa\rho_{\Lambda\left(B\right)}}{\left(\kappa\rho_{\left(AB\right)}\right)^{2}}-\frac{1}{2}\label{eq:Mdl_supLum_OBMatAccEos_line}
\end{equation}

Subdividing the allowed range of $\frac{\kappa\rho_{\Lambda\left(B\right)}}{\left(\kappa\rho_{\left(AB\right)}\right)^{2}}$:
For $0<\frac{\kappa\rho_{\Lambda\left(B\right)}}{\left(\kappa\rho_{\left(AB\right)}\right)^{2}}\le\frac{1}{4}$
the equation of state on brane~AB is confined to the range 
\[
-1\le-2\frac{\kappa\rho_{\Lambda\left(B\right)}}{\left(\kappa\rho_{\left(AB\right)}\right)^{2}}-\frac{1}{2}\le\frac{p_{\left(AB\right)}}{\rho_{\left(AB\right)}}\le0
\]
 with $\frac{p_{\left(AD\right)}}{\rho_{\left(AB\right)}}$ given
by 
\begin{equation}
\frac{p_{\left(AD\right)}}{\rho_{\left(AB\right)}}=-2\frac{\kappa\rho_{\Lambda\left(B\right)}}{\left(\kappa\rho_{\left(AB\right)}\right)^{2}}-\frac{1}{2}-\frac{p_{\left(AB\right)}}{\rho_{\left(AB\right)}}\label{eq:Mdl_supLum_OBMatAccEos_AD}
\end{equation}
 while for $\frac{1}{4}<\frac{\kappa\rho_{\Lambda\left(B\right)}}{\left(\kappa\rho_{\left(AB\right)}\right)^{2}}\le\frac{3}{4}$
the equation of state on brane~AB is confined to the range 
\[
-1\le\frac{p_{\left(AB\right)}}{\rho_{\left(AB\right)}}\le-2\frac{\kappa\rho_{\Lambda\left(B\right)}}{\left(\kappa\rho_{\left(AB\right)}\right)^{2}}+\frac{1}{2}\le0
\]
 with $\frac{p_{\left(AD\right)}}{\rho_{\left(AB\right)}}$ again
given by eq~\ref{eq:Mdl_supLum_OBMatAccEos_AD}. For $\frac{p_{\left(AB\right)}}{\rho_{\left(AB\right)}}>\frac{p_{\left(AD\right)}}{\rho_{\left(AB\right)}}$
the entire model will accelerate in the $s_{\left(AB2\right)}x^{1}$
direction, while for $\frac{p_{\left(AB\right)}}{\rho_{\left(AB\right)}}<\frac{p_{\left(AD\right)}}{\rho_{\left(AB\right)}}$
it will accelerate in the opposite direction and for $\frac{p_{\left(AB\right)}}{\rho_{\left(AB\right)}}=\frac{p_{\left(AD\right)}}{\rho_{\left(AB\right)}}$
the model will not accelerate. 

Once the equation of state $\frac{p_{\left(AB\right)}}{\rho_{\left(AB\right)}}$
is chosen, the entire model will have 
\[
\alpha_{\left(AB2\right)}=s_{\left(AB2\right)}\kappa\rho_{\left(AB\right)}\left(\frac{p_{\left(AB\right)}}{\rho_{\left(AB\right)}}+\frac{1}{4}+\frac{\kappa\rho_{\Lambda\left(B\right)}}{\left(\kappa\rho_{\left(AB\right)}\right)^{2}}\right)
\]
 and the equation of state on brane~AD will be determined by the
condition $\alpha_{\left(AD1\right)}=\alpha_{\left(AB2\right)}$:
\[
\frac{p_{\left(AD\right)}}{\rho_{\left(AB\right)}}=-2\frac{\kappa\rho_{\Lambda\left(B\right)}}{\left(\kappa\rho_{\left(AB\right)}\right)^{2}}-\frac{1}{2}-\frac{p_{\left(AB\right)}}{\rho_{\left(AB\right)}}
\]

Note that the $B_{B}$ of eq~\ref{eq:MnkLK_Bsoln} should be the
same value for branes $BC$ and $CD$. Since $K_{\left(B\right)}=K_{\left(D\right)}$,
$\kappa\rho_{\Lambda\left(B\right)}=\kappa\rho_{\Lambda\left(D\right)}$,
$\kappa\rho_{\left(BC\right)}=\kappa\rho_{\left(CD\right)}$ and $\phi_{\left(C\right)}=\phi_{\left(BC1\right)}=\phi_{\left(CD2\right)}$
this is constraint is satisfied.  A potential avenue by which this
model could fail would be if the required value of $B_{B}$ from eq~\ref{eq:MnkLK_Bsoln}
was inconsistent with the requirement that $\left|\chi^{0}\right|>\left|\chi^{1}\right|$
for brane~AB~or~BD. That is, if $B_{B}=B_{H}$ then for the entire
world-path of the brane $\left|\chi^{0}\right|=\left|\chi^{1}\right|$.
If $B_{B}<B_{H}$ then $\left|\chi^{1}\right|>\left|\chi^{0}\right|$.
It is then instructive to compute the ratio
\begin{equation}
\frac{B_{B}}{B_{H}}=\left(1+\frac{1}{\xi\eta}\frac{\left(N+1\right)}{2N}\frac{\left(\kappa\rho\right)^{2}}{\kappa\rho_{\Lambda}}\right)^{-\frac{2}{N+1}}\label{eq:Mdl_supLum_BB_div_BH}
\end{equation}
 Thus with the assumption $\kappa\rho_{\Lambda}>0$ and $\kappa\rho>0$,
one has $\xi\eta=+1\Rightarrow B_{B}<B_{H}$ and $\xi\eta=-1\Rightarrow B_{B}>B_{H}$.
This model passes this self-consistency test. Thus the conditions
of eq~\ref{eq:MnkLK_req_smry} are satisfied and this model is consistent
with all conditions discussed in subsection~\ref{subsec:MnkLK_constr}.

\subsubsection{Modifications to the Landau-Raychaudhuri Equation from the Singularity
at the Junctions: Evading Super-Luminal Warp-Drive No-Go Theorems\label{subsec:Mdl_supLum_LanRay_mod}}

Arguments have been put forth that super-luminal warp-drives require
the violation of various energy conditions. In \cite{bib:Olum_1998}
it is argued that super-luminal warp-drive requires the violation
of the null energy condition. The essence of the argument is that,
given a super-luminal warp-drive, there exists a null geodesic congruence~$k^{\mu}$
such that the warp-drive will de-focus (increase the expansion scalar~$\theta\equiv k_{\,;\alpha}^{\alpha}$)
of the congruence. Doing this necessarily requires a violation of
the null energy condition - so long as the region of space-time being
considered is free of singularities. There are two reasons why such
an argument does not apply to the models presented in this manuscript. 

First, the approach in \cite{bib:Olum_1998} requires that the region
of space-time that constitutes super-luminal warp-drive be compact
- isolated within flat space. Even though a model might allow for
indefinite operation of a warp-drive, there is a presumption that
the drive is turned on and off at finite times. To simply analysis,
the models presented in this manuscript were constructed to possess
translational and rotational symmetry in all but one of the spatial
directions. Because the membranes extend to infinity in all parallel
directions, the compactness assumption of \cite{bib:Olum_1998} does
not hold, and the arguments in that manuscript do not apply to these
models. However, that is a particularly unsatisfying means to evade
that no-go theorem, since the goal is to eventually use the chimeric
space-time approach to construct models that involve only compact
membranes. One should look for a means of evasion that would also
work with a spatial compact warp-drive model.

Second, the approach in \cite{bib:Olum_1998} assume that the region
of space-time under consideration is singularity-free. However, the
models presented in this manuscript do possess physical singularities.
A delta-function does appear in the full, global Riemann tensor of
the space-time (as opposed to a Riemann tensor that merely applies
to the regions between the membranes). The requirement to cancel this
singularity necessitates the singular stress-energy distribution present
in the membranes (as thin-shells), and the Israel junction conditions
are the requirement that this cancellation occur. This reason that
the arguments of \cite{bib:Olum_1998} do not apply to the model in
this section could potentially also apply to a super-luminal warp-drive
model that involved only compact membranes. 

A key component in the approach of \cite{bib:Olum_1998} is the Landau-Raychaudhuri
equation for a congruence of null geodesics~$k^{\alpha}$ (below
taken from \cite{bib:Poisson_Book_Relativists_Toolkit}, eq~2.39):
\begin{equation}
\frac{d\theta}{d\tau}=-\frac{1}{2}\theta^{2}-\sigma^{\alpha\beta}\sigma_{\alpha\beta}+\omega^{\alpha\beta}\omega_{\alpha\beta}-R_{\alpha\beta}k^{\alpha}k^{\beta}\label{eq:Mdl_suplum_RCHE_nonsing}
\end{equation}
where $\theta\equiv k_{\,;\alpha}^{\alpha}$ is the expansion scalar,
$\sigma_{\alpha\beta}$ is the shear tensor, $\omega_{\alpha\beta}$
is the rotation tensor, and $R_{\alpha\beta}$ is the Ricci tensor.
Because $k^{\alpha}$ is light-like, $R_{\alpha\beta}k^{\alpha}k^{\beta}$
is equal to the stress-energy tensor contracted with $k^{\alpha}k^{\beta}$,
and the null energy condition implies $R_{\alpha\beta}k^{\alpha}k^{\beta}\ge0$.
However, eq~\ref{eq:Mdl_suplum_RCHE_nonsing} is modified in the
presence of a thin-shell (membrane) singularity. That is, as the null
geodesic congruence $k^{\alpha}$ crosses the thin-shell singularity,
there is an abrupt change in $\theta$ - which manifests as a new
contribution (with delta-function support) to eq~\ref{eq:Mdl_suplum_RCHE_nonsing}.
To calculate this contribution, one may repeat the derivation of the
Landau-Raychaudhuri equation (subsection~2.3.4 of \cite{bib:Poisson_Book_Relativists_Toolkit})
(for a null geodesic congruence), employing the process used in \cite{bib:Poisson_Book_Relativists_Toolkit}
(section~3.7) to derive the Israel junction conditions. Note that
$k^{\alpha}$, $n^{\alpha}$ and $g_{\mu\nu}$ are continuous at the
membrane, have a discontinuity in the first derivative, which yields
a delta-function contribution to the second derivative. 
\[
\begin{array}{ccc}
k_{\alpha;\beta;\mu}k^{\mu} & = & k_{\alpha;\mu;\beta}k^{\mu}-k^{\rho}k^{\mu}R_{\left(NS\right)\alpha\rho\beta\mu}+\delta\left(\ell\right)k^{\mu}\left(k_{\rho}A_{\,\alpha\beta\mu}^{\rho}+B_{\,\alpha\beta\mu}\right)\\
 & = & -k_{\alpha;\mu}k_{\,;\beta}^{\mu}-k^{\rho}k^{\mu}R_{\left(NS\right)\alpha\rho\beta\mu}+\delta\left(\ell\right)k^{\mu}\left(-k^{\rho}A_{\alpha\rho\beta\mu}+B_{\,\alpha\beta\mu}\right)\\
\\
A_{\,\alpha\beta\mu}^{\rho} & \equiv & \left(n^{\sigma}n_{\sigma}\right)\left(\left[\Gamma_{\,\alpha\mu}^{\rho}\right]n_{\beta}-\left[\Gamma_{\,\alpha\beta}^{\rho}\right]n_{\mu}\right)\\
A_{\rho\alpha\beta\mu} & = & -A_{\alpha\rho\beta\mu}=-A_{\rho\alpha\mu\beta}=A_{\beta\mu\rho\alpha}\\
A_{\,\alpha\mu} & = & A_{\,\alpha\beta\mu}^{\beta}\ \ \ =\ \ \ \kappa S_{\alpha\mu}-\frac{1}{N}g_{\alpha\beta}\kappa S\\
B_{\,\alpha\beta\mu} & \equiv & \left(n^{\sigma}n_{\sigma}\right)\left(\left[k_{\alpha,\beta}\right]n_{\mu}-\left[k_{\alpha,\mu}\right]n_{\beta}\right)
\end{array}
\]
where $R_{\left(NS\right)\alpha\rho\beta\mu}$ is the non-singular
component of the Riemann tensor, and $\ell$ is the proper distance
from the membrane boundary in the direction of $n^{\mu}$. Note that
the source of $A_{\,\alpha\beta\mu}^{\rho}$ is the second derivative
of $g_{\mu\nu}$, and the source of $B_{\,\alpha\beta\mu}$ is the
non-commutation of the second derivatives of $k^{\mu}$. Contracting
with $g^{\alpha\beta}$ yields a modified Landau-Raychaudhuri equation
\[
\begin{array}{ccc}
\frac{d\theta}{d\tau} & = & g^{\alpha\beta}k_{\alpha;\beta;\mu}k^{\mu}\\
 & = & -k_{\,;\mu}^{\alpha}k_{\,;\alpha}^{\mu}-k^{\rho}k^{\mu}R_{\left(NS\right)\rho\mu}+\delta\left(\ell\right)k^{\mu}\left(-k^{\rho}g^{\alpha\beta}A_{\alpha\rho\beta\mu}+g^{\alpha\beta}B_{\,\alpha\beta\mu}\right)\\
 & = & \frac{1}{2}\theta^{2}-\sigma^{\alpha\beta}\sigma_{\alpha\beta}+\omega^{\alpha\beta}\omega_{\alpha\beta}-k^{\rho}k^{\mu}R_{\left(NS\right)\rho\mu}+\delta\left(\ell\right)k^{\mu}\left(-k^{\rho}A_{\rho\mu}+g^{\alpha\beta}B_{\,\alpha\beta\mu}\right)
\end{array}
\]
 where $R_{\left(NS\right)\rho\mu}$ is the non-singular component
of the Ricci tensor. Note that if the first derivative of $k^{\mu}$
is continuous (ie: $\left[k_{\alpha,\beta}\right]=0$) then $B_{\,\alpha\beta\mu}=0$
and the only new contribution to eq~\ref{eq:Mdl_suplum_RCHE_nonsing}
would be $-\delta\left(\ell\right)k^{\mu}k^{\rho}\kappa S_{\alpha\mu}$.
If the null energy condition is also satisfied by the membrane stress-energy
tensor, this would also only provide a non-positive contribution to
$\theta$ in the super-luminal case. Thus if the derivative of the
metric is discontinuous, but $\left[k_{\alpha,\beta}\right]=0$ then
the argument based on the Landau-Raychaudhuri equation that super-luminal
warp-drive requires null energy condition violation (or $k^{\mu}k^{\rho}\kappa T_{\alpha\mu}=0$
everywhere) would not be evaded. 

An obvious approach to computing $k^{\mu}g^{\alpha\beta}B_{\,\alpha\beta\mu}$
is to write $k_{\alpha,\beta}$ as $k_{\alpha;\beta}+k_{\rho}\Gamma_{\,\alpha\beta}^{\rho}$,
apply $g^{\alpha\beta}$, then use the geodesic equation for $k^{\mu}$.
One then obtains $k^{\mu}g^{\alpha\beta}B_{\,\alpha\beta\mu}=\left(n^{\sigma}n_{\sigma}\right)\left(k^{\mu}n_{\mu}\right)\left[k_{\,;\alpha}^{\alpha}\right]-k^{\mu}k_{\rho}g^{\alpha\beta}A_{\,\alpha\beta\mu}^{\rho}$,
yielding
\[
\begin{array}{ccc}
\frac{d\theta}{d\tau} & = & \frac{1}{2}\theta^{2}-\sigma^{\alpha\beta}\sigma_{\alpha\beta}+\omega^{\alpha\beta}\omega_{\alpha\beta}-k^{\rho}k^{\mu}R_{\left(NS\right)\rho\mu}+\delta\left(\ell\right)\left(n^{\sigma}n_{\sigma}\right)\left(k^{\mu}n_{\mu}\right)\left[k_{\,;\alpha}^{\alpha}\right]\\
 & = & \frac{1}{2}\theta^{2}-\sigma^{\alpha\beta}\sigma_{\alpha\beta}+\omega^{\alpha\beta}\omega_{\alpha\beta}-k^{\rho}k^{\mu}R_{\left(NS\right)\rho\mu}+\left(\frac{d}{d\tau}\varTheta\left(\ell\right)\right)\left[\theta\right]
\end{array}
\]
where $\varTheta\left(\ell\right)$ is the Heaviside step function.
While this result does not progress the computation, it does serve
as a sanity-check.

While a more intuitive expression for $k^{\mu}g^{\alpha\beta}B_{\,\alpha\beta\mu}$
is elusive in the general case, one may specialize to a particular
scenario as a demonstration of the effect of this additional contribution
to the Landau-Raychaudhuri equation. Consider a null geodesic congruence
$k^{\mu}$ passing from region~$\left(A\right)$ (Minkowski space)
to region~$\left(B\right)$ ($\Lambda-K$ space) in figs~\ref{fig:supLum_CrossSect},\ref{fig:supLum_ConfDiag},
such that $k^{\mu}$ has no component in any spatial parallel direction
(ie:~$k^{\underline{\mu}}=0$, which simplifies the computation).
This yields 
\[
\begin{array}{ccc}
k^{\mu}\left(-k^{\rho}A_{\rho\mu}+g^{\alpha\beta}B_{\,\alpha\beta\mu}\right) & = & \frac{N}{2}\left(n^{\sigma}n_{\sigma}\right)\left(k^{\mu}n_{\mu}\right)\left[\left(\ln B\right)_{,\alpha}n^{\alpha}\right]\\
 & = & \left(n^{\sigma}n_{\sigma}\right)^{2}\left(k^{\mu}n_{\mu}\right)^{2}\kappa\rho
\end{array}
\]
Recalling that $n^{\sigma}n_{\sigma}=-\xi\eta=+1$ in the super-luminal
case, the modified Landau-Raychaudhuri equation at the boundary becomes
\[
\begin{array}{ccccc}
\frac{d\theta}{d\tau} & = & \frac{1}{2}\theta^{2}-\sigma^{\alpha\beta}\sigma_{\alpha\beta}+\omega^{\alpha\beta}\omega_{\alpha\beta}-k^{\rho}k^{\mu}R_{\left(NS\right)\rho\mu}+\delta\left(\ell\right)\left(-\xi\eta\right)^{2}\left(k^{\mu}n_{\mu}\right)^{2}\kappa\rho\\
 & = & \frac{1}{2}\theta^{2}-\sigma^{\alpha\beta}\sigma_{\alpha\beta}+\omega^{\alpha\beta}\omega_{\alpha\beta}-k^{\rho}k^{\mu}R_{\left(NS\right)\rho\mu}+\left(\frac{d}{d\tau}\varTheta\left(\ell\right)\right)\left(-\xi\eta\right)\left.\left(k^{\mu}n_{\mu}\right)\right|_{\Sigma}\kappa\rho
\end{array}
\]
Since $n^{\mu}$ is time-like, and $k^{\mu}$ is passing from external
Minkowski space into the $\Lambda-K$ space, $k^{\mu}n_{\mu}>0$.
For $\kappa\rho>0$ on the boundary membrane, the additional contribution
to the Landau-Raychaudhuri equation in this scenario will be positive.
Thus $\theta$ increases as it crosses the boundary between the regions
- while the weak and null energy conditions are nowhere violated. 

Note that this result provides another perspective on why modified-Minkowski
space-time must be utilized in region~$\left(C\right)$. Had the
boundary layer not been used, then due to the opposite sign of $\left(\ln B\right)_{,\alpha}n^{\alpha}$,
in order for necessary shift in $\theta$ to happen at the boundary,
$\kappa\rho<0$ would have been required on branes~BC and CD. In
summary, the presence of the curvature singularities associated with
the thin-shell stress-energy on the boundary membranes modifies the
Landau-Raychaudhuri equation. The new terms allow for a positive contribution
to $\frac{d\theta}{d\tau}$ at these junctions, allowing for $\theta$
to both increase as well as decrease - without violating the null-energy
condition. The discontinuity in extrinsic curvature at the boundary
may de-focus a geodesic congruence. Future super-luminal warp-drive
models that involve such chimeric space-time construction (with their
associated curvature singularities at the boundaries) - even those
with compact membranes and finite extent - may potentially evade no-go
theorems that depend on this equation.

\subsubsection{Distance Between the Inner and Outer Branes }

A reasonable concern is that the proper distance between the outer
and inner branes might be increasing sufficiently rapidly so as to
ruin the utility of the propulsion model. One way to define such a
distance would be the proper distance of a line segment of constant
coordinate time $x^{0}$, and constant parallel directions coordinate
$x^{\underline{\alpha}}$, varying only in the $x^{1}$ direction,
from brane~BC at coordinate $x^{1}=x_{a}^{1}$ to brane~AB at coordinate
$x^{1}=x_{b}^{1}$. One would then consider how that distance changes
as the coordinate time of the model advances. Fix $x^{0}=const$,
$x^{\underline{\alpha}}=const$ and integrate proper distance $ds$
in the $x^{1}$ direction from $x_{a}^{1}$ to $x_{b}^{1}$ 
\[
\Delta s=\stackrel[x_{a}^{1}]{x_{b}^{1}}{\int}\sqrt{A}dx^{1}=\frac{1}{\left(-2\lambda\right)}\stackrel[y_{b}]{y_{a}}{\int}\frac{\sqrt{A}}{x^{1}}dy=\frac{1}{\left(-2\lambda\right)}\stackrel[y_{b}]{y_{a}}{\int}\frac{\sqrt{A}}{\sqrt{\left(x^{0}\right)^{2}-\frac{1}{\left(-2\lambda\right)}y}}dy
\]
Where $y_{a}=y\left(x^{0},x_{a}^{1}\right)$ and $y_{b}=y\left(x^{0},x_{b}^{1}\right)$.
Specializing the metric to region~$\left(B\right)$ and region~$\left(D\right)$
($\Lambda-K$ space) 
\[
\Delta s=\frac{1}{\sqrt{-2\lambda}}\stackrel[y_{b}]{y_{a}}{\int}\frac{\sqrt{N\sqrt{B_{H}}\mathcal{B}_{,y}}}{\sqrt{\left(-2\lambda\right)\left(x^{0}\right)^{2}-y}}dy=\frac{\sqrt{2N}}{\sqrt{\kappa\rho_{\Lambda}}}\stackrel[\mathcal{B}_{b}]{\mathcal{B}_{a}}{\int}\frac{1}{\sqrt{\mathcal{B}_{,y}\left(\left(-2\lambda\right)\left(x^{0}\right)^{2}-y\right)}}d\mathcal{B}
\]
Where $\mathcal{B}_{a}=\mathcal{B}\left(y_{a}\right)$ and $\mathcal{B}_{b}=\mathcal{B}\left(y_{b}\right)$.
It was shown in subsections~\ref{subsec:MnkLK_constr}~and~\ref{subsec:modMnkLK_constr}
that the brane world-paths are of constant $B$ (and hence constant~$y$).
In all cases $\Psi>0$ and thus $\chi^{1}>0\Rightarrow\dot{\chi}^{0}>0$.
That is, the world-path of the brane in the $\Lambda-K$ space-time
is such that $\chi^{0},\chi^{1}\rightarrow+\infty$ such that $\mathcal{Y}=constant$
as $\tau\rightarrow\infty$. At late time $\left(-2\lambda\right)\left(x^{0}\right)^{2}\gg y$
and thus one has $\sqrt{\left(-2\lambda\right)\left(x^{0}\right)^{2}-y}\simeq\sqrt{-2\lambda}\left|x^{0}\right|$.
The integrand is a function of $\mathcal{B}$ alone. Thus when distance
measurement is done at late coordinate time, the result is $\frac{1}{\left|x^{0}\right|}$
times a constant that depends on $\mathcal{B}_{b}$, $\mathcal{B}_{a}$
and fixed parameters in the model. 
\[
\begin{array}{c}
\Delta s\simeq\frac{\sqrt{N\sqrt{B_{H}}}}{\left(-2\lambda\right)\left|x^{0}\right|}\stackrel[\mathcal{B}_{b}]{\mathcal{B}_{a}}{\int}\frac{1}{\sqrt{\left(\mathcal{B}\right)_{,y}}}d\mathcal{B}\\
\mathcal{B}_{,y}=\frac{1}{\left(N+1\right)}\exp\left(-\mathcal{H}_{N}\left(\mathcal{B}\right)\right)\left(\frac{\stackrel[k=0]{N}{\sum}\mathcal{B}^{k}}{\mathcal{B}^{N-1}}\right)\\
\mathcal{H}_{N}\left(\mathcal{B}\right)=\int\frac{-\mathcal{B}^{N-1}+\stackrel[k=0]{N-2}{\sum}\left(k+1\right)\mathcal{B}^{k}}{\stackrel[k=0]{N}{\sum}\mathcal{B}^{k}}d\mathcal{B}
\end{array}
\]
With such a definition of proper distance between the outer and inner
branes, the distance decreases as time advances. A similar result
is obtained if one instead defines distance between the outer and
inner branes as a line segment of constant~$x^{1}$, varying in the
$x^{0}$ direction. 

Because the branes in this model extend infinitely far in the parallel
space-like directions~$\overrightarrow{x}$, it would not be feasible
to construct. Thus it is not productive at this stage to give actual
numbers to energy densities, nor to compute the total mass-energy
required. The purpose of the models presented in this paper is to
demonstrate the mechanisms that would be involved in a more realistic
model. Such a more realistic model would have compact branes - finite
surface area ``brane-bubbles'' - enclosing an interior bulk. Region~$\left(B\right)$
and region~$\left(D\right)$ would be identified, as would branes~AB,~AD
and also branes~BC,~CD. The symmetry would become axial. The perfect
fluid on the branes might flow - perhaps rotationally. There has been
no attempt in this paper to be any more specific about the nature
of the stress-energy on the brane than to describe it as a perfect
fluid that exhibits translational and rotational symmetries in the
parallel directions. The method of confinement of that stress-energy
to the brane has not been considered, nor has stability to deformations
away from that symmetry. In a more realistic and precise model such
matters could not be overlooked. One possible nature to consider for
the brane is that of a compact topological defect. However, ideally
one would prefer a generalization that is simultaneously stable and
allows greater freedom in the selection of the equation of state. 

\subsubsection{Speculation Concerning the Possibility of Transition from Sub-Luminality
to Super-Luminality}

\begin{figure}
\centering{}\includegraphics[viewport=318bp 74bp 789bp 608bp,clip,scale=0.78]{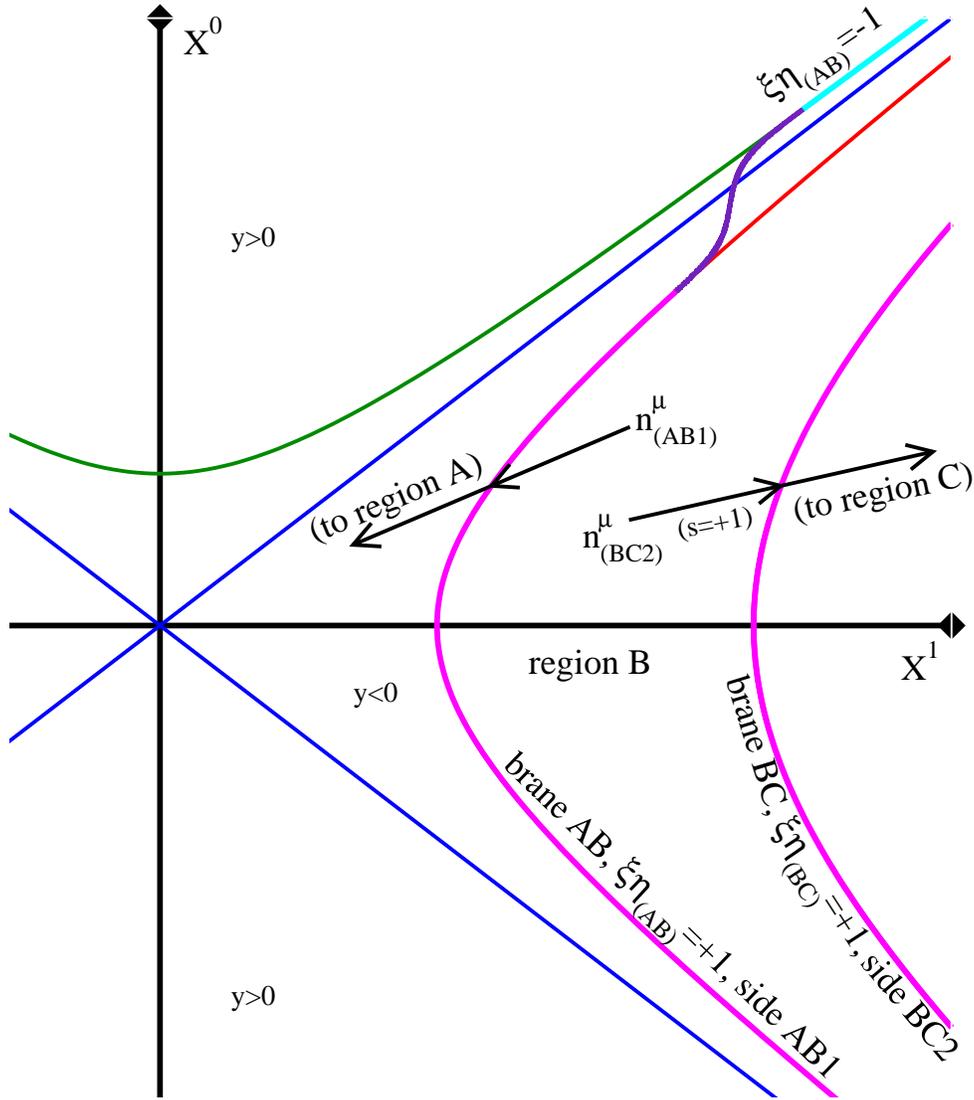}\caption{\label{fig:subLum_to_supLum_transition}Shown is a speculative representation
of a possible mechanism of transition from sub-luminal to super-luminal
warp. The world-path of brane~AB is time-like over the magenta segment
($B<B_{H}$ during this, where $B$ is the constant value along the
world-path of brane~AB) and is space-like over the cyan segment ($B>B_{H}$).
The transition happens over the violet segment. Note that the value
of $B$ along the entire trajectory of brane~AB must be constant
- because the Minkowski space of region~$\left(A\right)$ induces
a constant value of $B$ on that brane. What is speculated to occur
in the region of the violet segment is that the ratio~$\frac{-K}{\kappa\rho_{\Lambda}}$
decreases such that $B_{H}$ becomes less than the value of $B$ induced
on the brane. For convenience of display it is assumed - for simplicity
of this figure - that as $B_{H}$ decreases, $\lambda=-\frac{1}{4}\kappa\rho_{\Lambda}\sqrt{B_{H}}$
remains fixed - thus the $y$~value at a coordinate point~$\left(x^{0},x^{1}\right)$
remains fixed. $y=0$ and $B=B_{H}$ along the blue~$45^{\circ}$
lines. However, though $y=0$ is called a \textquotedblleft horizon\textquotedblright ,
this is merely the coordinate location where $B_{,z}=0$. This scenario
is extremely speculative because $K$ and $\kappa\rho_{\Lambda}$
are constants in all models considered here, and thus a possible variability
of those parameters is outside of the scope of this paper. }
\end{figure}

The similarity between the sub-luminal model of fig~\ref{fig:subLum_ConfDiag}
and the super-luminal model of fig~\ref{fig:supLum_ConfDiag} tempts
speculation concerning how the former might transition to the later.
Paths of constant~$y$ (which are paths of constant~$B$ when $B_{H}$
does vary) never cross the horizon at $B=B_{H}$. With the model parameters
fixed, the outer branes~AB~and~AD can not leave the $y=constant$
trajectories in $\Lambda-K$ space-time because the metric in the
outside Minkowski space is fixed, and the induced metric must match
across the brane. $B$ is fixed, hence $y$ and hence $A$. The next
question one might ask is if the $B$ on the world-path~AB1 is fixed
and thus can not change from less than $B_{H}$ to greater than~$B_{H}$,
can the $B_{H}$ be altered by changing parameters so that it is less
than the value of $B$ along world-path~AB1? Strictly speaking, not
within the frame-work of the models considered in this paper. This
is because the parameters whose ratio determines $B_{H}$ - $K$ and
$\kappa\rho_{\Lambda}$ - are constants. The Einstein field equations
in the bulk and the junction conditions were solved under the assumption
that these parameters do not vary as any of the coordinates~$x^{\mu}$
change. However, one could imagine the ratio $\frac{-K}{\kappa\rho_{\Lambda}}$
being changed sufficiently slowly in the hope that the resulting solution
of the Einstein equations in the $\Lambda-K$ bulk space-time would
be instantaneously closely approximated by the static solution of
subsection~\ref{subsubsec:LK_defn}. It is thus reasonable to speculate
that if one rewrote the metric of $\Lambda-K$ space-time such that
$\frac{-K}{\kappa\rho_{\Lambda}}$ became $\frac{-K}{\kappa\rho_{\Lambda}}\left(1-\epsilon\left(x^{0}-x_{start}^{0}\right)\right)$
with $\epsilon\ll1$ (and with the $\kappa\rho_{\Lambda}$ in the
bulk stress-energy tensor also slowly varying), one could solve the
Einstein field equations perturbatively, and thus obtain a result
where $B_{H}$ is slowly decreasing with coordinate time, causing
some trajectories of constant~$B$ close to the horizon to cross
the horizon (or, rather, the horizon to cross it). Such a scenario
is represented in fig~\ref{fig:subLum_to_supLum_transition}. Since
$\kappa\rho_{\Lambda}$ is presumably not the actual cosmological
vacuum energy (since the external space is being approximated as Minkowski,
and not Friedmann-Robertson-Walker-Lemaitre), it is merely effective
vacuum energy being generated in region~$\left(B\right)$ and region~$\left(D\right)$
by unspecified artificial physical processes, and is thus not truly
constant. Perhaps these same processes could very slowly change that
effective vacuum energy density, or its ratio to $K$ such as to cause
the $B_{H}$ to decrease to below the $B$ of the world-path~AB1. 

Recall from eq~\ref{eq:bulkLK_KScf_n2ly_soln} that $y=\left(\mathcal{B}-1\right)\exp\left(\mathcal{H}_{N}\left(\mathcal{B}\right)\right)$
where $\mathcal{B}\equiv\sqrt{\frac{B}{B_{H}}}$. Thus if $B_{H}$
decreases while $B$ along the world-path of the brain remains fixed,
$\mathcal{B}$ and hence $y$ will increase. For convenience of display
in fig~\ref{fig:subLum_to_supLum_transition} it is assumed that
$K$ and $\kappa\rho_{\Lambda}$ vary such that as $B_{H}$ decreases,
$\lambda=-\frac{1}{4}\kappa\rho_{\Lambda}\sqrt{B_{H}}$ remains constant,
and hence the value of $y$ (and hence~$\mathcal{B}$) at a given
point~$\left(x^{0},x^{1}\right)$ do not change. Thus, the $B=const$
trajectory of side~AB1 of brane~AB might look something like the
violet curve as $B_{H}$ decreases. When $B_{H}$ decreases below
$B$ on the brane, $\mathcal{B}$ increases to be greater than~$1$,
and $y$ transitions from negative to positive. Thus the world-path
of brane~AB on the $\Lambda-K$ side would cross to the other side
of the horizon. The ``horizon'' remains at $y=0$, the location
of the blue lines in fig~\ref{fig:subLum_to_supLum_transition}.
However, in this speculative example, the physical meaning of ``horizon
at $y=0$'' becomes simply the location at which~$B_{,z}=0$ - an
``instantaneous horizon'' in an intuitive sense. 

Note that eq~\ref{eq:Mdl_supLum_BB_div_BH} would not be valid in
a scenario such as this because the path of constant induced~$B$
is no longer a path of constant~$y$, as was assumed in the derivation
of that equation. The value of $B$ at the boundary is held fixed
via the junction conditions to match the external Minkowski space
(region~$\left(A\right)$). Thus $\mathcal{Y}\neq constant$ and
eq~\ref{eq:MnkLK_Bsoln} would need to be modified to reflect this
slow drift. If world-path~AB1 develops a $\xi_{\left(AB1\right)}=-1$,
and since $\eta_{\left(AB1\right)}=\eta_{\left(AB2\right)}=+1$ would
be maintained, then necessarily $\xi_{\left(AB2\right)}=-1$ also.
That is, because $\xi\eta$ must be the same for each side of the
brane, and $\eta=+1$ on the Minkowski space-time (external) side,
the trajectory of the brane in external Minkowski space-time must
become space-like - the model would transition from sub-luminal to
super-luminal. To transition in reverse it would be natural to imagine
the slow evolution of $\frac{-K}{\kappa\rho_{\Lambda}}$ could be
reversed. However, the presence of the horizon between the outer and
inner branes could complicate the matter on a technical level. Indeed,
the most difficult engineering challenge might not be achieving super-luminal
travel - it might be the return to sub-luminality.

\section{Conclusions: The Good, The Bad and The Future Work\label{sec:Conclusions}}

The primary achievement of the model presented in subsection~\ref{subsec:Mdl_supLum}
is that it achieves effective global super-luminal trajectory (with
respect to an external observer) while maintaining a local time-like
trajectory for a passenger - and while everywhere satisfying the weak
energy condition. Additionally, everywhere the equation of state of
stress-energy falls in the range~$\left[-1,1\right]$. One might
object to this claim because the $\Lambda-K$ space-time incorporated
in the models of subsections~\ref{subsec:Mdl_subLum}~and~\ref{subsec:Mdl_supLum}
uses an integration constant $K<0$, and it is an analogous integration
constant that is interpreted as the central mass in the Schwarzschild
solution. Thus one might argue that these models contain an 'effective
negative mass-energy'. However, this argument would not be correct
- note that the actual location of the curvature singularity (where~$B=0$)
does not appear in the resulting chimeric space-time. In general,
one can use a space-time solution that would normally contain an undesirable
feature, but not use the part of that space-time where that feature
is actually located - cutting that region out and discarding it. The
question remains if the boundary conditions necessary to initially
create such a space-time would need to contain a similar undesirable
feature. Never the less, the intent of this paper was not to present
a completely realistic model - but instead to present a toy model
as a proof-of-concept - that a chimeric space-time could be used to
construct a super-luminal warp-drive without the need to violate the
weak energy condition. Now that is has been shown to be true, it seems
likely that the same should be possible for a model with compact boundaries.
Another benefit of this approach - model-building with chimeric space-times
- is that the inner workings of the models are simple and well-enough
understood that one can predict the effects of the various components
in a straightforward manner. For example, components such as the $\Lambda-K$
horizon and the Modified-Minkowski space were included for specific,
foreseeable reasons - those reasons are clear and their effects are
understood. Insight gleaned from this approach suggests - as previously
explained - that the presence of a horizon might be unavoidable with
any model of super-luminal warp (this agrees with the results of previous
researchers~\cite{bib:Krasnikov_1998,bib:Low_1999,bib:Clark-Hiscock-Larson_1999}).
Certainly the presence of a horizon has the potential to be a troublesome
factor from an engineering viewpoint. However, future-pointing null-vectors
can still reach the the outer branes from the inner - so perhaps 'one-way'
control would be possible. Or, perhaps the parameters of the external
world-path could be pre-planned and pre-set, allowing a trip to be
conducted without the passenger receiving light-signals from the external
Minkowski space-time during that trip. 

Another approach considered in the literature~\cite{bib:Marolf_Yaida_EnCond_and_JuncConds_2005}
involves a speculation that negative energy density might allow the
induced metric to be discontinuous across the boundary. In the models
considered in this paper, the continuity of the induced metric across
the boundary forces~$\xi\eta$ to be the same on either side, thus
preventing the world-path of a brane from being space-like on one
side, and time-like on the other side. However, at the cost of invoking
negative energy density on the brane, if true, this conjecture might
allow~$\xi\eta=+1$ on one side and $\xi\eta=-1$ on the other. It
would be as if one removed the $\Lambda-K$ bulk from the model of
subsection~\ref{subsec:Mdl_supLum} and identified the inner and
outer branes into a single brane. Thus the horizon between those branes
would be removed from the model. Imagine brane~AB identified with
brane~BC in fig~\ref{fig:supLum_ConfDiag}. Thus, perhaps negative
energy density on the brane might allow the removal of the horizon
from the super-luminal warp model. 

Another advantageous feature of the models presented in subsections~\ref{subsec:Mdl_subLum}~and~\ref{subsec:Mdl_supLum}
is that in some formulas the dependence is on a ratio of energy densities
- ie: $\frac{\kappa\rho_{\Lambda}}{\left(\kappa\rho\right)^{2}}$
- instead of a ratio of one energy density to the Planck density.
For considerations involving such a ratio, there is no direct requirement
for large energy densities, such as at the Planck scale. While it
is true that the overall magnitude of the acceleration of the model
in external Minkowski space-time is set by $\kappa\rho$ on the brane,
this will be adjusted by an unseen unit-full factor originating from
the delta-function support of the brane stress-energy tensor. However,
there is little reason to attempt to determine a total energy requirement
for such a model until a realistic version of the model is designed
(that is, compact boundaries). 

The models presented in this paper are not entirely realistic because
the branes extend infinitely far in the parallel directions. In a
realistic model the branes would be compact - perhaps shaped something
like an ellipsoid in the external space. One might solve the Einstein
field equations for a $\Lambda-K$ space-time in an axially-symmetric
bulk, then compute and match both the intrinsic metric and the extrinsic
curvature (with a discontinuity caused by a brane) over the surface
of a deformed ellipsoid, containing the origin of that coordinate
system. One would have variation in three directions instead of two,
and the junction conditions would need to allow for a component of
world-path of a point on this surface to curve in a direction parallel
to the boundary surface. This is a topic of future work. A related
shortcoming of the models in subsections~\ref{subsec:Mdl_subLum}~and~\ref{subsec:Mdl_supLum}
is that these are steady-state solutions only (though a possible mechanism
for transitioning between sub-luminal and super-luminal warp is considered).
What is lacking is the specification of how such chimeric space-times
could be 'initialized'. That is, how one might begin with nothing
more than standard Minkowski (or other available one, such as Schwarzschild)
space-time, and then by manipulating a stress-energy tensor - both
in the bulk, and creating a membrane containing stress-energy with
delta-function support - could cause the model described in subsection~\ref{subsec:Mdl_subLum}
to arise. Correspondingly, how would one terminate such a model, returning
to standard Minkowski space-time? A feasible sub-luminal warp-drive
must have a means to be turned on and off. Other important topics
include a stability analysis of this model, and the role quantum effects
might play (no quantum effects were considered in this paper). These
shortcomings will be the subject of future work. 

In addition to satisfying the weak energy condition, the model of
subsection~\ref{subsec:Mdl_supLum} does not possess any super-luminal
flows in the parallel directions. It is true that the metric induced
on the outer branes is Euclidean in nature - having no direction with
a different sign in the metric than that of the spacial directions.
This is because the hypersurface is space-like - the time direction
is projected out when the external bulk metric is induced upon the
boundary. There is a frame in which an external observer would see
the world-path of the external brane as simultaneous - a constant-time
hypersurface. At first this might seem odd, but it is nothing more
than the result of a super-luminal boost. 

The $\Lambda-K$ bulk space-time was used in this paper because it
allowed the presence of a horizon without negative energy density
(ie:~$\kappa\rho_{\Lambda}>0$ and $K<0$). It could be replaced
by another space-time that allows for space-like and time-like $B=constant$
trajectories to coexist and not intersect. Another avenue of future
work that might prove interesting would be to replace the external
space-time with a Schwarzschild solution, and investigate if and under
what circumstances the resulting warp models might be able to exit
the event horizon. 

One extremely speculative consideration concerns the Resonant Cavity
Thruster (EM-Drive)~\cite{bib:RCT_EW_2017}. If this is found to
represent a real, physical effect, could a model based on the ideas
presented in this paper explain it's function, in tandem with dynamical
Casimir effects from the asymmetrical shape of the region boundaries?
The interior and exterior of the cavity would be modeled as two different
space-time bulk regions with the cavity walls playing the role of
a brane. In such a model momentum conservation would hold - the apparent
propulsive force would be akin to the propulsion mechanism suggested
in this paper - a geometric effect within the context of General Relativity.
Intuitively, one might expect such an effect to be too small to be
readily measured, and also be swamped by other effects. Never the
less, if the Resonant Cavity Thruster proves to be a real effect,
such an approach to understanding it might prove a fruitful avenue
of investigation. 

The results of this paper are consistent with those of the soliton
wave model of Erik~Lentz~\cite{bib:Lentz_soliton_SUPLUM_WEC} concerning
super-luminal warp not necessarily violating the Weak Energy Condition
within the context of classical General Relativity.
\begin{lyxlist}{00.00.0000}
\item [{}]~
\end{lyxlist}


\begin{thebibliography}{10}
\bibitem{bib:Alcub_1994}Alcubierre, M. : ``The Warp Drive: Hyper-fast
Travel within General Relativity'', Classical and Quantum Gravity,
11 (1994), L73-L77 

\bibitem{bib:Everett_1996}Everett, A. E. : ``Warp Drive and Causality'',
Phys. Rev. D, 53 (1996), 7365

\bibitem{bib:Pfenning-Ford_19967}Pfenning, M. J., Ford, L. H. : ``The
unphysical nature of `warp drive'{}'', Class. Quantum Grav., 14 (1997),
1743

\bibitem{bib:Olum_1998}Olum, K. : \textquotedblleft Superluminal
Travel Requires Negative Energy Density\textquotedblright , Phys.
Rev. Lett, 81 (1998), 3567-3570 

\bibitem{bib:Krasnikov_1998}Krasnikov, S. V., : ``Hyperfast Interstellar
Travel in General Relativity'', Phys. Rev. D, 57 (1998), 4760-4766

\bibitem{bib:Low_1999}Low, R. : ``Speed Limits in General Relativity'',
Class. Quant. Grav., 16 (1999), 543-54

\bibitem{bib:Clark-Hiscock-Larson_1999}Clark, C., Hiscock, W., Larson,
S. : ``Null Geodesics in the Alcubierre Warp Drive Spacetime: the
View from the Bridge'', Class. Quant. Grav., 16 (1999) 3965-3972

\bibitem{bib:VDBroeck_1999}Van Den Broeck, C. ``A `warp drive' with
more reasonable total energy requirements'', Class. Quantum Grav.,
16 (1999), 3973 

\bibitem{bib:Natario_2002}Natario, J. : ``Warp Drive With Zero Expansion'',
Class. Quant. Grav., 19 (2002), 1157-1166 

\bibitem{bib:Lobo-Crawford_2003}Lobo, F., Crawford, P. : ``Weak
Energy Condition Violation and Superluminal Travel'', Lect. Notes
Phys., 617 (2003), 277-291

\bibitem{bib:Lobo_2007}Lobo, F. S. N. : ``Exotic solutions in General
Relativity: Traversable wormholes and 'warp drive' spacetimes'',
Classical and Quantum Gravity Research, 1-78, (2008), Nova Sci. Pub.
ISBN 978-1-60456-366-5

\bibitem{bib:White-WFM101_2013}White, H. : ``Warp Field Mechanics
101'', Journal of the British Interplanetary Society, 66 (2013),
242-247

\bibitem{bib:Alcub_Lobo_2017}Alcubierre, M., Lobo, F. S. N. : ``Warp
Drive Basics'', Fundam. Theor. Phys., 189 (2017), 257-279

\bibitem{bib:DeBenedictis-Iliji=000107_2018}DeBenedictis, A., Iliji\'{c},
S. : ``Energy Condition Respecting Warp Drives: the Role of Spin
in Einstein--Cartan Theory'', Class. Quantum Grav., 35 (2018), 215001 

\bibitem{bib:bobrick_martire_subLum_WEC}Alexey Bobrick, Gianni Martire
: ``Introducing Physical Warp Drives'', Class. Quantum Grav., 38
(2021), 105009 

\bibitem{bib:Smolyaninov_Metamaterials}Smolyaninov, I. : ``Metamaterial-based
model of the Alcubierre warp drive'', Phys. Rev. B, 84 (2011) 113103 

\bibitem{bib:Drouhet_ExoticFluids}Béatrix-Drouhet, W. : ``Exotic
Fluids Matching the Stress-Energy Tensor of Alcubierre Warp Drive
Spacetimes'', arXiv:2012.09941 {[}gr-qc{]} 

\bibitem{bib:AbellanBolivarVasilev_AnisotropicMatter}Abellan, G.,
Bolivar, N., Vasilev, I., : ``Influence of anisotropic matter on
the Alcubierre metric and other related metrics: revisiting the problem
of negative energy'', General Relativity and Gravitation (2023) 55:60 

\bibitem{bib:Lentz_soliton_SUPLUM_WEC}Lentz, E. : ``Breaking the
Warp Barrier: Hyper-Fast Solitons in Einstein-Maxwell Plasma Theory'',
Class. Quantum Grav., 38 (2021), 075015

\bibitem{bib:IJC}Israel, W.\emph{ }:\emph{ }``Singular hypersurfaces
and thin shells in general relativity'', Nuovo Cimento B, 44 (1966),
1--14

\bibitem{bib:DJC}Darmois, G. : ``Mémorial des sciences mathématiques'',
Fascicule XXV (1927), Gauthier-Villars, Paris 

\bibitem{bib:Poisson_Book_Relativists_Toolkit}Poisson, E.: ``A Relativist's
Toolkit'', Cambridge University Press 2004, ISBN~978-0-521-83091-1

\bibitem{bib:Rindler_Relativity} Rindler, W. : ``Relativity: Special,
General and Cosmological, 2nd ed'', Oxford University Press 2006,
ISBN 0--19--856731--6 978--0--19--856731--8

\bibitem{bib:schw_soln}Schwarzschild, K.: ``On the Gravitational
Field of a Mass Point According to Einstein's Theory'', Sitzungsberichte
der Königlich Preussischen Akademie der Wissenschaften zu Berlin,
Phys.-Math. Klasse, 189-196 (1916)

\bibitem{bib:kruskal}Kruskal, M. D. : ``Maximal Extension of Schwarzschild
Metric'', Phys. Rev., 119 (1960), 1743

\bibitem{bib:Szekeres}Szekeres, G. : ``On the Singularities of a
Riemannian Manifold'', Publ. Math. Debrecen, 7 (1960), 285-301 

\bibitem{bib:RCT_EW_2017}White, H., March, P., Lawrence, J., Vera,
J., Sylvester, A., Brady, D., Bailey, P. : ``Measurement of Impulsive
Thrust from a Closed Radio-Frequency Cavity in Vacuum'', Journal
of Propulsion and Power, 33:4 (2017), 830-841 

\bibitem{bib:Marolf_Yaida_EnCond_and_JuncConds_2005}Marolf, D., Yaida,
S. : ''Energy Conditions and Junction Conditions'', Phys. Rev. D,
72 (2005), 044016
\end{thebibliography}
\end{document}